\documentclass[10pt,aps,prd,onecolumn,eqsecnum,amsmath,amssymb,nofootinbib,superscriptaddress]{revtex4-2}

\usepackage{xcolor}
\usepackage{mathtools}		
\usepackage{mathrsfs}
\usepackage{stmaryrd}
\SetSymbolFont{stmry}{bold}{U}{stmry}{m}{n}
\usepackage{longtable}
\usepackage{array} 
\usepackage[colorlinks=true,
            citecolor=red,
            linkcolor=blue,
            urlcolor=violet,
            filecolor=cyan,
            backref=false]{hyperref}
\definecolor{RED}{rgb}{1,0,0} 
\definecolor{BLUE}{rgb}{0,0,1} 

\newcommand{\lie}{\pounds}
\newcommand\bs{\boldsymbol}

\renewcommand{\sinh}{\operatorname{sh}}

\newcommand\feq{\mathrel{\phantom{=}}}

\begin{document}
\title{Quasi-topological gravity for 4-dimensional Taub--NUT, near-horizon extreme Kerr, and swirling symmetries}

\author{Aimeric Coll\'{e}aux}
\email{aimeric.colleaux@matfyz.cuni.cz}
\affiliation{Institute of Theoretical Physics, Faculty of Mathematics and Physics, Charles University, V Hole\v{s}ovi\v{c}k\'ach 2, Prague 180 00, Czech Republic}

\author{Ivan Kol\'a\v{r}}
\email{ivan.kolar@matfyz.cuni.cz}
\affiliation{Institute of Theoretical Physics, Faculty of Mathematics and Physics, Charles University, V Hole\v{s}ovi\v{c}k\'ach 2, Prague 180 00, Czech Republic}

\author{Tom\'a\v{s} M\'alek}
\email{malek@math.cas.cz}
\affiliation{Institute of Mathematics of the Czech Academy of Sciences, \v{Z}itn\'a 25, 115 67 Prague 1, Czech Republic}

\date{\today}

\begin{abstract}
We classify 4-dimensional gravitational theories with integrability properties analogous to quasi-topological gravity, but for metrics with the symmetries of spherical, hyperbolic, and planar Schwarzschild and Taub--NUT solutions, their double-Wick-rotated counterparts --- the B-metrics, the near-horizon extreme Kerr, and the swirling universe --- and the Eguchi--Hanson instanton. These are the symmetries that allow consistent reductions (principle of symmetric criticality) with 4 Killing vectors and 3-dimensional orbits. Considering theories depending only on the Riemann tensor, we show that, for these metrics, only those with third-order equations (second-order after trivial integration) can be analytic in the Riemann tensor. We show that there is a unique theory with first-order field equations (algebraic after trivial integration, with the same integrability as general relativity) at each order in curvature and construct regular static black holes from infinite towers of these high-energy corrections to general relativity. For these theories, we obtain closed-form solutions for all the symmetries listed above, which we analyze to ensure they have a clear physical interpretation.
\end{abstract}

\maketitle

\section{Introduction}

Quasi-topological gravities (QTG) can be defined as generally covariant metric theories, such that, in static spherical, hyperbolic, or planar (s./h./p.) symmetry, imposing the single-function (SF) ansatz, ${g_{tt} g_{rr} = -1}$  (in Schwarzschild coordinates), identically solves one of the two independent field equations. It then follows that the leftover field equation is at most of the third-order in derivatives of the metric and can be integrated once. Thus, they share the same type of integrability as general relativity (GR) for geometries with static s./h./p. symmetry.

Such theories have been studied in higher dimensions \cite{Oliva:2010eb,Myers:2010ru,Dehghani:2011vu,Oliva:2010zd,Cisterna:2017umf,Hennigar:2017ego,Bueno:2019ycr,Bueno:2019ltp,Bueno:2022res,Moreno:2023rfl,Bueno:2025qjk}. Most notably, a peculiar series of curvature invariants, polynomial in the Riemann tensor, has been shown to admit first-order and integrable field equation in static spherical symmetry and satisfy Birkhoff theorem in the time-dependent case, just like Lovelock-Lanczos gravity \cite{Maeda:2011ii}. Considering infinite series of such invariants, 5D regular black holes have been obtained \cite{Bueno:2024dgm}, and their formation have been studied in \cite{Bueno:2025gjg,Bueno:2024zsx}, see also \cite{Bueno:2025tli,Bueno:2026dln}.  

In four dimensions, theories with this property which are polynomial of the Riemann tensor yield instead third-order equations of motion and have been studied in \cite{Hennigar:2017ego,Bueno:2019ycr,Bueno:2019ltp,Bueno:2022res,Moreno:2023rfl}, where they were referred to as generalized QTG, or Einsteinian cubic gravity \cite{Bueno:2016xff} at the lowest order in the curvature. However, it is known that considering 4-dimensional non-analytic theories in the Riemann tensor (and possibly its derivatives) enables to obtain similar results as in the higher-dimensional case, such as second-order cosmological and static spherically symmetric equations \cite{Gao:2012fd,Deser:2007za,Colleaux:2015yta,Chinaglia:2017wim,Bellini:2010ar}. Furthermore, 4-dimensional QTG admitting Birkhoff theorem, first-order field equations in dynamical spherically symmetric spacetimes, as well as regular black hole and cosmological solutions, have been obtained in \cite{Colleaux:2017ibe} and \cite{Colleaux:2019ckh} and represent, as far as we know, the first examples of such exact regular solutions in pure gravity in any dimension $d\geq 4$.\footnote{Concerning $d\geq 4$ theories with these properties, reducing to GR-like equations or Lovelock-like generalised Wheeler-polynomial, well-known to yield regular black holes \cite{Kunstatter:2015raa,Kunstatter:2015vxa}, see \cite{Colleaux:2017ibe} and Chap.~II.C., III.A. and V. of \cite{Colleaux:2019ckh}. Remark that the same 2D Horndeski theory has recently been used in \cite{Bueno:2024eig} to study the formation of regular black holes and also appear in higher-dimensional polynomial QTG \cite{Bueno:2024dgm}.  More generally, it was shown in \cite{Colleaux:2017ibe} and Chap.~II.B.2. of \cite{Colleaux:2019ckh} that (almost) any 2-dimensional scalar-tensor theories can be seen as the dynamical spherically symmetric reductions of some non-analytic four (or higher) dimensional metric theories. This correspondence has been rediscovered recently in \cite{Borissova:2026krh}.} This non-polynomial approach to regular solutions has recently attracted some attention, following \cite{Frolov:2024hhe}, where it was shown that 4-dimensional theories depending (still non-analytically) on the Riemann tensor only (without its derivatives) is sufficient to obtain QTG admitting regular black holes, see also \cite{Bueno:2025zaj,Borissova:2026wmn,Borissova:2026klg} for applications in the context of regular black hole formation and regular cosmologies.

Given these successes, which follow from the s./h./p. symmetric SF ansatz, it is natural to wonder what are the most general 4-dimensional metric theories whose field equations are at most third-order for the largest possible proper subset of metric fields, and which share some integrability property with GR.\footnote{Otherwise, it has been shown by Lovelock \cite{Lovelock:1969vyr} (see also \cite{Crisostomi:2017ugk}) that the most general 4-dimensional polynomial theory yielding at most third-order field equations for all metric fields is given (up to boundary terms) by ${
\mathcal{L} = - 2 \Lambda + R + \alpha \star R_{\mu\nu}{}^{\rho\sigma}\star R_{\alpha\beta}{}^{\mu\nu}\star R_{\rho\sigma}{}^{\alpha\beta}}$, where $\star$ denotes the Hodge dual. This theory is a ``topological'' QTG in static spherical symmetry.} In this paper, we will commonly refer to such theories as QTG, irrespectively of the specific subsets of metric fields for which such integrability is preserved. Nevertheless, we will explicitly specify for which subsets said theory is a QTG.\footnote{As it is quite clear, the larger the subset of metric fields is, the lesser number of QTG theories there is.} Also, we will only occasionally use the word ``generalized'' when the order cannot be lower than third.

It was recently shown that generalized QTG for Taub--NUT geometries which are cubic and quartic in the Riemann tensor can be obtained and satisfy the SF ansatz (in the coordinates with the volume being proportional to ${r^2+n^2}$, like in the case of GR) \cite{Bueno:2018uoy},\footnote{Rotating near-horizon extreme geometries, related to hyperbolic Taub--NUT by a double Wick rotation \cite{Colleaux:2025uiw}, were also studied in Einsteinian cubic gravity \cite{Cano:2019ozf}.} providing the first examples of QTG property beyond (dynamical) s./h./p. symmetry. While the cubic one is simply Einsteinian cubic gravity, the quartic one differs from the quartic generalized QTG (adapted to s./h./p. symmetry) obtained in \cite{Moreno:2023rfl}, by curvature invariants whose field equations trivialize in static spherical symmetry when the SF ansatz is imposed. We will naturally call the latter theories ``topological" QTG, although it should be noted that they can have non-trivial field equations when SF ansatz is not imposed. Notice also that Taub--NUT solutions have been obtained in Lovelock gravity \cite{Dehghani:2005zm,Hendi:2008wq,Dehghani:2006aa,Corral:2019leh,Corral:2025yvr}, which is the simplest class QTG theories in higher dimensions.

\medskip

In this paper, we classify all 4-dimensional metric theories depending solely on the Riemann tensor (both polynomially or non-analytically), which are QTG for all geometries with the symmetries of (s./h./p.) Taub--NUT spacetime \cite{Taub:1950ez,Newman:1963yy,Griffiths:2009dfa}, near-horizon-extreme Kerr (NHEK) geometry \cite{Bardeen:1999px,Kunduri:2007vf}, the recently popular swirling universe \cite{Gibbons:2013yq,Astorino:2022aam}, or the Eguchi--Hanson instanton \cite{Eguchi:1978xp}. A fortiori, they are also QTG for all geometries with the symmetries of the  AI/AII/AIII metrics (static s./h./p. symmetry) as well as BI/BII/BIII-metrics \cite{EhlersKundt1962,Gott1974} (or Melvin spacetime \cite{Bonnor:1954tis,Melvin:1963qx}), which are obtained for vanishing NUT parameter.

This large set of metrics for which the QTG property holds can be understood from the general relations obtained in \cite{Colleaux:2025uiw}, consisting of mapping the isometries of these geometries into each other via double Wick rotations\footnote{Concerning the effect of double Wick rotations on field theories, and in particular Lorentz-violating ones, see \cite{Gotzberger:2026ujl}.}. In particular, this implies for our purpose that any theory which is a QTG for geometries with the symmetries of Taub--NUT, must also be so for NHEK and swirling symmetries. Given these close relations, we will commonly refer to these classes of metrics as \textit{Taub--NUT-type (TNT) geometries} and the corresponding QTG theories will be shorthanded as QTG-TNT.

Given the complexity of non-analytic Lagrangians in the Riemann tensor, we make extensive use of the principle of symmetric criticality (PSC) assessing under which conditions (on the spacetime symmetries) the field equations of the symmetry-reduced theory are equivalent to the symmetry-reduction of the complete field equations \cite{Fels:2001rv}. Surprisingly, it works for all TNT geometries \cite{Frausto:2024egp}.

The QTG-TNT we obtain can be organized according to the order of their field equations for TNT geometries. For instance, topological QTG-TNT, GR, and Einsteinian cubic gravity correspond to zeroth, first, and third order, respectively. Among these, we show that only those with the third-order field equations (second-order after one trivial integration) can be obtained from analytic (hence polynomial) actions in the Riemann tensor, while all the others must be non-analytic, like the (larger set of) theories considered in \cite{Frolov:2024hhe} (which are QTG for the restricted set of static spherically symmetric metrics). We conjecture that there is a unique QTG-TNT theory polynomial in the Riemann tensor at each order  of curvature. This would generalize the result of \cite{Moreno:2023rfl} obtained in static spherical symmetry.

Restricting to theories with first-order field equations (always for TNT geometries), we show that considering infinite towers of non-polynomial curvature invariants allows to construct theories with regular 4-dimensional static black holes. In these, we find exact closed-form solutions corresponding to the Taub--NUT-, NHEK-,  swirling-, as well as Eguchi--Hanson-type geometries.

\medskip

The paper is organized as follows: In Sec.~\ref{sc:TNTgeom}, we review various properties of geometries with the symmetries of s./h./p. Taub--NUT as well as its NHEK and swirling counterparts --- including their limit to static s./h./p. symmetries and BI/BII/BIII-metric counterparts. In particular, we describe their relations via double Wick rotations and discuss the regularity conditions and their relations. In Sec.~\ref{sc:effectiveAction}, we make use of the Zakhary--McIntosh invariants to identify invariant representatives\footnote{Given that many inequivalent curvature invariants become dependent for algebraically special metric configurations, there is an arbitrariness, or degeneracy in the choice of the curvature invariants used in any action designed to satisfy some specific properties (such as being a QTG) for such geometries (see chap.II.B.2. of \cite{Colleaux:2019ckh} for more details). Furthermore, as explained above, there exist some topological QTG, what introduces even more arbitrariness in the choice of the action.} of the independent components of the Riemann tensor for TNT geometries. This enables us to construct the most general covariant 4-dimensional gravitational effective action polynomial in these components for TNT geometries. In Sec.~\ref{sc:QTGprop}, we introduce the SF ansatz for TNT metrics and formulate the corresponding QTG-TNT property. In Sec.~\ref{sc:classQTG}, we classify 4-dimensional QTG-TNT obtained from the previous action which yields zeroth-, first-, second-, and third-order field equations for TNT geometries for SF ansatz. We show that among these, only the last ones can be expressed as polynomials of the Riemann tensor. In Sec.~\ref{sc:solAlgTh}, we focus on the unique class of theories with the first-order field equations (algebraic after one trivial integration --- the GR integrability) and obtain various static regular black hole solutions and other exact closed-form TNT solutions (including B-metric, Taub--NUT, NHEK, and swirling spacetimes). Finally, App.~\ref{ap:CurvOfTNT} gathers some material regarding the NP formalism for TNT geometries. App.~\ref{ap:ProofNAN} contains a proof of the non-analyticity of first- and second-order QTG-TNT. Consequently, any analytic QTG-TNT is necessarily of third order in derivatives; their covariant polynomial representatives are constructed in App.~\ref{ap:covrep}.

\paragraph*{Notation} We use the mostly-plus signature, i.e., ${(-,+,+,+)}$, for the Lorentzian metric and the following conventions for curvature: ${[\nabla_\mu,\nabla_\nu] v^\rho = R_{\mu\nu\sigma}{}^\rho\, v^\sigma}$ for the Riemann tensor, and ${R_{\mu\nu} = R_{\mu\rho\nu}{}^\rho}$ for the Ricci tensor. Tensors with suppressed indices are written in \textbf{boldface}. The symbol $\bs{\mathrm{d}}$ stands for the exterior derivative, while $\lie$ denotes the Lie derivative; $\vee$ and $\wedge$ indicate the symmetric and antisymmetric (exterior) products, respectively, with ${\bs{\alpha}\vee\bs{\beta} = \bs{\alpha}\bs{\beta} + \bs{\beta}\bs{\alpha}}$ and ${\bs{\alpha}\wedge\bs{\beta} = \bs{\alpha}\bs{\beta} - \bs{\beta}\bs{\alpha}}$ for 1-forms $\bs{\alpha}$ and $\bs{\beta}$.

\section{Taub--NUT-type geometries}\label{sc:TNTgeom}

In this section, we first present the most general s./h./p. Taub--NUT ansatz with some of its properties, as well as its limit leading to the general static metric s./h./p. symmetries. We then briefly summarize the results of \cite{Colleaux:2025uiw} in which is given a detailed account of its relation via double Wick rotations to further geometries enjoying the symmetries of, for example, the following GR solutions: BI/BII/BIII-metric (or Melvin spacetime), NHEK geometry, and swirling universe. We also comment on relations to symmetries of Bianchi class A cosmologies and Eguchi--Hanson instanton. As being closely related to the Taub--NUT, we refer to all spacetimes with such symmetries as the TNT geometries.

As we will be primarily interested in theories admitting regular solutions with physically interesting interpretation, we discuss several geometric properties the metrics with above symmetries should/could satisfy and their relation under the double Wick rotation. For example, it is interesting to analyze how the usual conditions on regular black-hole solutions [regular (A)dS cores, horizons, and Schwarzschild--(A)dS behavior at infinity] translate to conditions on all the other TNT geometries.

\subsection{Symmetries of Taub--NUT/Schwarzschild}

Let us consider the infinitesimal group action given by the following vector fields:
\begin{equation}\label{eq:iga}
\begin{aligned}
    \bs{X}_1 &= \sqrt{1-k\rho ^2} \left[\cos \varphi\bs{\partial}_\rho-\tfrac{  \sin \varphi}{\rho }\bs{\partial}_\varphi\right]-2n\tfrac{1-\sqrt{1-k\rho ^2} }{k}\tfrac{\sin \varphi}{\rho}\bs{\partial}_t\;, 
    \\ 
    \bs{X}_2 &=\sqrt{1-k\rho ^2}\left[ \sin \varphi\bs{\partial}_\rho+\tfrac{\cos \varphi }{\rho }\bs{\partial}_{\varphi}\right]+2n\tfrac{1-\sqrt{1-k\rho ^2} }{k}\tfrac{\cos \varphi}{\rho}\bs{\partial}_t\;, 
    \\
    \bs{X}_3 &=\bs{\partial}_{\varphi}\;,
    \\
    \bs{X}_4 &=\bs{\partial}_{t}\;,
\end{aligned}
\end{equation}
where ${k=\pm 1,0}$, ${n\in\mathbb{R}}$, and ${(t,r,\rho,\varphi)}$ are some coordinates on the manifold (more standard coordinates: ${\rho=\sin\vartheta}$ for ${k=1}$, and ${\rho=\sinh\vartheta}$ for ${k=-1}$). The case ${k=0}$ is meant in the limiting sense ${k\to0}$, where ${(1-\sqrt{1-k\rho ^2} )/k\to\rho^2/2}$.\footnote{In the notation from \cite{Hicks:thesis,Frausto:2024egp,Colleaux:2025uiw}, these infinitesimal group actions are together denoted by [4,3,\{1--6\}]. They belong to a larger class [4,3,-], which admit exactly 4 independent Killing vectors and have 3-dimensional orbits.}

The most general symmetry-invariant metric, ${\lie_{\bs{X}_i}\bs{g}=0}$, $\forall i$, i.e., the metric of which \eqref{eq:iga} is the Lie algebra of Killing vectors, can be written in the form
\begin{equation}\label{eq:ansatz}
  \bs{g} = - a(r) b(r)\left(\bs{\mathrm{d}}t +2n \bs{\omega}_k \right)^2   + \frac{\bs{\mathrm{d}}r^2}{a(r)} + c(r) \bs{q}_{k} +d(r)\left(\bs{\mathrm{d}}t + 2n \bs{\omega}_{k} \right)\vee\bs{\mathrm{d}}r\;,
\end{equation}
where $\bs{q}_{k}$ is the 2-dimensional Euclidean metric of constant curvature,
\begin{equation}
    \bs{q}_{k} = \frac{\bs{\mathrm{d}}\rho^2}{1-k \rho^2} + \rho^2 \bs{\mathrm{d}}\varphi^2 \;,
\end{equation}
and $\bs{\omega}_{k}$ is a 1-form given by
\begin{equation}
    \bs{\omega}_{k} = \frac{1 - \sqrt{1-k \rho^2}}{k} \bs{\mathrm{d}}\varphi \;.
\end{equation}
The metric has the Lorentzian signature whenever ${c>0}$ and ${\gamma>0}$, where ${\gamma = b + d^2}$.

Notice that the general symmetry-invariant metric is parametrized by four single-variable functions, which, however, can be gauge-fixed to two single-variable functions. Indeed, the gauge freedom of the metric ansatz \eqref{eq:ansatz} is captured by the residual diffeomorphism subgroup, which maps $\bs{g}$ to $\Phi_{\tau}^*\bs{g}$ while preserving the form of \eqref{eq:ansatz}. The generators of the flow $\Phi_{\tau}$ are vector fields $\bs{\mathcal{W}}$ satisfying ${[\bs{\mathcal{W}},\bs{X}_i]=\sum_{j=1}^{4} a_{ij} \bs{X}_j}$, ${\forall i=1,\dots,4}$,
for some constants $a_{ij}$. The non-Killing ones, which induce non-trivial transformations of $a$, $b$, $c$, $d$, are given by \cite{Colleaux:2025uiw}:
\begin{equation}\label{eq:resdifgen}
\begin{aligned}
    \bs{\mathcal{W}} &=\mathcal{F}_1(r)\, \bs{\partial}_{t} \allowbreak + \mathcal{F}_2(r)\, \bs{\partial}_{r}+\begin{cases}
        \mathcal{C}_1 t \,\bs{\partial}_{t}+ \mathcal{C}_2\rho \,\bs{\partial}_{\rho}\;, &n=0\;, \; k=0\;,
        \\
        \mathcal{C}_1 t\,\bs{\partial}_{t}\;, &n=0\;, \; k=\pm1\;,
        \\
        \mathcal{C}_1 (t\bs{\partial}_{t}+\tfrac{\rho}{2}\bs{\partial}_{\rho}) \;, &n\neq 0\;, \; k=0\;,
        \\
        \emptyset \;, & n\neq 0\;, \; k=\pm1\;,
    \end{cases}
\end{aligned}
\end{equation}
where $\mathcal{F}_{i}(r)$ and $\mathcal{C}_{i}$, ${i=1,2}$ are arbitrary functions and constants, respectively. The former allow us to gauge-fix two of the four functions $a$, $b$, $c$, $d$. For example, we will often work in the gauge
\begin{equation}\label{eq:gauge}
    c=r^2+n^2\;, \quad d=0\;,
\end{equation}
in which the metric is Lorentzian as long as ${b>0}$. Let us also recall that the GR Taub--NUT--(A)dS solution corresponds to \eqref{eq:gauge} with
\begin{equation}\label{eq:GRTNUT}
    a =\frac{k (r^2 {-} n^2) - 2 m r - \frac{\Lambda}{3} \left(r^4 {+} 6 n^2 r^2 {-} 3 n^4\right)}{r^2 + n^2}\;,
    \quad
    b=1\;,
\end{equation}
where Schwarzschild--(A)dS is recovered for ${n=0}$. For ${k=0}$, one can exploit an additional scaling freedom to fix one constant (e.g., $m$). This will turn out to be the case also in the theories studied below. 

A direct calculation shows that the independent components of the Riemann tensor of \eqref{eq:ansatz} are given by 
\begin{equation}\label{eq:R04comp1}
    \begin{aligned}
    \mathcal{R}_0 &= \frac{ab}{c^2} \left( \frac{(c')^2}{2\gamma} - 2n^2 \right) - \frac{ab'c'}{\gamma c} + b \frac{ \gamma'(ac') - 2\gamma (ac')'}{2\gamma^2 c} \;,\\
    \mathcal{R}_1 &= \frac{2n}{\sqrt\gamma} \left(\frac{ab}{c}\right)' \;, \\
    \mathcal{R}_2 &= - \frac{2\gamma(ab)'' - \gamma'(ab)'}{2\gamma^2} \;, \\
    \mathcal{R}_3 &= - \frac{ab}{c^2} \left( \frac{(c')^2}{2\gamma} - 6 n^2 \right) + \frac{2k}{c} \;, \\
    \mathcal{R}^2_4 &= \frac{a^2b^2}{c^4} \left( \frac{(\gamma cc')' - 3\gamma cc''}{4\gamma^2} + n^2 \right)^2 \;,   
    \end{aligned}
\end{equation}
where $\gamma = b + d^2$ and primes refer to derivatives with respect to the coordinate $r$.

\subsubsection{Regular static black holes}

The case ${n=0}$ corresponds to the static s./h./p. symmetry, i.e., for instance, the symmetries of the s./h./p. Schwarzschild also known as the AI/AII/AIII-metrics. In what follows, we consider the gauge \eqref{eq:gauge}, i.e., ${c=r^2}$, ${d=0}$. These metrics may represent static black holes if ${r\geq0}$, $b$ is positive, and $a$ has at least one zero, ${a(r_0)=0}$, ${r_0>0}$, corresponding to the horizon. We are, however, primarily interested in the regular black holes, for which, all polynomial scalar curvature invariants need to be finite. Assuming, $a$ and $b$ are analytic functions at ${r=0}$, the metric will have a regular core if
\begin{equation}\label{eq:regcore}
    a=k+a_2r^2+\mathcal{O}(r^3)\;, \quad b=b_0+b_2 r^2 +\mathcal{O}(r^3)\;,
\end{equation}
with ${b_0>0}$ to preserve Lorentzian signature near ${r=0}$.\footnote{There are other ways to regularize the singularity of static black holes besides having a regular central point, such as black-bounce geometries, which typically do not behave as (A)dS near ${r=0}$; see \cite{Simpson:2018tsi}.} For ${k=0}$, above conditions could be relaxed to include ${b_0=0}$ with ${b_1>0}$, however we omit this case as we typically demand regularity for all $k$ simultaneously. Although sometimes not imposed, one may also avoid divergence of scalar curvature-derivative invariants if $a$ and $b$ are even functions of $r$, ${a(-r)=a(r)}$ and ${b(-r)=b(r)}$  \cite{Antonelli:2025zxh,Giacchini:2021pmr}. It is often required that a regular black hole satisfies the limiting curvature condition, meaning that its polynomial scalar curvature invariants are bounded by a solution-independent constant (everywhere and for all values of the solution’s parameters). This often needs to be analyzed on a case-by-case basis \cite{Frolov:2016pav}.

We are mainly interested in the standard asymptotic behavior of s./h./p. Schwarzschild-(A)dS spacetime as ${r\to\infty}$
\begin{equation}\label{eq:asymtSchw}
    a = -\frac{\Lambda_{\text{eff.}}}{3}r^2 + k - \frac{2m}{r} +\mathcal{O}\left(\frac{1}{r^2}\right)\;, \quad b = 1 +\mathcal{O}\left(\frac{1}{r^2}\right)\;,
\end{equation}
where $m$ is a constant that represent the mass in some cases and $\Lambda_{\text{eff.}}$ is the effective cosmological constant corresponding to the local asymptotics of maximally symmetric spacetime. (Here, we omitted a possible $1/r$ term in $b$ for ${k=\Lambda_{\text{eff.}}=0}$.) For ${\Lambda_{\text{eff}}=0}$ and ${k=1}$, this reduces to standard asymptotic flatness and spherical symmetry, which, combined with regularity of the black holes, implies an even number of horizons (counted with multiplicities). For ${\Lambda_{\text{eff}}\neq0}$, however, we also allow different sub-leading terms, provided they have a smooth limit ${\Lambda_{\text{eff}}\to0}$.

If the horizon is compactified and typically ${\Lambda_{\text{eff.}}<0}$, the case ${k=0}$ describes static regular black holes with a toroidal horizon topology. The case ${k=-1}$ is locally hyperbolic; compact black-hole horizons of genus ${g>1}$ require a quotient of the hyperbolic plane. Nevertheless, the case ${k=-1}$ without compactification may be also interpreted as a gravitational field of tachyonic particle in the domain causally connected to the source \cite{Gott1974,Hruska:2018djo} but with regularized Mach--Cherenkov shockwave. The case ${k=0}$ without full compactification, can also be interpreted as regularization of either a gravitational field of a static infinite plane with negative mass or an infinite line. The latter is a special case of the static cylindrically symmetric spacetime known as the Levi--Civita metric (${\sigma=-1/2}$ in \cite{Griffiths:2009dfa}).

\subsubsection{Taub--NUT geometries}\label{sc:TNUT}
The case ${n\neq0}$ describes the symmetries of, e.g., s./h./p. Taub--NUT spacetime. At the level of symmetries, the specific value of $n$ is irrelevant; only two cases matter: ${n=0}$ and ${n\neq0}$. When specific field equations are imposed $n$ often becomes an independent geometrical parameter that is referred to as the NUT parameter. Again, let us consider the gauge \eqref{eq:gauge}. Regarding regularity, the only requirement at ${r=0}$ for analytic functions $a$ and $b$ is ${b_0>0}$; there are no further constraints. As the 2-surfaces at ${r=0}$ remain non-degenerate, there is no issue in extending the spacetime to ${r<0}$. Horizons may occur as well and correspond to zeros of $a$.

In the spherical case (${k=1}$), the coordinate ${\rho \in (0,1)}$ covers only half of the spacetime, with the regular semi-axis at ${\rho = 0}$. The other half can be obtained via the formal replacement ${\bs{\omega}_{1} \;\to\; \tilde{\bs{\omega}}_{1} =\big(1 {+} \sqrt{1{-}\rho^2}\big) \bs{\mathrm{d}}\varphi}$, where again ${\rho \in (0,1)}$, but now ${\rho = 0}$ corresponds to the opposite semi-axis containing the Misner string. To see this, we assume that the coordinate $\varphi$ is $2\pi$-periodic, which is equivalent to $\bs{\partial}{\varphi}$ being a Killing vector with closed orbits whose flow parameter has period $2\pi$. The region of closed timelike curves is given by $\bs{\partial}_{\varphi}^2=g_{\varphi\varphi}=-(2n)^2a(r)b(r)\big({1 \mp \sqrt{1{-}\rho^2}}\big)^2+(r^2+n^2)\rho^2<0$, where `$-$' refers to $\bs{\omega}_{1}$ and `$+$' to $\tilde{\bs{\omega}}_{1}$. The fact that this condition is satisfied for the $+$ sign near the axis ${\rho=0}$ in the stationary region ${a(r)>0}$, indicates the presence of the Misner string. This topological defect at the axis (quasi-regular singularity) is always present for ${k=1}$ irrespective of the particular form of $a$ and $b$.

As our ultimate goal is to fine-tune the theories to contain black holes with regular cores, we will demand that, in the limiting case of vanishing NUT parameter, ${n \to 0}$, the geometry reduces to the above regular static black holes with \eqref{eq:regcore} and \eqref{eq:asymtSchw} (possibly different sub-leading terms for ${\Lambda_{\text{eff.}}\neq0}$). We will not impose any specific $n$-dependent generalization of these formulas. Instead, we will analyze all Taub--NUT solutions only in theories admitting regular static black holes with above properties. Note that the behavior of GR Taub--NUT--(A)dS is
\begin{equation}\label{eq:asymtTNUT}
\begin{aligned}
    a&=\left(-k+n^2\Lambda\right)-\frac{2m}{n^2}r+\mathcal{O}\left(r^2\right)\;, & b &=1+\mathcal{O}\left(r^2\right)\;,
    \\
    a &= -\frac{\Lambda}{3}r^2 + \left(k-\frac{5\Lambda}{3}n^2\right) - \frac{2m}{r} +\mathcal{O}\left(\frac{1}{r^2}\right)\;, & b &= 1 +\mathcal{O}\left(\frac{1}{r^2}\right)\;.
\end{aligned}
\end{equation}
Although it is a geometry with finite polynomial scalar curvature invariants and is locally asymptotically maximally symmetric, we do not require this behavior for Taub--NUT solutions in theories below. Already the first line is not compatible with our requirement that the limit ${n\to0}$ leads to a regular static black hole.

\subsection{Double Wick rotations to other symmetries}\label{sc:wickrot}

Although \eqref{eq:ansatz} describes metrics that are invariant under the symmetries \eqref{eq:iga}, i.e. the symmetries of s./h./p. Taub--NUT/Schwarzschild, it effectively captures much larger class of physically interesting ansatzes. Specifically, these infinitesimal group actions are related, under double Wick rotations, to those of the NHEK geometry \cite{Bardeen:1999px,Kunduri:2007vf,Kunduri:2008tk}, swirling universe \cite{Gibbons:2013yq,Astorino:2022aam,Barrientos:2024pkt,DiPinto:2024axv}, or B-metrics \cite{EhlersKundt1962,Gott1974}.\footnote{They correspond to [4,3,\{8--11\}] using the notation from \cite{Hicks:thesis,Frausto:2024egp,Colleaux:2025uiw}.}

In what follows, we summarize the results of double Wick rotations of symmetries from \cite{Colleaux:2025uiw}. Since the global properties change significantly under these complex analytic continuations, we establish conditions under which the resulting metrics can be made regular (i.e., without curvature singularities and conical deficits) and represent geometries with desired interpretations. We will discuss how these conditions for the double-Wick-rotated metrics and above conditions for the original metrics affect each other.

Note that the Wick rotations from \cite{Colleaux:2025uiw} concern only the coordinates covering the orbits (not the space of orbits) so the form of functions $a$ and $b$ is preserved. This fact allows us to perform calculations below just for the metrics \eqref{eq:ansatz}, while ensuring that the resulting theories we construct exhibit the same properties also for the double-Wick-rotated metrics, i.e., for all TNT geometries. In fact, the invariants \eqref{eq:5Invariants} below remain unchanged when evaluated for the original and double-Wick-rotated ansatzes [see \eqref{eq:R04comp1} with \eqref{eq:gauge}]. However, maintaining a Lorentzian signature in the resulting metrics implies ${a>0}$; therefore, only stationary regions of the original spacetimes are relevant for the double Wick rotations. Let us also remark that the required complex analyticity in the coordinates covering the orbits does not prevent us from studying also the theories that are non-analytic in the curvature as only the coordinates from the space of orbits enter these invariants.

\subsubsection{Regular B-metrics}\label{sc:regbmet}

The symmetries of the BI/BII-metric, which are given by the same Killing vectors, can be obtained from \eqref{eq:iga} with ${n=0}$ and ${k=\pm1}$ by double Wick rotations of two of the three coordinates covering the orbit. The corresponding symmetry-invariant metric is given by
\begin{equation}\label{eq:BI/BII}
    \bs{g}_{\text{BI/BII}} = r^2 \check{\bs{q}}_{\pm1}  + a(r) b(r) \bs{\mathrm{d}}q^2+ \frac{\bs{\mathrm{d}}r^2}{a(r)}\;, 
\end{equation}
where $\check{\bs{q}}_{1}$ (obtained from $\bs{q}_{1}$ by taking ${\rho>1}$) and $\check{\bs{q}}_{-1}$ are the metrics of $\mathrm{dS_2}$ and $\mathrm{AdS_2}$, respectively,
\begin{equation}
    \check{\bs{q}}_1=-\frac{\bs{\mathrm{d}}\rho^2}{\rho^2-1} + \rho^2 \bs{\mathrm{d}}\varphi^2\;, \quad \check{\bs{q}}_{-1}=-\check{\bs{q}}_1\;.
\end{equation}
Clearly, we need to consider $r$ only in the range where ${a(r)>0}$. If there are disconnected domains of $r$, we need to treat them as separate manifolds. According to \eqref{eq:asymtSchw}, the manifold with ${r\to\infty}$ has local asymptotics of maximally symmetric spacetime while the manifold with ${r=0}$ (which need not be the same manifold) includes regular point due to \eqref{eq:regcore}. The metric with ${k=1}$ can be interpreted as the regularized gravitational field of a tachyonic particle not causally connected to the source \cite{Gott1974,Hruska:2018djo} or as an expanding `bubble of nothing' \cite{Witten:1981gj,Aharony:2002cx,Horowitz:2002cx}. The case ${k=-1}$ has no known interpretation.

The symmetries of the BIII-metric are obtainable from \eqref{eq:iga} with ${n=0}$ and ${k=0}$. Its symmetry-invariant metric reads
\begin{equation}\label{eq:BIII}
    \bs{g}_{\text{BIII}} = r^2 \check{\bs{q}}_{0} + a(r) b(r) \bs{\mathrm{d}}q^2+ \frac{\bs{\mathrm{d}}r^2}{a(r)} 
\end{equation}
where $\check{\bs{q}}_0$ denotes the metric of $\mathbb{M}^2$,
\begin{equation}
    \check{\bs{q}}_0=-\bs{\mathrm{d}}\tau^2+\bs{\mathrm{d}}x^2\;.
\end{equation}
The metric \eqref{eq:BIII} can be viewed as a regularization of a special case of the Levi--Civita metric (now with ${\sigma=1/4}$ in \cite{Griffiths:2009dfa}).\footnote{Note that the version with the cosmological constant is known as the Linet--Tian metric \cite{Podolsky:2018dpr}, while the charged cases is the Melvin spacetime \cite{Colleaux:2025uiw}.}

\subsubsection{NHEK and swirling geometries}\label{sc:nhekswirl}

The symmetries of the NHEK can be obtained from \eqref{eq:iga} with ${n\neq0}$ and ${k=\pm1}$. The symmetry-invariant metric is given by
\begin{equation}\label{eq:NHEK-sh}
    \bs{g}_{\text{NHEK-h/s} } = (r^2+n^2) \check{\bs{q}}_{\pm1} + a(r) b(r)\left(\bs{\mathrm{d}}q {-}2n\sqrt{\rho^2{-}1}\bs{\mathrm{d}}\varphi \right)^2 + \frac{\bs{\mathrm{d}}r^2}{a(r)}\;,
\end{equation}
where $\check{\bs{q}}_{-1}$ is the well-known $\mathrm{AdS_2}$ structure appearing in NHEK with the standard spherical horizon topology \cite{Bardeen:1999px,Kunduri:2007vf} while $\check{\bs{q}}_{1}$ is the $\mathrm{dS_2}$ structure that appears in its exotic version with the hyperbolic horizon topology (i.e., the near-horizon limit of a solution from \cite{Klemm:1997ea}). We will only discuss the former case below. As long as the $\mathrm{AdS_2}$ structure is present --- which is quite rigid across various theories \cite{Kunduri:2013gce} --- this is a general metric for the near-horizon limit of any extremal black hole provided that $a$ and $b$ satisfy specific conditions on regularity and compactness of the horizon.\footnote{If $a$ has no zeros, then the geometries may also be interpreted as wormholes connecting two locally AdS spacetimes \cite{Anabalon:2018rzq}.} Hence, it offers a direct insight into the structure of extremal rotating black holes, which are otherwise difficult to access in modified gravity. To arrive at the conditions, let us transform the metric to the form (using transformation analogous to \cite{Colleaux:2025uiw})
\begin{equation}\label{eq:NHEK}
    \bs{g}_{\text{NHEK-s}}=(r^2+n^2)\big(-w^2\bs{\mathrm{d}}u^2+\bs{\mathrm{d}}u\vee\bs{\mathrm{d}}w\big)+ a(r)b(r)\big(\bs{\mathrm{d}}\phi+2nw\bs{\mathrm{d}}u\big)^2+\frac{\bs{\mathrm{d}}r^2}{a(r)}\;,
\end{equation}
in which the horizon is located at ${w=0}$ with the induced metric (on arbitrary section) being ${a(r)b(r)
\bs{\mathrm{d}}\phi^2+{\bs{\mathrm{d}}r^2}/{a(r)}}$. Following \cite{Kunduri:2008rs}, the spherical horizon topology requires a smooth positive ${a>0}$ in ${r\in(r_-,r_+)}$ and to have the roots only at the endpoints ${a(r_{\pm})=0}$ while ${b(r_\pm)>0}$; here, $r$ plays a role of a latitudinal angular coordinate interpolating between two poles but with dimension of length. As ${a>0}$ elsewhere, then differentiability implies finite ${a'(r_-)\geq0}$, ${a'(r_+)\leq0}$, but with non-vanishing derivatives. Taking ${a=a'(r_\pm)(r-r_\pm) +\mathcal{O}\big((r-r_\pm)^2\big)}$ and ${b=b(r_\pm)+\mathcal{O}\big((r-r_\pm)^1\big)}$, the induced metric on the horizon becomes $\bs{\mathrm{d}}R^2+\big(a'(r_\pm)/2\big)^2 b(r_\pm)R^2\bs{\mathrm{d}}\phi^2+\mathcal{O}(R^4)$, upon coordinate transformation ${R=2\sqrt{(r-r_\pm)/a'(r_\pm)}}$. Denoting the periodicity of $\phi$ by ${2\pi\nu>0}$, the absence of conical deficits leads to the requirement
\begin{equation}\label{eq:cond}
    a=\mp\frac{2}{\nu\sqrt{b(r_{\pm})}} (r-r_\pm) +\mathcal{O}\big((r-r_\pm)^2\big)\;, \quad b=b(r_\pm)+\mathcal{O}\big((r-r_\pm)^1\big)\;,
\end{equation}
where ${b(r_\pm)>0}$. Clearly, any values of $a'(r_{\pm})$, satisfying ${a'(r_{-})\sqrt{b(r_-)}=-a'(r_{+})\sqrt{b(r_+)}>0}$, can be compensated by an appropriate choice of $\nu$. A sufficient condition is that $a$ and $b$ are even functions of $r$. We can also use the formalism of \cite{Kolar:2025kle} to show that this condition automatically guarantees the regular axis everywhere (not just at the horizon, ${w=0}$). The $2\pi\nu$ periodicity of $\phi$ is equivalent to ${\bs{Y}=\nu\bs{\partial}_{\phi}}$ having closed orbits with the parameter of its flow corresponding to the period $2\pi$. Now, for ${r=r_\pm}$ to be a regular axis, this spacelike Killing vector field needs to have a fixed point, ${\bs{Y}=0}$, at ${r=r_\pm}$. Then necessarily, its norm ${|\bs{Y}|=\sqrt{g_{\mu\nu}Y^{\mu}Y^{\nu}}}$ has to vanish, ${\lim_{r\to r_\pm}|\bs{Y}|=0}$, and satisfy the elementary flatness condition \cite{Mars:1992cm,Wilson_1996,Carot:1999zm,stephani2003,Kolar:2025kle}, ${\lim_{r\to r_\pm}\big|\bs{\mathrm{d}}|\bs{Y}|\big|=1}$; both are met by \eqref{eq:cond}. In retrospect, observe that the behavior \eqref{eq:cond} requires (some) solutions for the h. Taub--NUT to have (at least) two non-extreme horizons and a stationary regular region in between. For example, in GR, one may notice that $a$ and $b$ in \eqref{eq:GRTNUT} with ${k=-1}$ are even functions for ${m=0}$ (a similar property will remain also in theories below) and ${a>0}$ between two roots for ${\Lambda>-1/(4n^2)}$ (giving exactly the standard NHEK--(A)dS \cite{Kunduri:2008tk}).

The symmetries of the swirling universe are obtained from \eqref{eq:iga} with ${n\neq0}$ and ${k=0}$; the symmetry-invariant metrics
\begin{equation}\label{eq:swirl}
    \bs{g}_{\text{swirl}} = (r^2+n^2) \check{\bs{q}}_{0} +  a(r) b(r)\left(\bs{\mathrm{d}}q +2nx\bs{\mathrm{d}}\tau \right)^2  +\frac{\bs{\mathrm{d}}r^2}{a(r)}\;,
\end{equation}
The GR swirling--(A)dS spacetime admits a regular symmetry axis with no conical deficit \cite{Astorino:2022aam,Barrientos:2024pkt,DiPinto:2024axv}; it is given by \eqref{eq:GRTNUT} with ${k=0}$ after fixing one constant using the scaling freedom. [Remark that the regularity of the axis for ${\Lambda\neq0}$ was not mentioned in the previous literature (apart from a specific cancellation in Melvin--swirling--(A)dS \cite{Barrientos:2024pkt}), as it demanded the symmetry axis to be at ${r=0}$, although it naturally sits at ${r=r_0\neq0}$, where ${a(r_0)=0}$.] As there are no established criteria for swirling-type spacetimes beyond admitting this symmetry, we follow the GR example and require them to possess a regular axis and same global properties. It should then be interpreted as a gravitational whirlpool generated by a pair of counter-rotating sources at opposite infinities along the axis. Let us consider the surface ${\tau=x=0}$, which has again the induced metric ${a(r)b(r)
\bs{\mathrm{d}}q^2+{\bs{\mathrm{d}}r^2}/{a(r)}}$. The non-compact topology of this surface, occurring in the GR solution for ${\Lambda\leq0}$,\footnote{We will not discuss the case of compact topology for this surface, which occurs in the GR swirling--dS solution, ${\Lambda>0}$, as its interpretation is likely quite different.} requires smooth positive ${a>0}$ at ${r>r_0}$ and to have a root only at the endpoint ${a(r_0)=0}$ while ${b(r_0)>0}$. Here, $r$ now plays a role of the radial cylindrical coordinate going from the symmetry axis to infinity. Since ${a>0}$ elsewhere, then differentiability again means finite ${a'(r_0)\geq0}$, but we assume it to be non-vanishing. As above, taking ${a=a'(r_0)(r-r_0) +\mathcal{O}\big((r-r_0)^2\big)}$ and ${b=b(r_0)+\mathcal{O}\big((r-r_0)^1\big)}$, the induced metric turns into $\bs{\mathrm{d}}R^2+\big(a'(r_0)/2\big)^2 b(r_0)R^2\bs{\mathrm{d}}q^2+\mathcal{O}(R^4)$, upon analogous coordinate transformation. Again, considering the periodicity of $q$ to be ${2\pi\nu>0}$, we are led to the requirement
\begin{equation}\label{eq:cond2}
    a=\frac{2}{\nu\sqrt{b(r_0)}} (r-r_0) +\mathcal{O}\big((r-r_0)^2\big)\;, \quad b=b(r_0)+\mathcal{O}\big((r-r_0)^1\big)\;,
\end{equation}
where ${b(r_0)>0}$. Any value of ${a'(r_{0})>0}$ is allowed as long as we fix $\nu$ appropriately. As before, if $q$ is $2\pi\nu$-periodic coordinate along the cyclic Killing vector ${\bs{Y}=\nu\bs{\partial}_{q}}$, the two necessary conditions for the regular symmetry axis anywhere at ${r=r_0}$, ${\lim_{r\to r_0}|\bs{Y}|=0}$ and ${\lim_{r\to r_0}\big|\bs{\mathrm{d}}|\bs{Y}|\big|=1}$, are satisfied automatically by means of \eqref{eq:cond2}. Looking back at the p. Taub--NUT, the solutions should admit a non-extreme horizon with the stationary regular region extending above it to infinity. For instance, in GR, $a$ in \eqref{eq:GRTNUT} with ${k=0}$ satisfies this for ${m<0}$, and ${\Lambda\leq0}$ (corresponding to the standard swirling--AdS universe \cite{Astorino:2022aam,Barrientos:2024pkt,DiPinto:2024axv}).

\subsection{Bianchi class A cosmologies with extra symmetry}\label{sc:BianchiA}

It is worth noting that many of the symmetries above are special cases of Bianchi class A, which admit an extra Killing vector \cite{Frausto:2024egp}. Therefore, they can be also interpreted as homogeneous cosmologies. Specifically, the metric \eqref{eq:ansatz} with ${n=0}$, ${k=0}$, and the metric \eqref{eq:BIII} belong to Bianchi I. The ansatz \eqref{eq:ansatz} with ${n\neq0}$, ${k=0}$ as well as the ansatz \eqref{eq:swirl} fall into Bianchi II. The case \eqref{eq:ansatz} with ${n\neq0}$, ${k=-1}$, and the case \eqref{eq:NHEK-sh} is contained within Bianchi VIII. Finally, \eqref{eq:ansatz} with ${k=1}$, ${n\neq0}$ is part of Bianchi IX. The cosmological interpretation means that $r$ plays the role of time and requires a non-stationary region. This means ${a(r)<0}$ for \eqref{eq:ansatz}, and the $\mathrm{dS_2}$ structure for \eqref{eq:BI/BII} and \eqref{eq:NHEK-sh}; in the remaining cases there is no non-stationary region. Furthermore, as a consequence of our required regularity, the cosmologies are regular at ${r=0}$.

\subsection{Eguchi--Hanson instantons}

Let us consider the Euclidean metrics \eqref{eq:ansatz} with ${b<0}$ [still in ${\eqref{eq:gauge}}$]. The metric can be rewritten in more familiar coordinates by performing the following transformation for ${y>|n|}$:
\begin{equation}
    r =  \sqrt{y^2-n^2}\;, \quad t = n \psi\;.
\end{equation}
Then, we obtain the following class of Euclidean metrics,
\begin{equation}\label{eq:EHmet}
  \bs{g}_{\text{EH}}  = \frac{\bs{\mathrm{d}}y^2}{h(y)} +  y^2\left[ h(y)j(y)\left(\bs{\mathrm{d}}\psi +2 \bs{\omega}_k \right)^2 + \bs{q}_{k}\right] \;,
\end{equation}
where we introduced two auxiliary functions
\begin{equation}
    h = \frac{y^2-n^2}{y^2} a\left(\sqrt{y^2-n^2} \right)\;, \quad j= \frac{-n^2}{y^2-n^2} b\left(\sqrt{y^2-n^2}\right)\;.
\end{equation}
The metric can be extended to a larger $y$-domain as long as it remains regular and Euclidean. In the context of GR, the vacuum solutions is given by
\begin{equation}\label{eq:EH_GRab}
    a=\frac{k \left(n^2+r^2\right)}{4 r^2}-\frac{q}{4 r^2 \left(n^2+r^2\right)}-\frac{\Lambda  \left(n^2+r^2\right)^2}{6 r^2}\;, \quad b=-\frac{r^2}{n^2}\;,
\end{equation}
or, equivalently, by
\begin{equation}\label{eq:EH_GRhj}
    h=\frac{k}{4}-\frac{q}{4 y^4}-\frac{\Lambda  y^2}{6}\;,\quad j=1\;,
\end{equation}
where $q$ is a constant parameter. This metric is known as the Eguchi--Hanson instanton for ${k=1}$, ${q>0}$,  and ${\Lambda=0}$. The general metric \eqref{eq:EHmet} for ${k=1}$ can be interpreted as the Eguchi--Hanson-type geometry if $h$ admits a root ${h(y_0)=0}$, ${y_0>0}$, such that ${h>0}$ for ${y>y_0}$ with ${h'(y_0)>0}$ and ${j(y_0)>0}$. Assuming the expansions ${h=h'(y_0)(y-y_0) +\mathcal{O}\big((y-y_0)^2\big)}$ and ${j=j(y_0)+\mathcal{O}\big((y-y_0)^1\big)}$, we take the surface ${\rho=\text{const}}$ and ${\varphi=\text{const}}$, the induced metric on this surface, ${{\bs{\mathrm{d}}y^2}/{h(y)}+y^2h(y)j(y)
\bs{\mathrm{d}}\psi^2}$, becomes $\bs{\mathrm{d}}R^2+\big(h'(y_0)/2\big)^2 y_0^2j(y_0)R^2\bs{\mathrm{d}}\psi^2+\mathcal{O}(R^4)$ after the coordinate transformation ${R=2\sqrt{(y-y_0)/(h'(y_0))}}$. If $\psi$ is taken as a periodic coordinate with the period ${2\pi\nu>0}$, the conical deficit will vanish as long as
\begin{equation}\label{eq:regularityEH}
    h=\frac{2}{\nu y_0\sqrt{j(y_0)}} (y-y_0) +\mathcal{O}\big((y-y_0)^2\big)\;, \quad j=j(y_0)+\mathcal{O}\big((y-y_0)^1\big)\;,
\end{equation}
where ${y_0>0}$ and  ${j(y_0)>0}$. As before, any value of ${h'(y_{0})>0}$ is allowed if $\nu$ is fixed accordingly.

\section{Gravitational effective action for TNT geometries}\label{sc:effectiveAction}

In this section, we first construct a set of curvature invariants which yield the independent components of the Riemann tensor for TNT geometries \eqref{eq:R04comp1}, i.e., of the metric \eqref{eq:ansatz} or its double Wick rotations \eqref{eq:BI/BII},  \eqref{eq:BIII}, \eqref{eq:NHEK-sh}, and  \eqref{eq:swirl}. Considering a general gravitational theory depending on the Riemann tensor (not its derivatives), i.e.,
\begin{equation}\label{eq:FR2}
    I\left[\bs{g}\right] = \int d^4 x \sqrt{-g}\, \mathcal{L}\left( R_{\mu\nu\rho}{}^\sigma \right)\;,
\end{equation}
we will derive a parametrization of its effective action, which is equivalent when evaluated on TNT geometries. Given the difficulty to classify all independent theories \eqref{eq:FR2}, this is a very useful simplification which enables us to study all such theories within the TNT sector, subject to covariance.

\subsection{Zakhary--McIntosh invariants \& representatives}

The procedure by which one obtains curvature invariants yielding independent components of the Riemann tensor in some metrics is highly non-unique. Indeed, many invariants which are inequivalent for generic metric fields can be equal for a restricted set of geometries. However, this issue arises mainly when investigating broader sets of metric configurations or perturbations, which is beyond the scope of the present work.
 
Thus, we present here a concrete example of curvature invariants which achieve this aim. The method can likely be applied to other geometries beyond TNT, provided they have nontrivial scalar invariants. It is based on the Zakhary--McIntosh invariants \cite{Zakhary:1997xas} --- a complete polynomial generating set of scalar invariants constructed from the Riemann tensor in four dimensions. In the generic case only 14 combinations are algebraically independent,\footnote{Indeed, the number of independent non-differential curvature invariants in 4-dimensions is ${14=21-1-6}$, where $21$ refers to the independent components of any tensor with the ${(2,2)}$ symmetries of the (metric) Riemann tensor, $1$ refers to the Bianchi identity and $6$ to the dimension of the Lorentz group.} while 17 are required to obtain a set that remains complete for all algebraic types. Any algebraic (polynomial or not) scalar curvature invariant can be expressed as an algebraic function of these 17 invariants.

Given that we are interested in TNT geometries, it is clear that we need only five independent invariants. We consider the following ones:
\begin{equation}\label{eq:ZMInv}
\begin{aligned}
    \mathscr{I}_{1} &= C_{\mu\nu\kappa\lambda} C^{\mu\nu\kappa\lambda}\;, \\
    \mathscr{I}_{6} &= S_{\mu\nu} S^{\mu\nu}\;, \\
    \mathscr{I}_9 &= S^{\mu\kappa} S^{\nu\lambda} C_{\mu\nu\kappa\lambda}\;, \\
    \mathscr{I}_{10} &= S^{\mu\kappa} S^{\nu\lambda} \tilde{C}_{\mu\nu\kappa\lambda}\;, \\
    \mathscr{I}_{11} &= S^{\mu\nu} S_{\kappa\lambda}(C_{\mu\alpha\beta\nu} C^{\kappa\alpha\beta\lambda} - \tilde{C}_{\mu\alpha\beta\nu} \tilde{C}^{\kappa\alpha\beta\lambda})\;,
\end{aligned}
\end{equation}
where $\bs{C}$ is the Weyl tensor, $\bs{S}$ is the trace-free (TF) Ricci tensor, and $\tilde{\bs{C}}$, given by ${\tilde{C}_{\mu\nu\kappa\lambda} = \frac12 \varepsilon_{\mu\nu\alpha\beta}C^{\alpha\beta}{}_{\kappa\lambda}}$, is the dual Weyl tensor.

In Appendix \ref{ap:CurvOfTNT}, we show using the Newman--Penrose (NP) formalism that the following non-analytic invariants in the Riemann tensor, given by 
\begin{equation}\label{eq:5Invariants}
    \begin{aligned}
    \mathcal{R}_0 &= \frac{R}{3} + \frac{\mathscr{I}_1 \mathscr{I}_9}{\mathscr{I}_1 \mathscr{I}_6 - 6 \mathscr{I}_{11}}\;, \\
    \mathcal{R}_1 &= - \frac{\mathscr{I}_1 \mathscr{I}_{10}}{\mathscr{I}_1 \mathscr{I}_6 - 6 \mathscr{I}_{11}}\;, \\
    \mathcal{R}_2 &=  \frac{R}{6} - 2 \sqrt{\frac{\mathscr{I}_{11}}{\mathscr{I}_1} - \frac{\mathscr{I}_6}{12}} - \frac{\mathscr{I}_1 \mathscr{I}_9}{\mathscr{I}_1 \mathscr{I}_6 - 6 \mathscr{I}_{11}}\;, \\
    \mathcal{R}_3 &= \frac{R}{6} + 2 \sqrt{\frac{\mathscr{I}_{11}}{\mathscr{I}_1} - \frac{\mathscr{I}_6}{12}} - \frac{\mathscr{I}_1 \mathscr{I}_9}{\mathscr{I}_1 \mathscr{I}_6 - 6 \mathscr{I}_{11}}\;, \\
    \mathcal{R}_4 &=  \sqrt{\frac{2 \mathscr{I}_6}{3} - \frac{2 \mathscr{I}_{11}}{\mathscr{I}_1}}  \;, 
    \end{aligned} 
\end{equation}
precisely reduce to the independent components of the Riemann tensor \eqref{eq:R04comp1}. By abuse of notations, we use the same name for both objects, as they are designed to match each other. Notice that $\mathcal{R}_1$ is the only pseudo-scalar and that the Ricci scalar can be written as 
\begin{equation}\label{eq:RicciScalarinRi}
    R= 2 \mathcal{R}_0 + \mathcal{R}_2 + \mathcal{R}_3 \;. 
\end{equation}
Also, observe that the invariants \eqref{eq:5Invariants} are ill-defined whenever ${\mathscr{I}_1 \mathscr{I}_6 - 6 \mathscr{I}_{11}=0}$ and ${\mathscr{I}_1=0}$; this happens, for instance, for Schwarzschild-de-Sitter and conformally flat metrics ${\bs{C}=0}$. This already shows that the theories we construct do not admit maximally symmetric vacua in the usual sense, as one encounters undefined expressions such as $0/0$. Still, one may view them as arising in an appropriate limiting sense, for instance, within the symmetry-reduced framework below. As we will see, even there the result depends on how the limit is taken.

Although our representatives are taken to be non-analytic in the Riemann tensor to access all independent components linearly, polynomial representatives should be preferred whenever they exist.

\subsection{Most general effective action for TNT geometries}

The most general gravitational action of the form \eqref{eq:FR2} (in general non-analytic in the curvature) that reduces to an analytic function in the independent components of the Riemann tensor for the TNT ansatz can be parametrized by the covariant action:
\begin{equation}\label{eq:ActionNP}
    I[\bs{g}] =\int d^4 x \sqrt{-g} \, \mathcal{L}\left( \mathcal{R}_0, \mathcal{R}_1, \mathcal{R}_2, \mathcal{R}_3 , \mathcal{R}_4 \right)\;,
\end{equation}
where
\begin{equation}
    \mathcal{L}\left( \mathcal{R}_0, \mathcal{R}_1, \mathcal{R}_2, \mathcal{R}_3 , \mathcal{R}_4 \right) = \sum_{p=0}^\infty  \sum_{\sum\limits_{j=0}^4 i_j =p }  \alpha_{\left(p,\{i_j\}\right)}  \prod_{j=0}^4  \mathcal{R}_j^{i_j}
\end{equation}
and $ \alpha_{(p,\{i_j\})}= \alpha_{\left(p,i_0,i_1,i_2,i_3,i_4 \right)}$ are coupling constants, while $p$ is the order of the polynomial. By construction, any other inequivalent action (for a generic metric) reduces to this one after reduction to the ansatz \eqref{eq:ansatz}.

\subsection{Symmetry reduction of Lagrangians --- PSC}\label{sc:QTprop}

Our aim is to investigate the integrability properties of the field equations of \eqref{eq:ActionNP} for TNT geometries. Given the very complicated dependence on a generic metric field, we will rather use the symmetry-reduction of Lagrangians. As demonstrated in \cite{Frausto:2024egp}, this procedure is mathematically well defined for TNT geometries owing to the fact that they satisfy the principle of symmetric criticality (PSC) \cite{Fels:2001rv,Palais:1979rca,Anderson1997,Anderson:1999cn,Anderson:1999cm,Torre:2010xa}. Recall that PSC states that the field equations of a symmetry-reduced theory are fully equivalent to the symmetry reduction of the field equations of the original theory, and that this equivalence must hold for all generally covariant gravitational Lagrangians.

In contrast to reducing the field equations, which merely requires substituting the symmetry-invariant metric ansatz \eqref{eq:ansatz}, the proper reduction of the Lagrangian --- viewed as a 4-form rather than as an integrable density, needs a simultaneous reduction of its form degree. Specifically, the original Lagrangian 4-form $\underline{L}=L[\bs{g}]\bs{\mathrm{d}}x^1\wedge\bs{\mathrm{d}}x^2\wedge\bs{\mathrm{d}}x^3\wedge\bs{\mathrm{d}}x^4$, where ${L[\bs{g}]=\sqrt{-g}\mathcal{L}\left( \mathcal{R}_0, \mathcal{R}_1, \mathcal{R}_2, \mathcal{R}_3 , \mathcal{R}_4 \right)}$, must be mapped [upon substituting the ansatz \eqref{eq:ansatz}] to a 1-form ${\underline{\hat{L}}=\hat{L}[a,b,c,d]\bs{\mathrm{d}}r}$, as the reduced theory lives on the space of orbits parametrized by coordinate $r$. Rather than performing the often ill-defined integration over the symmetry orbits, one can rigorously implement this by contraction with the symmetry-invariant antisymmetric $(3,0)$-tensor ${\bs{\chi}=\sum_{{i<j<k}}\chi^{ijk} \bs{X}_{i}\wedge\bs{X}_{j}\wedge\bs{X}_{k}}$, as ${\underline{\hat{L}}=\underline{L}|_{\eqref{eq:ansatz}}\bullet\bs{\chi}}$, where $\bullet$ denotes the contraction in all indices. The tensor $\bs{\chi}$ is called the $3$-chain and is unique (up to an overall constant) for PSC-compatible symmetries. Here, it reads
\begin{equation}\label{eq:lchain}
    \bs{\chi} =\text{const}\frac{\sqrt{1-k\rho^2}}{\rho }\bs{\partial}_t\wedge\bs{\partial}_\rho\wedge\bs{\partial}_\varphi\;,
\end{equation}
and the reduced Lagrangian $\hat{L}$ is given by
\begin{equation}
    \hat{L}[a,b,c,d] =  c \sqrt{\gamma}  \mathcal{L}\left( \mathcal{R}_0, \mathcal{R}_1, \mathcal{R}_2, \mathcal{R}_3 , \mathcal{R}_4 \right)\;,
\end{equation}
where ${\gamma=b + d^2}$ and $\mathcal{R}_A$ with ${A=0,...,4}$ are specific expressions \eqref{eq:R04comp1}; we have omitted the overall constant. Since the PSC is satisfied, the corresponding field equations obtained by variation with respect to $a$, $b$, $c$, and $d$,
\begin{equation}\label{eq:feq}
    \frac{\delta \hat{L}}{\delta a}=0\;, \quad \frac{\delta \hat{L}}{\delta b}=0\;, \quad \frac{\delta \hat{L}}{\delta c}=0\;, \quad \frac{\delta \hat{L}}{\delta d}=0\;,
\end{equation}
are fully equivalent to the symmetry reduction of the field equations of the original theory, ${({\delta L}/{\delta \bs{g})}|_{\eqref{eq:ansatz}}=0}$. 

As mentioned above, the symmetry-invariant metric \eqref{eq:ansatz} contains a residual diffeomorphism freedom \eqref{eq:resdifgen}, which allows one to impose, e.g., the gauge \eqref{eq:gauge}. Although this is always allowed at the level of the field equations (after variation), doing so at the level of the reduced Lagrangian (before variation) may, in principle, cause loss of essential field equations (even if PSC is satisfied). By inspection of the Noether identities associated with the infinite-dimensional part of the residual diffeomorphism group [first two terms in \eqref{eq:resdifgen}] through the second Noether's theorem (i.e., the symmetry-reduced Bianchi-type identities) simplified by means of the finite-dimensional part [other terms in \eqref{eq:resdifgen}], one may often argue that a specific gauge fixing does not break the equivalence of field equations \cite{Anderson:1999cn,Frausto:2024egp}. Although the equivalence seems to hold in general, above arguments do not apply in all situations.\footnote{Specifically, for ${n\neq0}$ and ${k=\pm1}$, the arguments from \cite{Anderson:1999cn,Frausto:2024egp} cannot be used in theories that do not admit any vacuum solution with such symmetries.} We explicitly verified, by direct comparison for a large class of theories, that imposing \eqref{eq:gauge} in the reduced Lagrangian always correctly reproduces the reduced field equations (where this gauge is imposed). Specifically, the last two equations in \eqref{eq:feq} are identically satisfied when the first two are. Therefore, we only need to analyze the two equations
\begin{equation}\label{eq:feq_gf}
    \frac{\delta \big(\hat{L}|_{\text{GF}}\big)}{\delta a} 
    =0\;, \quad \frac{\delta \big(\hat{L}|_{\text{GF}}\big)}{\delta b} 
    =0\;,
\end{equation}
where `GF' stands for fixing the gauge \eqref{eq:gauge}; in what follows the subscript `GF' will be omitted for brevity.

\section{QTG property for TNT geometries}\label{sc:QTGprop}

\subsection{SF ansatz and QTG-TNT definition}\label{sc:QTGpropdef}

Let us state precisely what we mean by quasi-topological gravity. In the following, we will refer to the subclass of metrics \eqref{eq:ansatz} in the gauge \eqref{eq:gauge} with
\begin{equation}\label{eq:SFansatz}
    b = 1\;, 
\end{equation}
which are then characterized only by the metric function $a$, as the \textit{single-function (SF) ansatz}. Following the well-known integrability property of Einstein's field equations for TNT geometries, as well as the results of \cite{Bueno:2018uoy} on Einsteinian cubic gravity, a theory will be said to be a \textit{quasi-topological gravity for Taub--NUT-type geometries (QTG-TNT)} when the following equation is identically satisfied
\begin{equation}\label{eq:Ea}
     \frac{\delta \hat{L}}{\delta a} \bigg|_{\text{SF}} = 0\;, 
    \quad \forall \;\; a(r), r, n, k \;.
\end{equation}
As we will show, it always implies that the leftover (independent) equation 
\begin{equation}\label{eq:leftovereq}
    \frac{\delta \hat{L}}{\delta b} \bigg|_{\text{SF}} = 0 
\end{equation}
can be integrated once as follows
\begin{equation}\label{eq:Q}
    \int  \frac{dr \left(r^2+n^2\right)}{r^2} \frac{\delta \hat{L}}{\delta b} \bigg|_{\text{SF}}  = Q\left( a'', a', a ,r,n,k \right)= 2 m
\end{equation}
for some function $Q$, where $m$ is an integration constant typically related to the mass of the system. As a concrete example, note that in GR, the two independent field equations \eqref{eq:feq_gf} are of the first order in the derivatives of the metric functions $a$ and $b$. Thus, the $Q$-function of GR is algebraic, i.e., it does not depend on radial derivatives.

For later use, notice that the components of the Riemann tensor \eqref{eq:R04comp1} reduce for the SF ansatz to 
\begin{equation}
    \mathcal{R}_0\big|_{\text{SF}} = -\frac{2}{r^2+n^2} \left( \frac{2 n^2 a}{r^2+n^2} + r a' \right) \;, \;\;\;
    \mathcal{R}_1\big|_{\text{SF}} = -\frac{2n}{r^2+n^2} \left( \frac{2 r a}{r^2+n^2} - a'  \right)  \;, \;\;\; 
    \mathcal{R}_2\big|_{\text{SF}} = - a''  \;, \label{R04compSF1}
\end{equation}
which involve derivatives of $a$, and
\begin{equation}
    \mathcal{R}_3 \big|_{\text{SF}}= \frac{2}{r^2+n^2} \left( k +\frac{ \left( 3 n^2 - r^2 \right)a}{r^2+n^2} \right)  \;, \;\;\;
    \mathcal{R}_4 \big|_{\text{SF}}= 0 \;, \label{R04compSF2}
\end{equation}
which do not. (Note that ${{\mathscr{I}_1 \mathscr{I}_6 - 6 \mathscr{I}_{11}}\big|_{\text{SF}}\neq0}$ and ${\mathscr{I}_1\big|_{\text{SF}}\neq0}$.)

Remark that being a QTG-TNT is a property which follows from some fine-tuning between invariants at a given order $p$ in the components of the Riemann tensor $\mathcal{R}_A$.  Therefore, it is sufficient to restrict our study to a generic order $p$ Lagrangian, 
\begin{equation}\label{eq:EffLagOp}
    \mathcal{L}_{(p)} \left( \mathcal{R}_0, \mathcal{R}_1, \mathcal{R}_2, \mathcal{R}_3 , \mathcal{R}_4 \right) = \sum_{\sum\limits_{j=0}^4 i_j =p }  \alpha_{\left(p,\{i_j\}\right)}  \prod_{j=0}^4  \mathcal{R}_j^{i_j} \;.
\end{equation}
This can be understood by a simple scaling argument and \eqref{eq:Ea}, which must be satisfied for any metric function, radius, NUT parameter, and topology. 

\subsection{Covariant SF ansatz}\label{sc:SFcov}
It can be shown that the SF ansatz \eqref{eq:SFansatz} is geometrically distinguished, i.e., covariantly defined. Consider the subclass of TNT geometries for which the TF Ricci tensor specializes to Segre type \{(1,1),(11)\} (or equivalently to null alignment type D) while the Weyl tensor remains of Petrov type D; this is further equivalent to 
\begin{equation}\label{eq:CovQTGProp}
    \mathcal{R}_4 = 0\;,
\end{equation}
see \eqref{eq:R_AinNP}. Indeed, using \eqref{eq:R04comp1} with \eqref{eq:gauge}, this can be readily solved and yields ${ b = {r^2}/{(r^2 + c_0(r^2 + n^2))}}$, where $c_0$ is an integration constant. Without loss of generality, it can be brought to the form\footnote{One uses the coordinate transformation ${t= w_0 \tilde{t}}$, ${r= \sqrt{\tilde{r}^2+\left(1-w_0^2\right)\tilde{n}^2}}$, the redefinition ${n = w_0 \tilde{n}}$, ${a=\tilde{a}\tilde{r}^2/(\tilde{r}^2+(1-w_0^2)\tilde{n}^2)}$, where ${w_0=\sqrt{-(c_0+1)/\varepsilon}}$, and then drops tildes.}
\begin{equation}\label{eq:bvareps}
    b =- \frac{r^2}{\varepsilon (r^2+n^2) + n^2}\;,
\end{equation}
with ${\varepsilon=0,\pm 1}$, while remaining in the gauge \eqref{eq:gauge}. Clearly, the metric is Lorentzian only for ${\varepsilon=-1}$, which corresponds to ${b=1}$. Hence, as long as the Lorentzian signature is considered ${b>0}$, the covariant definition \eqref{eq:CovQTGProp} (i.e., specialization of TF Ricci tensor) is equivalent to SF ansatz \eqref{eq:SFansatz}.

The case ${\varepsilon=0}$ is a metric with Euclidean signature ${(+,+,+,+)}$ for ${a>0}$; it contains, e.g., the Eguchi--Hanson instanton in GR. The case ${\varepsilon=1}$ is either Euclidean for ${a>0}$ or it has the signature ${(-,-,+,+)}$ for ${a<0}$. The latter can be continued to the Lorentzian-signature metric by Wick rotation ${t= i \bar{t}}$ accompanied by redefinition ${n = i \bar{n}}$,
\begin{equation}
  \bs{g} = - \frac{a(r)r^2}{r^2-2 \bar{n}^2} \left(\bs{\mathrm{d}}\bar{t} +2\bar{n} \bs{\omega}_k \right)^2   + \frac{\bs{\mathrm{d}}r^2}{a(r)} + \left(  r^2-\bar{n}^2\right) \bs{q}_{k} \;,
\end{equation}
provided that ${r^2 \geq 2 \bar{n}^2}$. Nevertheless, this metric is again just the standard SF ansatz \eqref{eq:SFansatz} (i.e., ${\varepsilon=-1}$), as can be seen by the coordinate transformation ${r = \sqrt{\bar{r}^2+2\bar{n}^2}}$, the redefinition ${a=\bar{a}\bar{r}^2/(2\bar{n}^2+\bar{r}^2)}$, and then dropping the bars.

Remark that all the QTG-TNT of Class I obtained in the next section actually admit the solution ${\mathcal{R}_4=0}$ for arbitrary~${\varepsilon}$. As for the remaining ones, we checked that this is also true up to high (finite) orders in $p$. Therefore, \eqref{eq:Ea} with `SF' replaced by ${\mathcal{R}_4 = 0}$ can be seen as a covariant definition of QTG-TNT. The generalization of the integrability of the leftover equation \eqref{eq:Q} for arbitrary ${\varepsilon=0,\pm 1}$ is given by
\begin{equation}\label{eq:EqAlga}
    \int  dr \, \frac{r^2  \left(r^2+n^2\right)}{\left(\varepsilon \left(r^2+n^2\right)+n^2\right)^2}  \frac{\delta \hat{L}_1}{\delta b} \bigg|_{\mathcal{R}_4=0}  = Q\left(a'',a',a,r,n,k \right)= 2 m \;.
\end{equation}

\section{Classification of QTG-TNT theories}\label{sc:classQTG}

In this section, we focus on the 4-dimensional gravitational theories polynomial in the components of the Riemann tensor for TNT geometries given by \eqref{eq:ActionNP} (not necessarily polynomial in the Riemann tensor for generic metrics). We find all such QTG-TNT theories, i.e., all theories satisfying the property \eqref{eq:Ea}. These are classified according to the number $N$ of derivatives of the metric appearing in their leftover equation \eqref{eq:leftovereq}, 
\begin{equation}\label{eq:I0I1I2I3}
    I_{\text{QTG-TNT}} = \sum_{N=0}^3 I_{N}\;.
\end{equation} 
Here, $I_{0}$ corresponds to topological QTG-TNT, which are trivial for the SF ansatz; $I_{1}$ is the class to which GR belongs; and $I_{3}$ encompasses theories like the Einsteinian cubic gravity and the generalized QTG-TNT. Indeed, the leftover equation of the theory can be once-integrated, showing that the full theory admits TNT solutions [i.e., of the form \eqref{eq:ansatz} and their double-Wick-rotated counterparts  \eqref{eq:BI/BII}, \eqref{eq:BIII}, \eqref{eq:NHEK-sh}, and  \eqref{eq:swirl}] within the SF ansatz \eqref{eq:SFansatz}. The only metric function $a$ satisfies the differential equation 
\begin{equation}\label{eq:Q2}
    Q_{\text{1}}\left(a ,r,n,k \right) + Q_{\text{2}}\left(a', a ,r,n,k \right)  + Q_{\text{3}}\left( a'', a', a ,r,n,k \right)  = 2 m \;,
\end{equation}
where $m$ is an integration constant related to the mass of the system. It will be useful to split the effective action \eqref{eq:ActionNP} into two classes, in which we search for QTG-TNT geometries.

\subsection{Class I}

Let us start by considering Class I theories, given by 
\begin{equation}\label{ActionTop}
    I^{\text{(I)}} =\sum_{A}\int d^4 x \sqrt{-g} \; \mathcal{A}^A \left( \mathcal{R}_0, \mathcal{R}_1, \mathcal{R}_2, \mathcal{R}_3  \right) \mathcal{P}_A\left(\mathcal{R}_4\right)\;,
\end{equation}
where $\mathcal{A}^A$ are analytic functions and $\mathcal{P}_A$ are non-constant monomials labeled by index $A$. Theories of this form are special because ${\mathcal{R}_4=0}$ for the SF ansatz, see \eqref{R04compSF2}.

\subsubsection{Topological QTG-TNT}

 Using the standard big-$O$ notation,
\begin{equation}
    \mathcal{P}_A\left(\mathcal{R}_4\right) = O\left( \mathcal{R}_4^2 \right) \quad \implies \quad \frac{\delta \hat{L}^{\text{(I)}}}{\delta a} \bigg|_{\text{SF}} = \frac{\delta \hat{L}^{\text{(I)}}}{\delta b} \bigg|_{\text{SF}} =0 \;,
\end{equation}
identically. This means that the theory $I^{\text{(I)}}$ is a topological QTG-TNT, which we denote $I^{\text{(I)}}_{0}$, unless it is linear in $\mathcal{R}_4$, in which case we denote it by $I^{\text{(I)}}_{1,2,3}$. Thus, the effective gravitational action \eqref{eq:ActionNP} splits into
\begin{equation}
    I = I^{\text{(I)}} + I^{\text{(II)}} = I^{\text{(I)}}_{0} + I^{\text{(I)}}_{1,2,3}+I^{\text{(II)}}\;,
\end{equation}
where $I^{\text{(II)}}$ will be referred to as the Class II theories, and
\begin{equation}\label{eq:CICII}
\begin{aligned}
    I^{\text{(I)}}_{0} &= \sum_A\int d^4 x \sqrt{-g} \; \mathcal{A}_0^A \left( \mathcal{R}_0, \mathcal{R}_1, \mathcal{R}_2, \mathcal{R}_3  \right) \mathcal{P}_A\left(\mathcal{R}_4\right)  \;,  \quad \mathcal{P}_A\left(\mathcal{R}_4\right) = O\left( \mathcal{R}_4^2 \right)\;,\\
    I^{\text{(I)}}_{1,2,3} &=   \int d^4 x \sqrt{-g} \; \mathcal{A}_{1,2,3} \left( \mathcal{R}_0, \mathcal{R}_1, \mathcal{R}_2, \mathcal{R}_3  \right) \mathcal{R}_4  \;,\\
    I^{\text{(II)}} &= \int d^4 x \sqrt{-g} \; \mathcal{L}^{\text{(II)}} \left( \mathcal{R}_0, \mathcal{R}_1, \mathcal{R}_2, \mathcal{R}_3  \right) \;.
\end{aligned}
\end{equation}

\subsubsection{First- and second-order QTG-TNT}

It is straightforward to show that $I^{\text{(I)}}_{1,2,3}$ are QTG-TNT as well, albeit non-trivial, so that all the Class I theories \eqref{ActionTop} are QTG-TNT. Their leftover field equations can be integrated once and their associated $Q$-function [see \eqref{eq:Q} and \eqref{eq:Q2}] is simply given by 
\begin{equation}\label{eq:QNP}
\begin{aligned}
    Q^{\text{(I)}}_{1,2,3} &= \frac{1}{8 n r} \left( r^2+n^2 \right)^2 \mathcal{A}_{1,2,3}\left( \mathcal{R}_0, \mathcal{R}_1, \mathcal{R}_2, \mathcal{R}_3  \right) \left( n \mathcal{R}_0 + r \mathcal{R}_1 \right) \\
    &= - \frac{a}{2 r} \left( r^2+n^2 \right) \mathcal{A}_{1,2,3}\left( \mathcal{R}_0, \mathcal{R}_1, \mathcal{R}_2, \mathcal{R}_3  \right)\;,
\end{aligned}
\end{equation}
where we used \eqref{R04compSF1}. Using \eqref{R04compSF2} as well, we obtain
\begin{equation}
 I^{\text{(I)}}_{1,2,3} =  I^{\text{(I)}}_{1} +  I^{\text{(I)}}_{2} +  I^{\text{(I)}}_{3}
\end{equation}
in terms of the following Class I QTG-TNT whose field equations involve respectively first, second, and third derivatives of the metric field \eqref{eq:ansatz}:
\begin{equation}\label{eq:II1II2II3}
\begin{aligned}
    I^{\text{(I)}}_{1} &=   \int d^4 x \sqrt{-g}\,  \mathcal{A}_1\left( \mathcal{R}_3  \right) \mathcal{R}_4      \;, \\
    I^{\text{(I)}}_{2} &=    \int d^4 x \sqrt{-g} \; \mathcal{A}_2 \left( \mathcal{R}_0, \mathcal{R}_1,  \mathcal{R}_3  \right) \mathcal{R}_4      \;, \\
    I^{\text{(I)}}_{3} &=     \int d^4 x \sqrt{-g} \; \mathcal{A}_3 \left( \mathcal{R}_0, \mathcal{R}_1, \mathcal{R}_2, \mathcal{R}_3  \right) \mathcal{R}_4     \;,  
\end{aligned}
\end{equation}
where
\begin{equation}\label{eq:II22}
    \frac{\partial \mathcal{A}_2 }{\partial \mathcal{R}_0} \neq 0 \quad \text{or} \quad \frac{\partial \mathcal{A}_2 }{\partial \mathcal{R}_1} \neq 0  \;, 
\end{equation}
and 
\begin{equation}\label{eq:II32}
    \frac{\partial \mathcal{A}_3 }{\partial \mathcal{R}_2} \neq 0
\;. 
\end{equation}
Interestingly, other than general relativity which will appear later in the analysis, the unique new 4-dimensional QTG-TNT with algebraic (and non-vanishing) $Q$-function, meaning with first-order field equations for \eqref{eq:ansatz}, is given by $I^{\text{(I)}}_{1}$, as the completion of the classification will show.

Furthermore, in App.~\ref{ap:ProofNAN}, we show that a Lagrangian analytic (i.e., also polynomial) in the Riemann tensor cannot depend linearly on $\mathcal{R}_4$ when evaluated in TNT geometries. Consequently, all non-topological QTG-TNT of Class I theories are necessarily non-analytic. The analytic (hence polynomial) non-topological QTG-TNT  can only occur in Class II.
\medskip

For static s./h./p. symmetric spacetimes, i.e. setting ${n=0}$ in \eqref{eq:QNP}, this algebraic $Q$-function reduces to 
\begin{equation}
    Q_{\text{1}}^{\text{(I)}}\left(a ,r,0,k \right) = - \frac{r a}{2}\, \mathcal{A}_1\left( \frac{2\left( k - a \right)}{r^2}  \right)\;,
\end{equation}
which, although very similar to the Wheeler polynomial appearing in QTG and Lovelock--Lanczos gravity, differs from these by the pre-factor $r a$. Given the unicity of this theory, this proves that QTG-TNT constructed out of the Riemann tensor only cannot yield a Wheeler polynomial in four dimensions. In particular, there is no  $\mathcal{L}\left(g_{\mu\nu},R_{\mu\nu\rho\sigma}\right)$ QTG-TNT theory admitting a generalisation of the Lovelock-type regular black holes obtained in \cite{Kunstatter:2015raa,Kunstatter:2015vxa,Colleaux:2019ckh,Bueno:2025zaj} with non-vanishing NUT, and similarly for the associated swirling-type backgrounds and most notably near-horizon extremal rotating black holes. This strongly indicates a fortiori that there should be no $\mathcal{L}\left(g_{\mu\nu},R_{\mu\nu\rho\sigma}\right)$ QTG compatible with stationarity and axisymmetry (if they even exist) admitting rotating generalisation of these Lovelock-type regular black holes.

\subsection{Class II}

We now focus on finding all the QTG-TNT in the Class II of theories given by \eqref{eq:CICII}. As explained around \eqref{eq:EffLagOp}, we can restrict our study to a fix order $p$ of the polynomials in the components of Riemann. Thus, we consider the action 
\begin{equation}\label{eq:IClassII}
    I^{\text{(II)}}_{(p)} = \int d^4 x \sqrt{-g} \; \mathcal{L}_{(p)}^{\text{(II)}} \left( \mathcal{R}_0, \mathcal{R}_1, \mathcal{R}_2, \mathcal{R}_3  \right)  
\end{equation}
such that 
\begin{equation}
    I^{\text{(II)}} = \sum_{p=1}^\infty I^{\text{(II)}}_{(p)}\;.
\end{equation}
Given that all theories of Class I are already QTG-TNT, they cannot  be used to help in the fine tuning of the theories of Class II to enforce the QTG-TNT property \eqref{eq:Ea}. Therefore, obtaining all QTG-TNT in Class II will complete the classification.

\subsubsection{General relativity and  topological invariants}\label{sc:GRandtopolinv}

Clearly, at linear order, $p=1$, the only QTG-TNT of this class is general relativity, given by
\begin{equation}
    R= 2 \mathcal{R}_0 + \mathcal{R}_2+ \mathcal{R}_3\;,
\end{equation} 
while at quadratic order, $p=2$, we obtain two theories given by 
\begin{equation}
    \mathcal{R}_0^2 - 3 \mathcal{R}_1^2 + 2 \mathcal{R}_2 \mathcal{R}_3 \;, \quad
    \mathcal{R}_1 \left( - \mathcal{R}_0 + \mathcal{R}_2 + \mathcal{R}_3 \right) \;,
\end{equation}
which are both topological QTG-TNT. Regarding the first invariant, notice that
\begin{equation}
    R^2  -4 R_{\mu\nu}R^{\mu\nu}+R_{\mu\nu\rho\sigma}R^{\mu\nu\rho\sigma} = \mathcal{R}_0^2 - 3 \mathcal{R}_1^2 + 2 \mathcal{R}_2\mathcal{R}_3 - 4 \mathcal{R}_4^2 \;,
\end{equation}
so it is just the Gauss-Bonnet invariant modulo Class I topological QTG-TNT, $I^{\text{(I)}}_{0}$, defined by \eqref{eq:CICII}. Similarly, the second theory is the Chern-Pontryagin invariant,
\begin{equation}
    \frac{1}{4} \varepsilon_{\mu\nu\sigma\rho} R_{\alpha\beta}{}^{\mu\nu}R^{\alpha\beta\sigma\rho} =  \mathcal{R}_1 \left( - \mathcal{R}_0 + \mathcal{R}_2 + \mathcal{R}_3 \right)\;.
\end{equation}
These two are the last remaining topological QTG-TNT and, more generally, are well-known topological invariants.

\subsubsection{Analytic third-order (generalized) QTG-TNT}\label{sc:SecGQTG}

Let us now focus on theories of Class II involving higher-order polynomials in the components of the Riemann tensor, $p \geq 3$. Any such polynomial theory which is not at most linear in $\mathcal{R}_2$ cannot be QTG-TNT, because the reduced Lagrangian with non-linear dependence on $a''$ [which can only come from $\mathcal{R}_2$, see \eqref{eq:R04comp1} with \eqref{eq:gauge}] necessarily produce $a''''$ in the field equations. From the form of general polynomial invariant \eqref{eq:IuR}, it then follows that the linearity in $\mathcal{R}_2$ implies the linearity in $\mathcal{R}_3$, if one restricts to polynomial theories. Even though we consider more general non-analytic theories, we have checked up to high-order $p$ that QTG-TNT of this class are linear in $\mathcal{R}_3$. Although this might hint that all theories of this class have polynomial representatives, we have been able to prove this only up to order $p=5$, see \eqref{eq:covrep3}, \eqref{eq:covrep4}, and \eqref{eq:covrep5}.  Thus, the minimal action relevant to obtain in a covariant way the Class II theories enjoying QTG-TNT property is given by
\begin{equation}\label{eq:LagGQTG}
    \mathcal{L}_{(p)}^{\text{(II)}}  = \gamma_0\left(\mathcal{R}_0,\mathcal{R}_1 \right)+ \gamma_1\left(\mathcal{R}_0,\mathcal{R}_1 \right) \mathcal{R}_2 + \gamma_2\left(\mathcal{R}_0,\mathcal{R}_1 \right) \mathcal{R}_3+ \gamma_3\left(\mathcal{R}_0,\mathcal{R}_1 \right) \mathcal{R}_2 \mathcal{R}_3\;, 
\end{equation}
where
\begin{equation}\label{eq:Polygamma}
    \gamma_0 = \sum_{j=0}^p \alpha_{0 j} \,  \mathcal{R}_0^j \mathcal{R}_1^{p-j}\;, \quad
    \gamma_1 = \sum_{j=0}^{p-1} \alpha_{1 j} \,  \mathcal{R}_0^j \mathcal{R}_1^{p-1-j}\;, \quad
    \gamma_2 = \sum_{j=0}^{p-1} \alpha_{2 j}  \, \mathcal{R}_0^j \mathcal{R}_1^{p-1-j}\;, \quad
    \gamma_3 = \sum_{j=0}^{p-2} \alpha_{3 j}   \, \mathcal{R}_0^j \mathcal{R}_1^{p-2-j}\;,  
\end{equation}
where the $\alpha_{Ij}$ for $I\in \{0,...,3\}$ are constants.

Deriving the reduced field equation with respect to $a(r)$ and imposing the SF ansatz \eqref{eq:SFansatz}, the conditions for which the equation identically vanishes for any $a(r)$, $k$, $r$ and $n$ can be shown to be 
\begin{equation}
\begin{aligned}
\partial^2_1 \gamma_2 &= \frac{1}{2} \left( -\mathcal{R}_0 \partial^2_1  + 5 \partial_0 + 3 \mathcal{R}_1 \partial_0\partial_1  \right)\gamma_3\;,
\\
\partial^2_0 \gamma_2 &= -\frac{1}{2} \left( \partial_0 +3 \mathcal{R}_1 \partial_0\partial_1 + \mathcal{R}_0 \partial^2_0  \right) \gamma_3 \;,
\\
\partial_0 \partial_1 \gamma_2 &=- \frac{1}{4} \left( 6 \partial_1 +3 \mathcal{R}_1 \partial^2_1 + 2 \mathcal{R}_0 \partial_0 \partial_1  -3 \mathcal{R}_1 \partial^2_0 \right) \gamma_3\;,
\\
\partial^2_1 \gamma_0 &=- \frac{1}{2} \left( \left( 6  + 3 \mathcal{R}_1 \partial_1  \right) \gamma_3 +\left( \mathcal{R}_0  \partial^2_1 - 5 \partial_0 - 3 \mathcal{R}_1 \partial_0\partial_1  \right)  \gamma_1 \right)\;,
\\
\partial^2_0 \gamma_0 &= \frac{1}{2} \left( \left( 2 + \mathcal{R}_0 \partial_0 \right) \gamma_3 +\left(\partial_0  -3  \mathcal{R}_1 \partial_0\partial_1 - \mathcal{R}_0 \partial^2_0 \right)  \gamma_1 \right)  \;,
\\
\partial_0 \partial_1 \gamma_0 &= \frac{1}{4} \left( \left( \mathcal{R}_0 \partial_1 - 3 \mathcal{R}_1 \partial_0 \right) \gamma_3 +\left( -4\partial_1 -3 \mathcal{R}_1 \partial^2_1 -2 \mathcal{R}_0 \partial_0\partial_1 + 3 \mathcal{R}_1 \partial^2_0 \right)  \gamma_1 \right) \;,
\end{aligned}
\end{equation}
where ${\partial_I=\partial/\partial \mathcal{R}_I}$. Inserting the expressions of $\gamma_I$ in terms of polynomials of $\mathcal{R}_0$ and $\mathcal{R}_1$, given by \eqref{eq:Polygamma}, these differential equations can be reduced to difference ones. Solving the equations we obtain
\begin{equation}
\begin{aligned}
    \alpha_{0j} &= - \frac{6 + 9 p^2 + 12 j + 4 j^2 - 3 p (5+4j)}{4(p-j-1)(p-j)}\, \alpha_j \;,
    \\
    \alpha_{1j} &=  \alpha_{2j} = \frac{4-3p +2j}{2j} \, \alpha_{j-1} \;,
    \\
    \alpha_{3j}  &=    \alpha_{j} \;,
\end{aligned}
\end{equation}
in terms of the following series of real constants
\begin{equation}
\alpha_{j} =\frac{1}{2} e^{\frac{i p \pi}{2}} \left(z_p +(-1)^p \bar{z}_p \right) \binom{p-2}{j} \;,
\end{equation}
where 
\begin{equation}\label{eq:couplingconsts}
    z_p = \zeta_p + i \xi_p 
\end{equation}
is a complex number with $\bar{z}_p$ its conjugate, $i$ is the imaginary unit, while $\zeta_p$ and $\xi_p$ are real coupling constants.\footnote{Notice that the introduction of complex numbers is used to take into account both odd and even $j$ at the same time.} This means that there are two (non-trivial) theories per curvature order which admit the QTG-TNT property. 

Performing the summations \eqref{eq:Polygamma} we finally get the expressions of the $\gamma_I$, 
\begin{equation}
\begin{aligned}
    \gamma_0 \left( \mathcal{R}_0,\mathcal{R}_1\right) &= - \frac{z_p}{8(p-1)}   \left( \frac{6}{p} \mathcal{P}_{(1,1)}^2 +\mathcal{P}_{(3,1)} \mathcal{P}_{(3p-5,p-3)}  \right)\mathcal{P}_{(1,1)}^{p-2} + \text{c.c.}\;, \\
    \gamma_1 \left( \mathcal{R}_0,\mathcal{R}_1\right) &= \gamma_2 \left( \mathcal{R}_0,\mathcal{R}_1\right)=  \frac{i z_p}{4(p-1)} \mathcal{P}_{(3p-4,p-2)} \mathcal{P}_{(1,1)}^{p-2} + \text{c.c.}\;,  \\
    \gamma_3 \left( \mathcal{R}_0,\mathcal{R}_1\right) &= \frac{1}{2} z_p \mathcal{P}_{(1,1)}^{p-2} + \text{c.c.}\;, 
\end{aligned}\label{CouplingFunctTNUT}
\end{equation}
where c.c. denotes the complex conjugate and
\begin{equation}
    \mathcal{P}_{(\lambda_p,\delta_p)}= \lambda_p \mathcal{R}_1 + i \delta_p \mathcal{R}_0\;, 
\end{equation}
for some series of constants $\lambda_p$ and $\delta_p$. The leftover equation can be integrated once and yields
\begin{equation}\label{eq:IntEqTaubNUT}
    Q^{\text{(II)}}_3\left( a'', a', a ,r,n,k \right) =  \sum_{p=3}^\infty  \frac{p-2}{2 r^2}  \left[\bar{z}_p \frac{(n-i r)^{3-2p}}{(n+i r)^{p}}   Z^{p-3} \left( - i a X Y + \frac{1}{8 p(p-1)} V_p \right) + \text{c.c.} \right]   \;,
\end{equation}
where the terms involving first and second derivatives of $a$ are given by
\begin{equation}\label{eq:XYZ}
\begin{aligned}
    Z &= 4 i n a + 2(n^2+r^2) a' \;,
    \\
    X &= \left(8n^2 - 2 r^2 + 6  i n r  \right) a + \left(r^2+n^2\right)\left( 2 k + (r - 3 i n) a' \right) \;,
    \\
    Y &= - 8 n^2 r a + \left(r^2+n^2\right) \left( \left(3 n^2-r^2\right) a' + r \left(r^2+n^2\right) a'' \right)\;,
\end{aligned}
\end{equation}
while the following are polynomials in $a$ and depend on its first derivative only through $Z$,
\begin{equation}\label{eq:Vp}
\begin{aligned}
    V_p &= (p-1) (3n + ir) r Z^3 + 96 n p(p-1) a^2 \left( k (n-i r)^3 (n+i r) + \left(r^2+n^2\right)^2 a \right) 
    \\
    &\feq- 24 i n p(p-1) (n-i r) a  \left( -k \left(r^2+n^2\right) + 2 \left( n^2 - 2 i n r + r^2 \right) a \right)Z  \;,
    \\
    &\feq+ \left( 4 i k p r \left(r^2+n^2\right) + 2p \left( 9 n^3 (p-1) + 2 i n^2 (4-3p) r + n (3p-7) r^2 - 2 i r^3 \right) a \right)  Z^2 \;.
\end{aligned}
\end{equation}

When the NUT parameter vanishes, ${n=0}$, one can see that the field equation reduces to that of 4-dimensional generalized QTG \cite{Moreno:2023rfl} (extended to arbitrary order $p$) given that in this case the summand in \eqref{eq:IntEqTaubNUT} satisfies that the coefficients of $\xi_{2q}$ and $\zeta_{2q+1}$ vanish for ${q\geq 1}$, so that the terms proportional to $\xi_{2q+1}$ give the odd orders of generalized QTG while those proportional to $\zeta_{2q}$ give the even ones. Thus, one out of the two families of Class II QTG-TNT theories \eqref{CouplingFunctTNUT} is trivial when reduced to static s./h./p. symmetry. It corresponds to odd theories under parity. 

For ${p=3}$ and ${p=4}$, the even-parity theories should match to \cite{Bueno:2018uoy}, corresponding to Einsteinian cubic and a quartic theory; both are polynomial in the Riemann tensor. Our theories extend these to arbitrary order $p$ and also to odd parity.

Since there are no other Class II QTG-TNT that are analytic in the components of the Riemann tensor, any QTG-TNT theory that is analytic (inc. polynomial) must necessarily be third-order (i.e., second-order after one integration). In App.~\ref{ap:covrep}, we construct covariant polynomial representatives of all these third-order Class II QTG-TNT up to order ${p=5}$; the results are summarized in Tab.~\ref{tab:covrep}.


\begin{table}[!ht]
\begin{tabular}{l|l|l}
$p$ & even parity & odd parity \\ \hline

1 &
$R$ &
-- \\ 

2 &
$\displaystyle
\begin{aligned}[t]
\tfrac{R^2}{12} - S_{ab} S^{ab} + \tfrac12 C_{abcd} C^{abcd}
\end{aligned}
$
&
$\displaystyle
-\tfrac12 \tilde{C}_{abcd} C^{abcd}
$
\\ 

3 &
$\displaystyle
\begin{aligned}[t]
&-\tfrac{R^3}{108} + 12 R S_{ab}S^{ab} - 36 S^{ac} S^{bd} C_{abcd} - 6 R C_{abcd} C^{abcd}\\
&+ 42 C_{ab}{}^{cd} C_{cd}{}^{ef} C_{ef}{}^{ab}\;,
\end{aligned}
$

&
$\displaystyle
\begin{aligned}[t]
&- \tfrac{R}3 \tilde{C}^{abcd} C_{abcd} - 2 S^{ac} S^{bd} \tilde{C}_{abcd} + \tfrac73 \tilde{C}_{ab}{}^{cd} C_{cd}{}^{ef} C_{ef}{}^{ab}\;, \\
\end{aligned}
$
\\

4 &
$\displaystyle
\begin{aligned}[t]
&- \tfrac{R^4}{648}
+ \tfrac{R^2}{9} S_{a}{}^{b} S_{b}{}^{a}
- \tfrac{R^2}{18} C_{abcd} C^{abcd}
 \\
& - \tfrac{2R}{3} S^{ac} S^{bd} C_{abcd}
+ \tfrac{7R}{9} C_{ab}{}^{cd} C_{cd}{}^{ef} C_{ef}{}^{ab}
\\
&+ 4 S_{ab} S_{cd} C^{acef} C^{bd}{}_{ef}
- \tfrac{5}{6} C_{ab}{}^{cd} C_{cd}{}^{ef} C_{ef}{}^{gh} C_{gh}{}^{ab}
\end{aligned}
$
&
$\displaystyle
\begin{aligned}[t]
&\tfrac{R^2}{18} \tilde{C}^{abcd} C_{abcd}
+ \tfrac{2R}{3} S^{ac} S^{bd} \tilde{C}_{abcd}
- \tfrac{7R}{9} \tilde{C}_{ab}{}^{cd} C_{cd}{}^{ef} C_{ef}{}^{ab} \\
&- 2 S^{ab} S_{cd} \tilde{C}^{cedf} C_{aebf}
+ \tfrac56 \tilde{C}_{ab}{}^{cd} C_{cd}{}^{ef} C_{ef}{}^{gh} C_{gh}{}^{ab}
\end{aligned}
$
\\ 

5 &
$\displaystyle
\begin{aligned}[t]
&\tfrac{R^5}{3240}
- \tfrac{R^3 }{27} S_{ab} S^{ab}
+ \tfrac{R^3 }{54} C_{ab}{}^{cd} C_{cd}{}^{ab}
+ \tfrac{R^2}{3} S^{ac} S^{bd} C_{abcd} \\
&- \tfrac{7R^2}{18} C_{ab}{}^{cd} C_{cd}{}^{ef} C_{ef}{}^{ab}
- \tfrac{R}{3} S_{gh} S^{gh} C_{ab}{}^{cd} C_{cd}{}^{ab} \\
&+ \tfrac{5R}{6} C_{ab}{}^{cd} C_{cd}{}^{ef} C_{ef}{}^{gh} C_{gh}{}^{ab}
+ \tfrac23 S_{gh} S^{gh} C_{ab}{}^{cd} C_{cd}{}^{ef} C_{ef}{}^{ab} \\
&- \tfrac{26}{25} C_{ab}{}^{cd} C_{cd}{}^{ef} C_{ef}{}^{gh} C_{gh}{}^{ij} C_{ij}{}^{ab}
\end{aligned}
$
&
$\displaystyle
\begin{aligned}[t]
&\tfrac{R^3 }{54} \tilde{C}^{abcd} C_{abcd}
+ \tfrac{R^2}{3} S^{ac} S^{bd} \tilde{C}_{abcd}
- \tfrac{7 R^2}{18} \tilde{C}_{ab}{}^{cd} C_{cd}{}^{ef} C_{ef}{}^{ab} \\
&- \tfrac{R}{3} S_{gh} S^{gh} \tilde{C}_{ab}{}^{cd} C_{cd}{}^{ab}
+ \tfrac{5R}{6} \tilde{C}_{ab}{}^{cd} C_{cd}{}^{ef} C_{ef}{}^{gh} C_{gh}{}^{ab} \\
&+ \tfrac23 S_{gh} S^{gh} \tilde{C}_{ab}{}^{cd} C_{cd}{}^{ef} C_{ef}{}^{ab}
- \tfrac{26}{25} \tilde{C}_{ab}{}^{cd} C_{cd}{}^{ef} C_{ef}{}^{gh} C_{gh}{}^{ij} C_{ij}{}^{ab}
\end{aligned}
$
\\ 
\end{tabular}
\caption{Covariant polynomial representatives of third-order Class II QTG–TNT up to order $p=5$. At quadratic and cubic order ($p=2, 3$), the covariant representatives are unique in the sense that every term in the reduced theory \eqref{eq:RTirrcompform} can be associated with exactly one invariant from the complete set of independent curvature invariants. At higher orders ($p=4,5$), due to increased ambiguity, only a single explicit covariant representative is presented for each theory. See Appendix~\ref{ap:covrep} for further details.}\label{tab:covrep}
\end{table}

\subsection{Hybrid Class I \& II second-order QTG-TNT}

The remaining step is to investigate the possible existence of at most second-order QTG-TNT ($N\leq 2$) constructed from a combination of Class I and Class II theories, so that we can obtain the explicit expressions of the QTG-TNT actions $I_N$ [see \eqref{eq:I0I1I2I3}] and Q-functions \eqref{eq:Q2} for any order $N$ in the derivatives of the metric field appearing in the field equations for TNT geometries.

In order to cancel the (linear) third-order derivatives appearing in the field equation of the Class II theory $I^{\text{(II)}}$, given by \eqref{eq:IClassII}, \eqref{eq:LagGQTG}, and \eqref{CouplingFunctTNUT}, the only possible Class I theories to consider are those among $I^{\text{(I)}}_{3}$, given by \eqref{eq:II1II2II3} and \eqref{eq:II32}, which are linear in  ${\mathcal{R}_2 = -a''}$, because of \eqref{eq:QNP}. Therefore, we consider the theory
\begin{equation}
    I^{\text{(I+II)}}_{2} = I^{\text{(I)}}_{3\text{,lin}} + I^{\text{(II)}} \;,
\end{equation}
where
\begin{equation}
    I^{\text{(I)}}_{3\text{,lin}} =     \int d^4 x \sqrt{-g} \; \mathcal{H} \left( \mathcal{R}_0, \mathcal{R}_1, \mathcal{R}_3  \right)\mathcal{R}_2 \mathcal{R}_4   \;.
\end{equation}
Using their respective $Q$-functions obtained previously [see \eqref{eq:QNP} and \eqref{eq:IntEqTaubNUT}], it is straightforward to obtain the fine-tuning for which the terms involving $a''$ identically vanish. It is given by
\begin{equation}\label{eq:Sterm}
    \mathcal{H}\left( \mathcal{R}_0, \mathcal{R}_1, \mathcal{R}_3  \right) = -2 \left( \partial_1 \gamma_1\left(\mathcal{R}_0,\mathcal{R}_1\right) +\mathcal{R}_3 \partial_1 \gamma_3\left(\mathcal{R}_0,\mathcal{R}_1\right)  \right) \;, 
\end{equation}
where $\gamma_I$ are given by \eqref{CouplingFunctTNUT}. Indeed the leftover integrated equation reads 
\begin{equation}
    Q^{\text{(I+II)}}_2\left( a', a ,r,n,k \right) =  \sum_{p=3}^\infty  \frac{p-2}{2 r^2}  \left[\bar{z}_p \frac{(n-i r)^{3-2p}}{(n+i r)^{p}}   Z^{p-3} \left( - i a X \tilde{Y} + \frac{1}{8 p(p-1)} V_p \right) + \text{c.c.} \right] \;,   
\end{equation}
where $X$ and $Z$ are given by \eqref{eq:XYZ}, $V_p$ is given by \eqref{eq:Vp} while 
\begin{equation}
    \tilde{Y}=- 8 n^2 r a + \left(r^2+n^2\right)  \left(3 n^2-r^2\right) a'\;.
\end{equation}
Therefore, in addition to $I^{\text{(I)}}_{2}$ defined by \eqref{eq:II1II2II3} and \eqref{eq:II22}, there is a two-parameter-per-order-$p$ family of theories with second-order field equations (i.e. first-order $Q$-function) given by $I^{\text{(I+II)}}_{2}$ with \eqref{eq:Sterm}. Furthermore, we have checked that adding $I^{\text{(I)}}_{2}$ to this action is not sufficient to remove the second-order derivatives, at least up to high (but finite) order $p$.

\section{Exact solutions in first-order QTG-TNT}\label{sc:solAlgTh}

In this section, we investigate the implications of the integrability conditions \eqref{eq:Ea} and \eqref{eq:Q} by finding exact solutions in the simplest class of 4-dimensional QTG-TNT given by those with algebraic $Q$-functions, i.e., those with first-order (non-redundant) field equations and GR-like integrability for the SF TNT geometries. They are given by the following action containing an infinite tower of (non-analytic) curvature invariants:
\begin{equation}\label{eq:I1s4}
    I_{\text{1}} =\int d^4 x \sqrt{-g}  \left( -2 \Lambda + R +  \mathcal{A}\left(\ell^{2} \, \mathcal{R}_3\right)\mathcal{R}_4 \right)\;, 
\end{equation}
where
\begin{equation}
    \mathcal{A}\left(\ell^{2} \, \mathcal{R}_3\right)= \sum_{p=1}^\infty \alpha_p \, \left( \ell^{2} \, \mathcal{R}_3 \right)^p \;,
\end{equation}
where $\ell$ is a length scale, while $\alpha_p$ are dimensionless coupling constants. Recall that the curvature invariants $\mathcal{R}_3$ and $\mathcal{R}_4$ are given in terms of the Zakhary--McIntosh invariants \eqref{eq:ZMInv} by \eqref{eq:5Invariants}. Therefore, this theory is non-analytic in curvature. Due to the absence of the pseudoscalar $\mathcal{R}_1$, this theory is parity invariant.

In particular, we first construct theories where the constants $\alpha_p$ are chosen such that the static s./h./p black-hole solutions are regular at the origin \eqref{eq:regcore} and take a closed-form. We then investigate the other TNT solutions of such theories, i.e., the spacetimes with symmetries of s./h./p. Taub--NUT, B-metric, swirling universe, NHEK, and also discuss the Euclidean solutions of Eguchi--Hanson type; all these solutions will be exact and of a closed form (like in GR). As noted in Sec.~\ref{sc:BianchiA}, some solutions also admit a Bianchi class A interpretation, which we will not pursue.

\subsection{Field equations and factorization}

In order to uncover all the TNT (not necessarily SF) solutions in these models, we derive the general reduced field equations. Assuming the gauge \eqref{eq:gauge}, the first field equation in \eqref{eq:feq_gf} takes the factorized form,\footnote{Remark that adding the topological QTG-TNT, given by ${I^{\text{(I)}}_{0}=  \int d^4 x \sqrt{-g} \; \mathcal{A}_0 \left( \mathcal{R}_0, \mathcal{R}_1, \mathcal{R}_2, \mathcal{R}_3  \right) \mathcal{P}\left(\mathcal{R}_4\right)}$, where ${\partial_4^2 \mathcal{P} \neq 0}$, and $\mathcal{A}_0$ and $\mathcal{P}$ are polynomials, to the action \eqref{eq:I1s4} preserves the factorization \eqref{eq:Facto}, but with different expression for $P$. This gives a theory admitting the same ${\mathcal{R}_4=0}$ solutions as those of \eqref{eq:I1s4} but different non-SF branch of solutions, ${P=0}$. It might be an interesting way to investigate deviations from \eqref{eq:I1s4}, but it is outside the scope of this paper.}
\begin{equation}\label{eq:Facto}
    \frac{\delta \hat{L}_1}{\delta a} = \sqrt{b} (r^2+n^2) P \mathcal{R}_4 = 0\;, 
\end{equation}
where
\begin{equation}\label{eq:PSbR3R4}
\begin{aligned}
    P &= \frac{2}{a}\left( (1+\frac{1}{2}\mathcal{A}(\ell^2\mathcal{R}_3)) + \frac{3n^2b-r^2}{(r^2 + n^2)^2} \ell^2 a \mathcal{A}'(\ell^2\mathcal{R}_3) \right)=\frac{2}{a} \left( 1 +\ell^2  \, \sum_{p=1}^\infty \alpha_p \, S_p \left(\ell^{2} \, \mathcal{R}_3\right)^{p-1}  \right) \;, 
    \\
    S_p &= \frac{k}{r^2+n^2} - (p+1) a \, \frac{r^2- 3 n^2 b}{\left(r^2+n^2\right)^2}\;,
    \\
    \mathcal{R}_3  &= S_1  +\frac{k}{r^2+n^2}\;, \quad \mathcal{R}_4   =\frac{a}{2\left(r^2+n^2\right)} \left( \frac{2n^2 (b-1)}{r^2+n^2} + \frac{r b'}{b} \right)\;.
\end{aligned}
\end{equation}
The second field equation reads
\begin{equation}\label{eq:varb}
\begin{aligned}
    \frac{\delta \hat{L}_1}{\delta b} &= \frac{1}{\sqrt{b}} \left[ k - \Lambda(r^2+n^2) - \left(ra' + a\frac{r^2 - 3n^2b + 2n^2}{r^2+n^2}\right)\left(1+\frac{1}{2}\mathcal{A}(\ell^2 \mathcal{R}_3)\right) \right.
    \\
    &\feq\left.+ \left( r \frac{2kr + a'(r^2 - 3 n^2 b)}{(r^2+n^2)^2} -2a \frac{r^4 - n^2 r^2 (1 + 6b) + 3n^4b(1-b)}{(r^2+n^2)^3} \right) \ell^2 a \mathcal{A}'(\ell^2 \mathcal{R}_3) \right]=0\;.
\end{aligned}
\end{equation}

If ${\mathcal{R}_4=0}$, the first field equation \eqref{eq:Facto} is trivially satisfied and one finds that $b$ is given by \eqref{eq:bvareps}, where we consider ${\epsilon=-1,0}$. The leftover field equation \eqref{eq:varb} is then integrated once as explained in Sec.~\ref{sc:QTGprop}. It gives rise to the equation \eqref{eq:EqAlga} with the $Q$-function given by
\begin{equation}\label{eq:Q1fun}
    Q_1 = - \frac{b\sqrt{b}\left(r^2+n^2\right)}{r} \left[  \left( 1 + \frac{1}{2} \mathcal{A}\left(\ell^{2} \, \mathcal{R}_3\right) \right) a  - T\right]\;, 
\end{equation}
where $b(r)$ is given by \eqref{eq:bvareps} and
\begin{equation}
    \mathcal{R}_3= \frac{2 k}{r^2+n^2} - \frac{2 r^2}{\left(r^2+n^2\right)^2} \left( 1 + \frac{3 n^2}{s^2} \right) a\;,
\end{equation}
with ${s^2= \varepsilon \left(r^2+n^2\right)+n^2}$, and
\begin{equation}
T=\frac{s^2}{\varepsilon^3 r^2 \left( r^2+n^2 \right)} \left(  \varepsilon k \left( s^2+n^2 \right) + \Lambda\left(n^4+2n^2 s^2 - \frac{s^4}{3} \right)  \right)\;.
\end{equation}
In the SF ansatz \eqref{eq:SFansatz}, ${b=1}$, (i.e., ${\varepsilon=-1}$), the equation \eqref{eq:EqAlga} with \eqref{eq:Q1fun} further reduces to
\begin{equation}\label{eq:feqSFA-alg}
    - \left(1+ \frac{n^2  }{ r^2} \right)\left(1+ \frac12\mathcal{A}\left(\ell^2 \,\mathcal{R}_3 \right)\right)a-\frac{k n^2}{r^2}+k+\frac{\Lambda  n^4}{r^2}-2 \Lambda  n^2-\frac{\Lambda  r^2}{3}=\frac{2 m}{r}\;.
    \end{equation}
The strict limit ${\varepsilon \to 0}$, corresponding to Euclidean metrics with ${b=-r^2/n^2<0}$, of \eqref{eq:EqAlga} is ill-defined (when ${\Lambda\neq 0}$ and ${k \neq 0}$). Nevertheless, it actually differs from a well-defined expression just by divergent constant terms, which can be reabsorbed into the integration constant. Alternatively, the same result can be obtained by setting ${\varepsilon=0}$ in ${{\delta \hat{L}_1}/{\delta b} \big|_{\mathcal{R}_4=0}=0}$ and subsequent integration. This yields
\begin{equation}\label{eq:feqalgeps0}
    -  r^2 \left( 1 + \frac{1}{2} \mathcal{A}\left(\ell^{2} \, \mathcal{R}_3\right)   \right) a+\left( r^2 + n^2\right) \left( \frac{k}{4} - \frac{\Lambda}{6} \left( r^2 + n^2\right)  \right)  =  \frac{q}{4\left( r^2 + n^2\right)}\;,
\end{equation}
where $q$ is an integration constant.

In the next section, we first identify the theories with the regular black holes among the SF ansatz, ${b=1}$, (i.e., Lorentzian metrics with ${\mathcal{R}_4=0}$ and ${\epsilon=-1}$). Then we explore various SF closed-form TNT solutions within these theories and also comment on the Euclidean solutions (within ${\mathcal{R}_4=0}$ but with ${\varepsilon=0}$). Finally, we move our focus towards theories admitting non-SF solutions ${\mathcal{R}_4\neq0}$, i.e., ${P=0}$. 

Before doing so, let us also comment on the maximally symmetric vacua of the theories. These can be obtained by inserting ${a=k-{\Lambda_{\text{eff.}} r^2}/{3}}$ into \eqref{eq:feqSFA-alg}. Clearly, it can be a solution only if ${m=n=0}$ and either ${k=0}$ or ${\Lambda_{\text{eff.}}=\Lambda=0}$. In the former, the non-trivial effective cosmological constant ${\Lambda_{\text{eff.}}}$ is then determined by solving the equation
\begin{equation}\label{eq:vacuaeq}
    \left(1+\frac12\mathcal{A}\left(\frac{2 \Lambda_{\text{eff.}} \ell ^2}{3}\right)\right)\Lambda_{\text{eff.}} =\Lambda\;.
\end{equation}
The fact that (A)dS spacetime solves the equation only when expressed in p. slicing ${k=0}$ and not in the s./h. slicing ${k=\pm1}$ may seem contradictory. However, this precisely reflects the fact that a maximally symmetric spacetime can only be recovered as a solution in a limiting sense. The limit towards (A)dS produces a valid solution within the p. symmetry reduction, but not within the s./h. symmetry reductions.

\subsection{Regular static black holes and B-metrics --- \texorpdfstring{${n=0}$}{n=0}}\label{sc:RBH}

To identify some theories with static regular black holes (${n=0}$) in the SF ansatz ${b=1}$ (i.e., ${\epsilon=-1}$),  we restrict our attention to the case ${\Lambda=0}$ and s. topology ${k=1}$. Most of the relevant properties persist (or generalize straightforwardly) when extending to the more general setting with ${\Lambda\in\mathbb{R}}$ and ${k=\pm1,0}$. The equation \eqref{eq:feqSFA-alg} simplifies to 
\begin{equation}\label{eq:feqSFA-alg-sss}
    - \left(1+\frac12\mathcal{A}(\ell^2\mathcal{R}_3)\right)a+1=\frac{2 m}{r}\;, 
\end{equation}
while $\mathcal{R}_3$ takes the form
\begin{equation}\label{eq:R3eq}
    \mathcal{R}_3=\frac{2 (1-a)}{r^2}\;.
\end{equation}
We will consider theories $\mathcal{A}$ for which at least one solution $a$ exists.

Interestingly, some properties of the horizon structure are shared across all theories \eqref{eq:I1s4}. Assuming that $r_0$ is a horizon ${a(r_0)=0}$ and $a$ is analytic there, there are two options: Either ${\mathcal{A}(\ell^2\mathcal{R}_3(r))}$ is finite at ${r=r_0}$ and then necessarily ${r_0=2m}$, or ${\mathcal{A}(\ell^2\mathcal{R}_3(r))}$ diverges at ${r=r_0}$ and the field equation is satisfied in the limiting sense. The divergence points ${x_0}$, ${\mathcal{A}(x_0)=\pm\infty}$, however, originate solely from $\mathcal{A}$ whose structure is agnostic of $m$ (it depends only on $\alpha_p$ and $\ell$), so these horizons ${r_0=\sqrt{2}\ell/\sqrt{x_0}}$ must be independent of $m$. There can be as many of them as there are positive $x_0$; however, they may belong to different solution branches. On the other hand, ${r=2m}$ is always present because the divergence-point horizons can never coincide with all values of $2m$. As the static non-extreme s. symmetric asymptotically flat regular black hole must have an even number of horizons, one horizon is always at ${r=2m}$ and the others are at fixed positions given by the theory. Although, for a sufficiently large $m$, the solutions we present below can be interpreted as regular black holes, this horizon $m$-(in)dependence typically produces singularities (and solution branching) once $m$ is decreased. Consequently, the limiting-curvature condition typically fails to hold in the original sense, but may still be satisfied in the restricted sense for large $m$. This represents a somewhat milder departure than the typical large-$m$ breakdown observed in many regular black holes in the literature \cite{Maeda:2021jdc}.

Let us now impose the regular-core conditions at ${r=0}$ given in \eqref{eq:regcore}. First, we observe that $a$ cannot be an even function of ${r}$ when ${m \neq 0}$. [If it were, the left-hand side of \eqref{eq:feqSFA-alg-sss} would be even, while the right-hand side is odd unless ${m=0}$.] Consequently, although all curvature invariants remain finite, some curvature-derivative invariants must necessarily diverge --- complete smoothness is impossible. Nevertheless, we will see that the lowest order at which such divergences appear can be controlled by the form of $\mathcal{A}$.

First, we assume just the standard regularity with the (A)dS core \eqref{eq:regcore} but diverging curvature-derivative invariants with two covariant derivatives (e.g., $\Box R$), which means that ${a=1+a_2 r^2+a_3 r^3 + \mathcal{O}(r^4)}$, where ${a_2\neq0}$ and ${a_3\neq0}$. We find from \eqref{eq:R3eq} that ${\mathcal{R}_3=-2 a_2 -2  a_3 r+O\left(r^2\right)}$. Denoting ${x=\ell^2\mathcal{R}_3}$, we invert the series to yield
\begin{equation}\label{eq:rusingR3}
    r=-\frac{1}{2 \ell^2a_3 }(x+2 \ell^2a_2)+O\left(\left(x+2 \ell^2 a_2 \right)^2\right)\;.
\end{equation}
Denoting ${x=\ell^2\mathcal{R}_3}$, expressing $\mathcal{A}(x)$ from \eqref{eq:feqSFA-alg-sss}, and inserting \eqref{eq:rusingR3}, we obtain
\begin{equation}\label{eq:reg0}
    \mathcal{A}_{\text{0-reg}}=\frac{8 \ell^2 a_3  m }{x+2 \ell^2 a_2 }+O\left(\left(x+2 \ell^2 a_2 \right)^0\right)\;.
\end{equation}
Remark that $\mathcal{A}$, as a function of $x$, encodes the information about the theory and only depends on dimensionless constants $\alpha_p$. Hence, the apparent $m$-dependence is merely an artifact, fully compensated by the corresponding $m$-dependence of $a_k$. More importantly, this equation tells us that $\mathcal{A}$ needs to have a simple pole at ${x=-2 \ell^2 a_2}$ in order for the black hole to have standard regularity at the origin ${r=0}$. Clearly, the same analysis can be repeated for higher-order regularity. For example, ${a=1+a_2 r^2 + a_4 r^4+ O(r^5)}$, ${a_2\neq0}$, ${a_4\neq0}$, leads to the square-root singularity at ${x=-2 \ell^2 a_2}$ instead,
\begin{equation}\label{eq:reg2}
    \mathcal{A}_{\text{2-reg}}=
        \pm\frac{4 \sqrt{2} \ell m \sqrt{|a_4| }}{\sqrt{|x+2 \ell^2 a_2 |}}+O\left(|x+2 \ell^2 a_2|^0\right)\;,
\end{equation}
while ${a=1+a_2 r^2 + a_4r^4+\mathcal{O}(r^6)}$, ${a_2\neq0}$, ${a_4\neq0}$, implies that the next-to-leading term must behave as $\sqrt{|\mathcal{R}_3+2 a_2|}$ rather than a constant,
\begin{equation}\label{eq:reg4}
    \mathcal{A}_{\text{4-reg}}=
        \pm\frac{4 \sqrt{2}\ell m \sqrt{|a_4| }}{\sqrt{|x+2 \ell^2 a_2 |}}+O\left(\sqrt{|x+2 \ell^2a_2|}\right)\;.
\end{equation}
This analysis also tells us that the regularity (of any type) cannot occur at finite order at $p$, i.e., for polynomial $\mathcal{A}$. In fact, it is evident already from \eqref{eq:feqSFA-alg-sss} as its left-hand side is analytic at ${r=0}$ [by \eqref{eq:regcore} and \eqref{eq:R3eq}, $a$ and $\mathcal{R}_3$ are analytic, and polynomials preserve analyticity], while the right-hand side is not for ${m\neq0}$. Only if $\mathcal{A}$ is a non-polynomial function (still analytic at ${x=0}$), then its possible non-analyticity (at ${x=-2 \ell^2a_2\neq0}$) may generate the desired $1/r$ singularity to match the right-hand side of \eqref{eq:feqSFA-alg-sss}. The above requirements \eqref{eq:reg0}, \eqref{eq:reg2}, and \eqref{eq:reg4} are only necessary conditions for regularity. In fact, the algebraic equation \eqref{eq:feqSFA-alg-sss} may admit multiple solutions $a$ only some of which may be regular at ${r=0}$.

\begin{figure}
    \centering
    \includegraphics[width=0.45\linewidth]{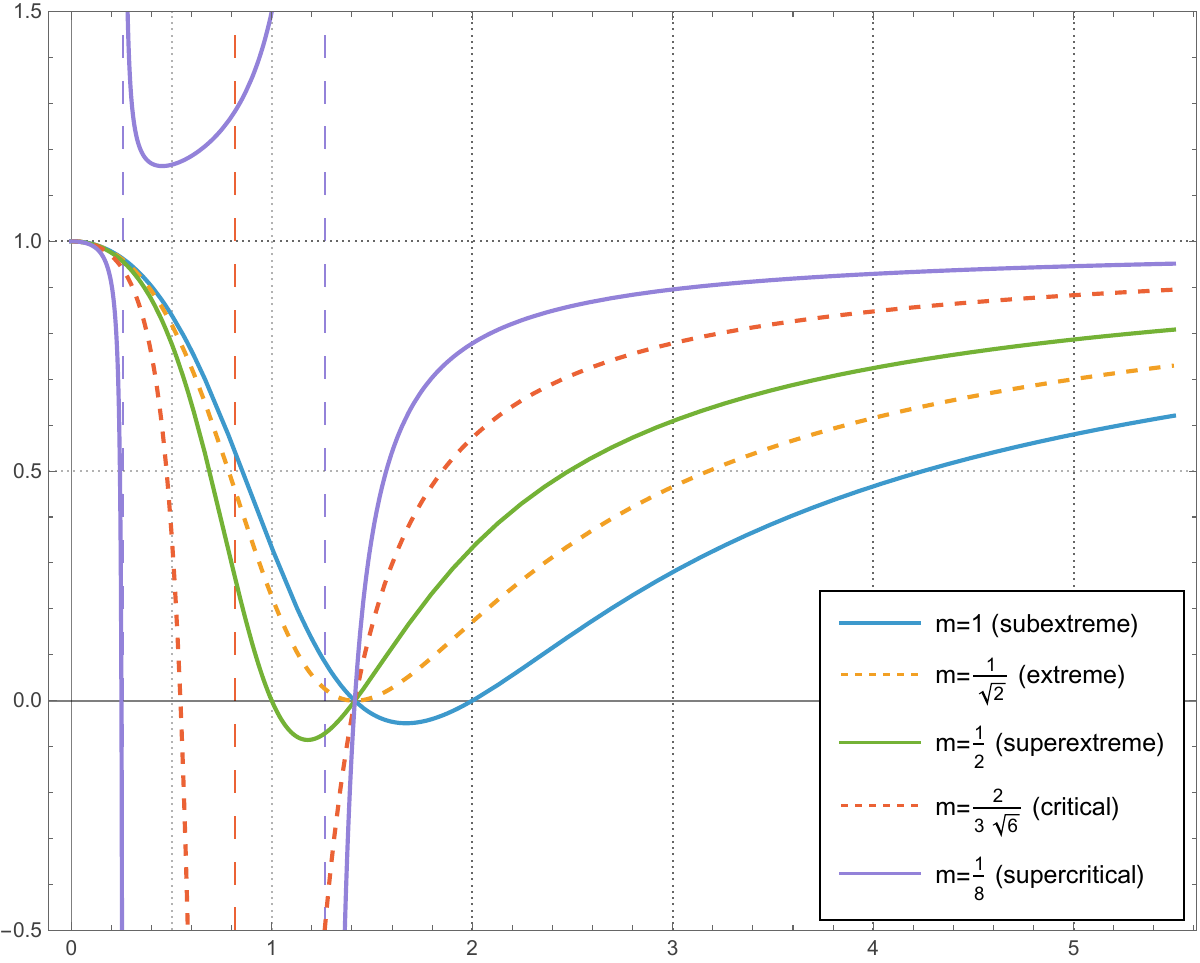}
    \caption{Static s. symmetric black hole solution given by \eqref{eq:solbh} for the first-derivative QTG-TNT \eqref{eq:simpletheory}. The plot is shown for ${\ell=1}$. The solution represents a regular black hole with two horizons not only in the subextreme case ${m>\ell/\sqrt{2}}$ but also in the superextreme case ${2\ell/(3\sqrt{6})<m<\ell/\sqrt{2}}$. Beyond the critical value ${m<2\ell/(3\sqrt{6})}$ the solution describes three distinct spacetimes (depending on the range of $r$) all of which are singular. The branch extending to ${r\to\infty}$ corresponds to an asymptotically flat singular black hole with a single horizon.}
    \label{fig:plot1}
\end{figure}

Although we named several conditions on $\mathcal{A}$, there exist many such theories. A very simple example of a theory admitting a solution with the standard regularity \eqref{eq:reg0}, for which \eqref{eq:feqSFA-alg-sss} becomes linear in $a$, is
\begin{equation}\label{eq:simpletheory}
    \mathcal{A}=\frac{2x}{1-x}\;, \quad \alpha_p=2\;,
\end{equation}
whose only solution is
\begin{equation}\label{eq:solbh}
    a=\frac{(r-2 m) \left(r^2-2 \ell ^2\right)}{r^3-2 r \ell ^2+4 m \ell ^2}\;,
\end{equation}
which is shown in Fig.~\ref{fig:plot1}. As long as ${m>2\ell/(3\sqrt{6})}$, the solution represents a regular static s. symmetric asymptotically flat black hole with two horizons located at ${r=2m}$ and ${r=\sqrt{2}\ell}$. The fact that one horizon is fixed by the theory while the other is $m$-dependent and both are always present (even below extremality) causes singular behavior at lower values of $m$. Nevertheless, the extremality ${m=\ell/\sqrt{2}}$ occurs still within the well-behaved range of $m$. Also, the curvature invariants tend to be bounded by a solution-independent constant for $m$ in the range ${m\geq 2\ell/(3\sqrt{6}) +\epsilon}$, with ${\epsilon>0}$; hence, the limiting curvature is satisfied in this restricted sense.

The unboundedness of curvature invariants already appears for ${m=2\ell/(3\sqrt{6})}$ at the root of the denominator. If ${0<m<2\ell/(3\sqrt{6})}$, then the above function diverges at the two positive roots $r_i$ of the denominator, ${0<2m<r_1<r_2<\sqrt{2}\ell}$,\footnote{This follows from positivity of denominator at ${r=2m}$ and $r=\sqrt{2}\ell$ and negativity at its minima ${r=\sqrt{2/3}\ell}$.} which correspond to curvature singularities and split the solution in three separate spacetimes. The part ${r\in(0,r_1)}$ describes a solution with a regular center ${r=0}$ and a horizon at ${r=2m}$, but singularity at ${r=r_1}$. The part ${r\in(r_1,r_2)}$ is a static horizonless solution with two singularities ${r=r_{1,2}}$. The part ${r\in(r_2,\infty)}$ is an asymptotically flat black hole with one horizon at ${r=\sqrt{2}\ell}$ and a singularity hidden behind it at ${r=r_2}$. When taking the limit ${m\to0}$, one finds that ${r_1\to0}$ and ${r_2\to\sqrt{2} \ell}$; the solutions become flat.

The above solution can be generalized to arbitrary $k$ and $\Lambda$ by solving \eqref{eq:feqSFA-alg} with \eqref{eq:simpletheory} and ${n=0}$,
\begin{equation}\label{eq:statRBHsimpleth}
    a=\frac{ \left(k r-2 m-\frac{\Lambda}{3}  r^3\right)\left(r^2-2 k \ell ^2\right)}{r^3 \left(1+\frac{2}{3}\Lambda  \ell ^2\right)-2\left( k r -2 m \right)\ell ^2}\;,
\end{equation}
whose GR limit, ${\ell\to0}$, recovers the s./h./p. Schwarzschild--(A)dS black holes \eqref{eq:GRTNUT} (with ${n=0}$). Introducing the effective cosmological constant ${\Lambda_{\text{eff.}}=\Lambda/(1+\tfrac23\Lambda\ell^2)}$, the asymptotic expansion takes the form
\begin{equation}
    a=-\frac{\Lambda_{\text{eff.}} r^2}{3}+\left(1-\frac{2}{3} \Lambda_{\text{eff.}} \ell ^2+\frac{4}{9} \Lambda_{\text{eff.}}^2 \ell ^4\right)k -\left(1-\frac{2}{3} \Lambda_{\text{eff.}} \ell ^2\right)^2\frac{2 m }{r}+O\left(\frac{1}{r^2}\right)\;,
\end{equation}
which matches \eqref{eq:asymtSchw} for ${\Lambda_{\text{eff.}}=0}$ and differs in the sub-leading terms for ${\Lambda_{\text{eff.}}\neq0}$. The value of $\Lambda_{\text{eff.}}$ corresponds to the unique maximally symmetric vacuum of the theory, i.e., the only solution of \eqref{eq:vacuaeq} in the limiting sense through the ${k=0}$ slicing [\eqref{eq:statRBHsimpleth} with ${m=0}$ is not maximally symmetric]. The position of horizons corresponds to the GR ones, ${\Lambda  r^3-3 k r+6 m=0}$, and those at ${r=\sqrt{2}\ell}$ for ${k=1}$. The scalar curvature invariants are finite at ${r=0}$ due to the regular dS core,
\begin{equation}
    a=k-\frac{r^2}{2 \ell ^2}-\frac{k r^3}{4 m \ell ^2}+O\left(r^4\right)\;,
\end{equation}
but has diverging curvature-derivative invariants as expected. If ${k=0}$, the metric possesses an additional scaling freedom, ${t\to t/S}$, ${r\to Sr}$, ${\rho\to \rho/S}$, ${m\to S^3m}$, which allows $m$ to be set to any nonzero value, as in GR.

As mentioned in Sec.~\ref{sc:regbmet}, the same function \eqref{eq:statRBHsimpleth} can also describe the regular B-metric solution when considered within the ansatz \eqref{eq:BI/BII} or \eqref{eq:BIII}. This is true as long as ${r>0}$ is within the range given by ${a>0}$ and parameters are chosen such that the denominator of \eqref{eq:statRBHsimpleth} has no positive roots in that range. For example, the regular BI-metric (${k=1}$) for ${\Lambda=0}$ splits into the solution ${r>\max(\sqrt{2}\ell,2m)}$ and ${0<r<\min(\sqrt{2}\ell,2m)}$ if ${m>0}$, while it splits into ${r>r_0>\sqrt{2}\ell}$ and ${0<r<\sqrt{2}\ell}$ if ${m<0}$; here, $r_0$ is the positive root of denominator.

Apart from \eqref{eq:simpletheory}, one may identify many other theories with closed-form solutions admitting static regular black holes. We list a few together with their static solutions in Tab.~\ref{tab:regBH}. Although algebraic, the equation \eqref{eq:feqSFA-alg} (with ${n=0}$) is non-linear and may lead to multiple branches of the solutions. As in the above example, these solutions (for suitable combinations of the $\pm$ branches and different domains) yield regular black holes at large $m$, while exhibiting singular behavior at low mass. In fact, even \eqref{eq:vacuaeq} may admit several maximally symmetric vacua when accessed through the ${k=0}$ slicing, see Tab.~\ref{tab:vac}. These vacua typically cannot be identified with one another because the corresponding equations have manifestly non-zero discriminants. 

\begin{table}[!ht]
    \centering
    \begin{tabular}{c|c|c|c}
    $\mathcal{A}$ & $\alpha_p$ & reg. & $a$\\
    \hline
    $\frac{2x}{1-x}$ & $2$ & 0 & $\frac{ \left(k r-2 m-\frac{\Lambda}{3}  r^3\right)\left(r^2-2 k \ell ^2\right)}{r^3 \left(1+\frac{2}{3}\Lambda  \ell ^2\right)-2\left( k r -2 m \right)\ell ^2}$\\
    $\frac{2x^2}{1-x^2}$ & $(-1)^{-p}+1$ & 0 & $\frac{-r^5+8 k \ell ^4 \left(k r-2 m-\frac{\Lambda  }{3}r^3\right)\pm r^2\sqrt{r^{6}-16  \ell ^4 \left(2 m+\frac{\Lambda  }{3}r^3\right) \left(k r-2 m-\frac{\Lambda  }{3}r^3\right)}}{8 \ell ^4 \left(k r-2 m-\frac{\Lambda  }{3}r^3\right)}$\\
    $\frac{2 \left(\sqrt{2}-\sqrt{2-x}\right)}{\sqrt{2-x}}$ & $(-1)^p 2^{1-p} \binom{-\frac{1}{2}}{p}$ & 2 & $\frac{\left(k r-2 m-\frac{\Lambda}{3}r^3\right) \left( \left(k r-2 m-\frac{\Lambda}{3}r^3\right)\ell ^2\pm\sqrt{4 r^4 \left(r^2-k \ell ^2\right)+ \left(k r-2 m-\frac{\Lambda}{3}r^3\right)^2\ell ^4}\right)}{2 r^4}$\\
    $\frac{2 x}{\sqrt{1-x}}$ & $-2 (-1)^p \binom{-\frac{1}{2}}{p-1}$ & 4 & long expression
    \end{tabular}
    \caption{Theories with static black holes of various degrees of regularity at ${r=0}$. (The third line solves the equations only for a restricted subset of parameters and coordinates, though algebraically well-defined more generally.)}
    \label{tab:regBH}
\end{table}

\begin{table}[!ht]
    \centering
    \begin{tabular}{c|c}
    $\mathcal{A}$ & $\Lambda_{\text{eff.}}$\\
    \hline
    $\frac{2x}{1-x}$ & $\frac{ \Lambda }{1+\frac{2}{3} \Lambda  \ell ^2}$\\
    $\frac{2x^2}{1-x^2}$ & $\frac{2 \Lambda }{1\pm \sqrt{1+\frac{16 }{9}\Lambda ^2 l^4}}$\\
    $\frac{2 \left(\sqrt{2}-\sqrt{2-x}\right)}{\sqrt{2-x}}$ & $\frac{\Lambda}{\sqrt{1+\frac{ \Lambda^2}{36}\ell^4}+\frac{\Lambda}{6}\ell^2}$\\
    $\frac{2 x}{\sqrt{1-x}}$ & long expression
    \end{tabular}
    \caption{Maximally symmetric vacua of the theories from Tab.~\ref{tab:regBH}.}
    \label{tab:vac}
\end{table}

\subsection{Taub--NUT, NHEK, and swirling --- \texorpdfstring{${n\neq 0}$}{n≠0}}

\begin{figure}
    \centering
    \includegraphics[width=0.45\linewidth]{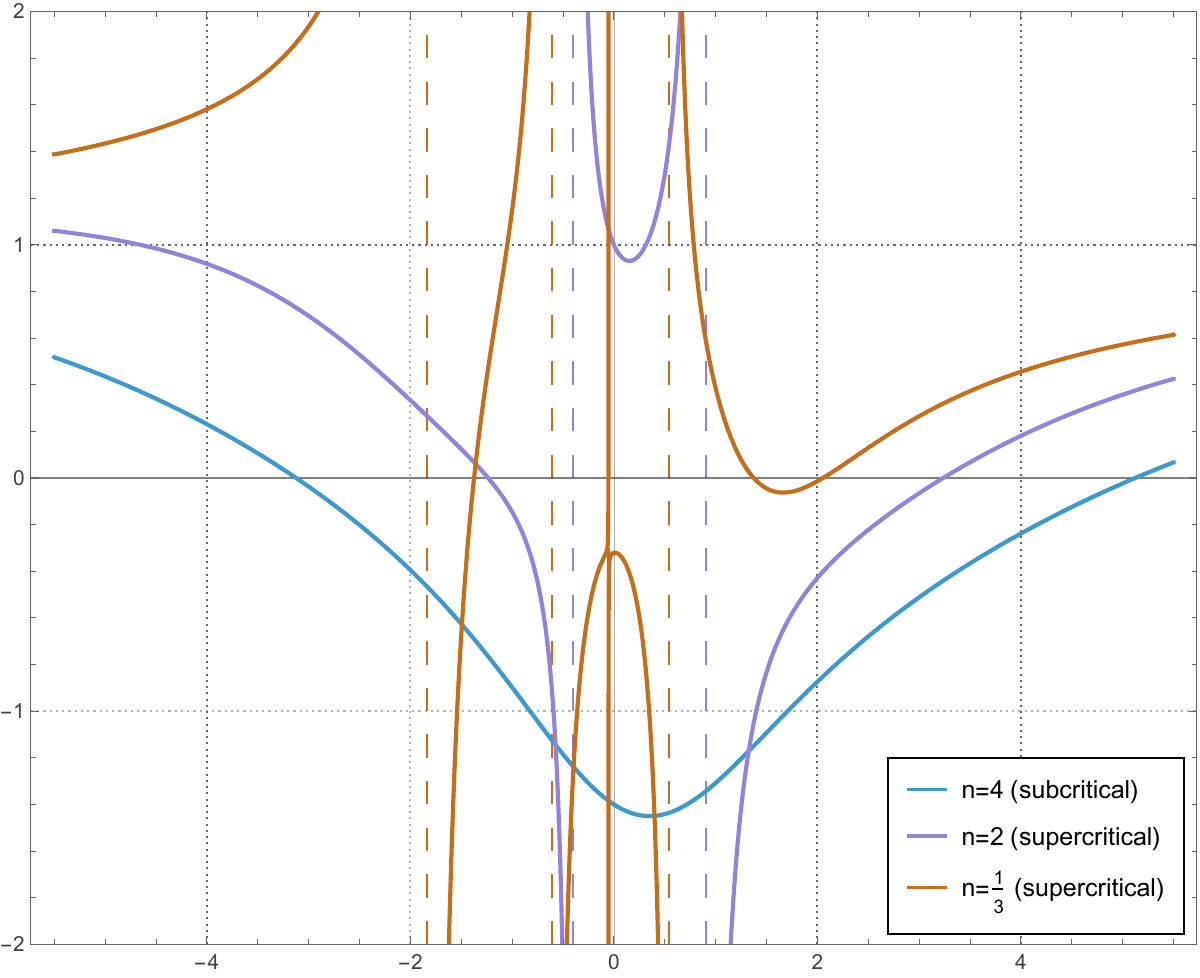}
    \caption{S. Taub--NUT solution given by \eqref{eq:TaubNUTsol} for the first-derivative QTG-TNT \eqref{eq:simpletheory}. The plot is shown for ${\ell=1}$ and ${m=1}$. The solution represents a regular Taub--NUT-type geometry at least for ${n^2}/\big(8 ({| m| }/{| n|}+1)\big)>\ell ^2$ as illustrated for ${n=4}$. For smaller $|n|$ the solution develops singularities and describes several distinct spacetimes (depending on the range of~$r$). The overall profile is sensitive to the root structure of the numerator and denominator of \eqref{eq:TaubNUTsol}. Here we show only two qualitatively different representatives, ${n=2}$ and ${n=1/3}$.}
    \label{fig:plot2}
\end{figure}

Let us consider the s. topology ${k=1}$ and set ${\Lambda=0}$ for simplicity. Then \eqref{eq:gensol_simpletheory} in the simple theory \eqref{eq:simpletheory} takes the form
\begin{equation}\label{eq:TaubNUTsol}
    a=\frac{\left(r^2+n^2\right) \left( r^2-n^2-2 m r\right) \left(r^2+n^2-2 \ell ^2\right)}{ \left(r^2+n^2\right)^3-2 \ell ^2 \left(r^2-3 n^2\right) (r^2-n^2-2 m r)}\;,
\end{equation}
which is depicted in Fig.~\ref{fig:plot2}. Notice that the denominator is negative at ${r=0}$ for ${\sqrt{6}\ell>|n|}$ while it remains positive at ${r\to\pm\infty}$, implying that there always exists a real root of the denominator for small $|n|$ in contrast to GR ($\ell=0$); unless it accidentally matches the numerator roots when a specific relation between $m$ and $n$ (for some $\ell$) is met. On the other hand, it is possible to show that ${{n^2}/\big(8 ({| m| }/{| n|}+1)\big)>\ell ^2}$ implies the strictly positive denominator.\footnote{Indeed, one can derive the bound $\left(r^2+n^2\right)^3-2 \ell ^2 \left(r^2-3 n^2\right) \left(r^2-n^2-2 m r\right)\geq  \left(n^2+r^2\right)^3-2 \ell ^2 \left| r^2-3 n^2\right|  \left| r^2-n^2-2 m r\right| \geq \left(r^2+n^2\right)^2 \left( r^2+n^2-8 \ell ^2 \left(\frac{| m| }{| n|}+1\right)\right)\geq n^4 \left( n^2-8 \ell ^2 \left(\frac{| m| }{| n|}+1\right)\right)>0$, where we made us of $\left| r^2-3 n^2\right| \leq r^2+3 n^2\leq 4 \left(r^2+n^2\right)$, ${\left| r^2-n^2-2 m r\right| \leq r^2 +n^2+ 2 | m|  | r|\leq \left(r^2+n^2\right) \left(\frac{| m| }{| n|}+1\right)}$ implying $\left| r^2-3 n^2\right|  \left| r^2-n^2-2 m r\right| \leq 4 \left(r^2+n^2\right)^2 \left(\frac{| m| }{| n|}+1\right)$.} As the root of the denominator can be shown to correspond to the scalar curvature singularity, the regular s. Taub--NUT-type geometry cannot generically occur for small values of $|n|$, but it can occur, for example, for ${{n^2}/\big(8 ({| m| }/{| n|}+1)\big)>\ell ^2}$, which we will assume below. Observe that there are only two horizons, since ${n^2-2\ell^2>0}$ by this assumption. They are located at the GR positions ${r=m\pm\sqrt{m^2+n^2}}$. As expected, the closed time-like curves are still present near the symmetry axis with the Misner string (${\rho=0}$ with ${{\bs{\omega}}_{1}\to\tilde{\bs{\omega}}_{1}}$) and do not extend to infinite $y$ for any $x$, where ${x=r\sqrt{1-\rho^2}}$, ${y=r\rho}$. 

Taub--NUT-type solutions can be also extended to other topologies and an arbitrary cosmological constant as a solution of \eqref{eq:feqSFA-alg} with ${n\neq0}$ by 
\begin{equation}\label{eq:gensol_simpletheory}
    a=\frac{  \left(r^2+n^2\right) \left(r^2+n^2-2 k \ell ^2\right)\Upsilon}{\left(n^2+r^2\right)^3+2   \ell ^2 \left(3 n^2-r^2\right)\Upsilon}\;,
\end{equation}
where we introduced 
\begin{equation}
    \Upsilon=k \left(r^2-n^2\right)-2 m r-\tfrac{\Lambda}{3}  \left(r^4+6 n^2 r^2-3 n^4\right)\;.
\end{equation}
Analogous solutions to the other theories mentioned above, which share qualitatively similar properties, are summarized in Tab.~\ref{tab:sols} and will not be discussed further. The GR limit of \eqref{eq:gensol_simpletheory}, ${\ell\to0}$, recovers the s./h./p. Taub--NUT--(A)dS solutions \eqref{eq:GRTNUT}. The asymptotic expansion reads
\begin{equation}
    a=-\frac{\Lambda_{\text{eff.}} }{3}r^2+\left(\left(1-\frac{2}{3} \Lambda_{\text{eff.}} \ell ^2+\frac{4}{9} \Lambda_{\text{eff.}}^2 \ell ^4\right)k-\frac{5}{3}\Lambda_{\text{eff.}}n^2\right) -\left(1-\frac{2}{3} \Lambda_{\text{eff.}} \ell ^2\right)^2\frac{2 m }{r}+O\left(\frac{1}{r^2}\right)\;.
\end{equation}
Similarly to the static case, it matches the expansion \eqref{eq:asymtTNUT} for ${\Lambda_{\text{eff.}}=0}$ and differs in the sub-leading terms for ${\Lambda_{\text{eff.}}\neq0}$. The horizons are again located at the GR values ${\Upsilon=0}$ and at ${r=\pm\sqrt{2k\ell^2-n^2}}$ for ${k>0}$. In contrast to the ${n=0}$ case, the latter is no longer fixed only by the theory, but depends also on the solution parameter $n$.

\begin{table}[!ht]
    \centering
    \begin{tabular}{c|c}
    $\mathcal{A}$ & $a$\\
    \hline
    $\frac{2x}{1-x}$ & $
    \frac{  \left(r^2+n^2\right) \left(r^2+n^2-2 k \ell ^2\right)\Upsilon}{\left(n^2+r^2\right)^3+2   \ell ^2 \left(3 n^2-r^2\right)\Upsilon}
    $\\
    $\frac{2x^2}{1-x^2}$ & $\frac{-\left(r^2+n^2\right)^5+ 8 k   \ell ^4 \left(r^2-3 n^2\right)\left(r^2+n^2\right)\Upsilon\pm\left(r^2+n^2\right)^2 \sqrt{\left(r^2+n^2\right)^6+16   \ell ^4 \left(3 n^2-r^2\right) \left(k \left(r^2+n^2\right)^2+  \left(3 n^2-r^2\right)\Upsilon\right)\Upsilon}}{8   \ell ^4 \left(r^2-3 n^2\right)^2\Upsilon}$
    \\
    $\frac{2 \left(\sqrt{2}-\sqrt{2-x}\right)}{\sqrt{2-x}}$ & $\frac{ \ell ^2 \left(r^2-3 n^2\right)\Upsilon ^2\pm\sqrt{4  \left(r^2+n^2\right)^5 \left(r^2+n^2-k \ell ^2\right)\Upsilon ^2+ \ell ^4 \left(r^2-3 n^2\right)^2\Upsilon ^4}}{2 \left(r^2+n^2\right)^4}$
    \\
    $\frac{2 x}{\sqrt{1-x}}$ & long expression
    \end{tabular}
    \caption{TNT SF solutions of theories from Tab.~\ref{tab:regBH}. (The third line solves the equations only for a restricted subset of parameters and coordinates, though algebraically well-defined more generally.)}
    \label{tab:sols}
\end{table}

As explained in Sec.~\ref{sc:nhekswirl}, the same function \eqref{eq:gensol_simpletheory} (for some parameter choices) also describes the geometries arising as near-horizon limits of extremal rotating black holes (assuming the same enhanced $\mathrm{AdS}_2$ symmetry as in GR). Similarly, the function \eqref{eq:gensol_simpletheory} also contains a case corresponding to the swirling universe, interpretable as a rotating gravitational field generated by counter-rotating sources at opposite ends of the axis.

\begin{figure}
    \centering
    \includegraphics[width=0.45\linewidth]{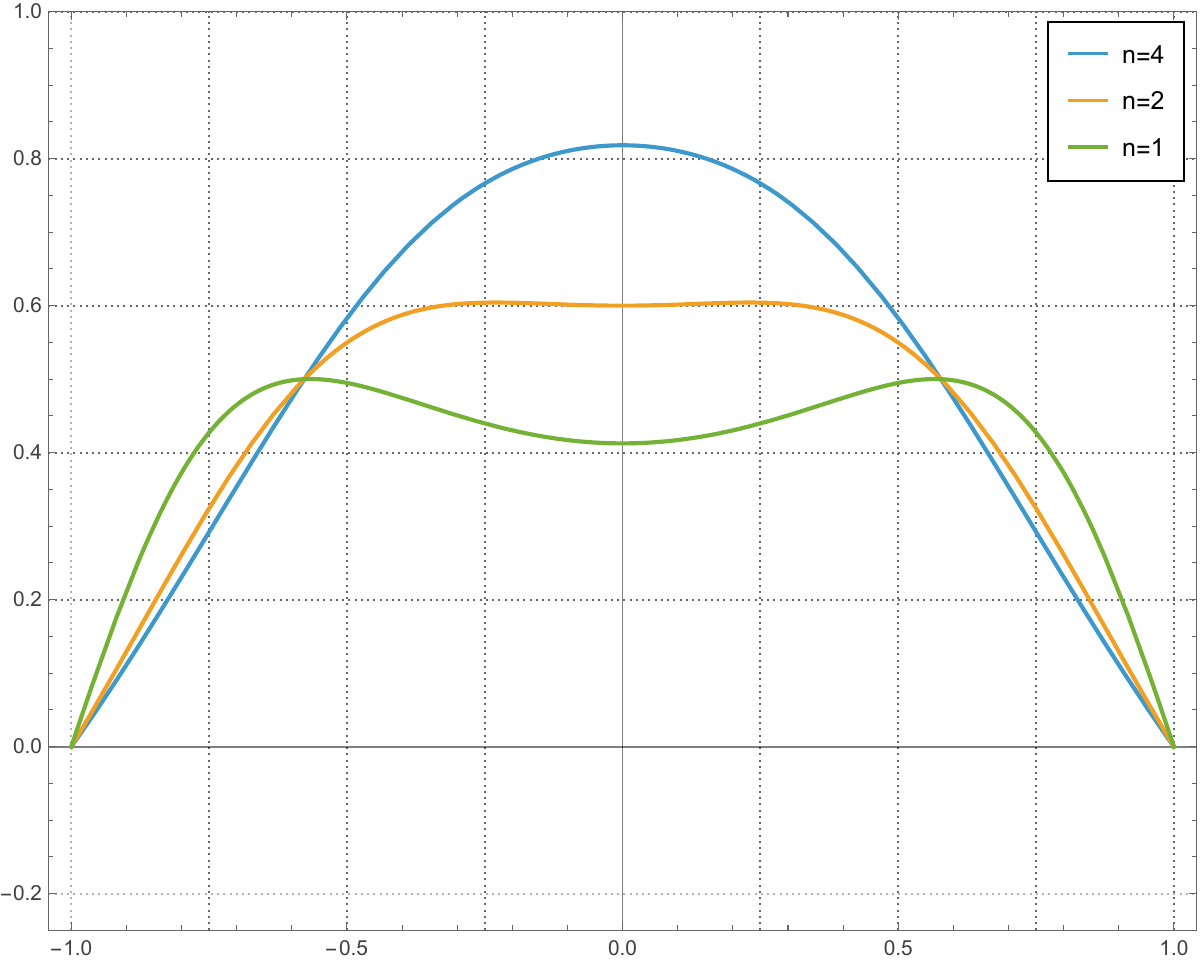}
    \includegraphics[width=0.45\linewidth]{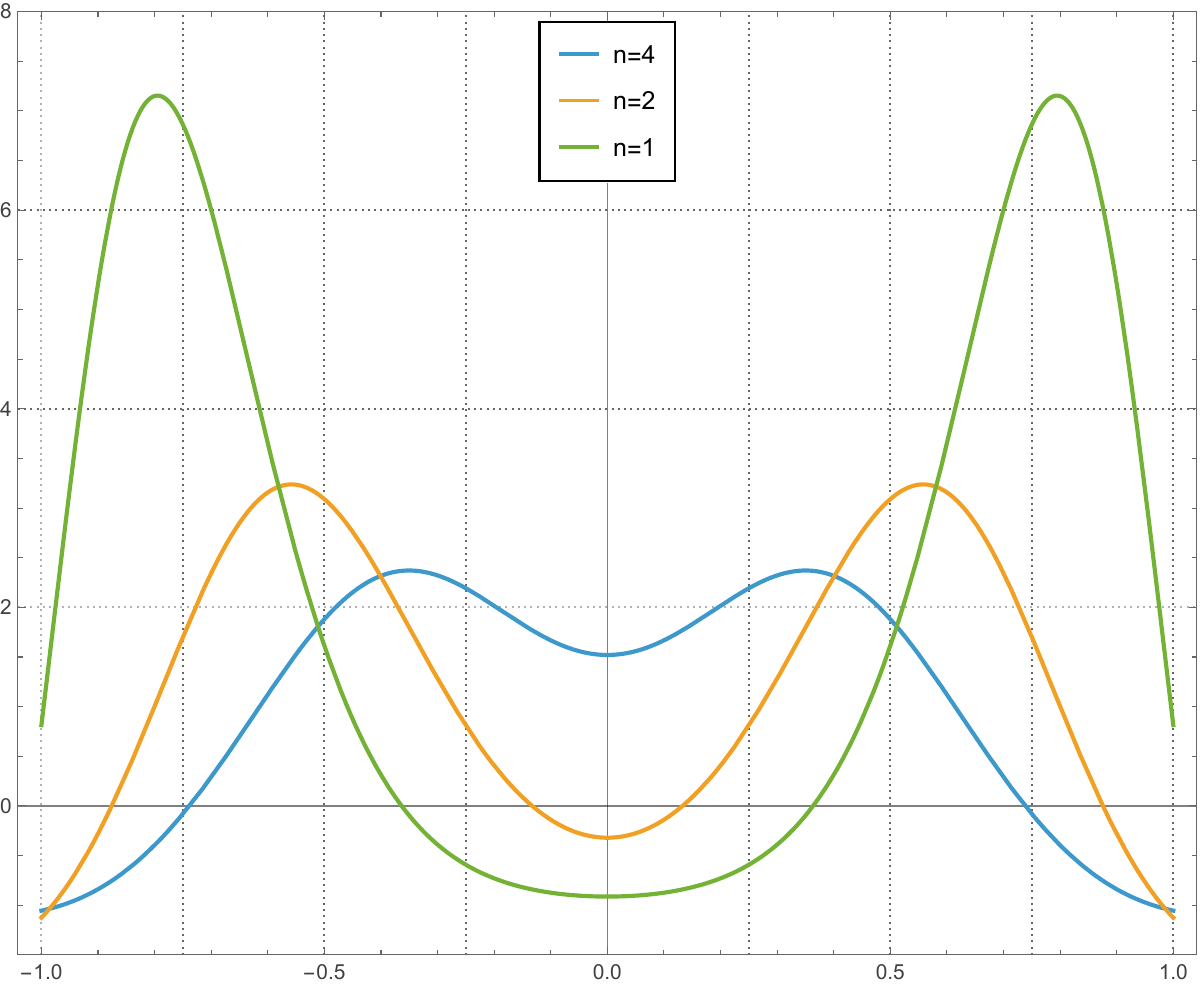}
    \caption{NHEK solution $a(n\breve{r})$ given by \eqref{eq:NHEKsol} (left) and the corresponding scalar curvature  of the horizon ${n^2R_{\mathcal{H}}(n\breve{r})}$ (right) for the first-derivative QTG-TNT \eqref{eq:simpletheory}. The plot is shown for ${\ell=1}$. The solution is plotted for ${n=4,2,1}$ to show all possibilities for sign of ${R_{\mathcal{H}}}$ at the horizons and equator. In particular, ${R_{\mathcal{H}}|_{r=0}<0}$ for ${n^2<2 \left(\sqrt{3}+1\right) \ell ^2}$ and ${R_{\mathcal{H}}|_{r=\pm n}>0}$ for ${n^2<\ell^2}$. }
    \label{fig:plot3plot4}
\end{figure}

The NHEK-type solution is given by \eqref{eq:NHEK} [or \eqref{eq:NHEK-sh}] with \eqref{eq:gensol_simpletheory} where we set ${k=-1}$ (as it corresponds to the double-Wick-rotated h. Taub--NUT). Although one could be more general at this point, we will also set ${m=0}$, which implies evenness, ${a(-r)=a(r)}$. This then automatically guarantees \eqref{eq:cond} (making the conical defects equal on both poles, so a specific $\phi$-periodicity can remove them). As a consequence of $a$ being even, the only condition we seek is smooth positive ${a>0}$ between roots ${a(r_+)=a(r_-)=0}$, ${r_+=-r_->0}$, with non-zero derivative at the roots. This is precisely the condition on the massless h. Taub--NUT admitting two non-extreme horizons with stationary regular region in between. If we also set ${\Lambda=0}$, for simplicity, we get
\begin{equation}\label{eq:NHEKsol}
    a=\frac{\left(n^4-r^4\right) \left(r^2+n^2+2 \ell ^2\right)}{\left(r^2+n^2\right)^3+2 \ell ^2 \left(3 n^2-r^2\right) (n^2-r^2)}\;,
\end{equation}
which it reproduces the usual NHEK in the GR limit ${\ell\to0}$. Even for ${\ell>0}$, as illustrated in Fig.~\ref{fig:plot3plot4}, the solution can be interpreted as a regular near-horizon extreme geometry for any ${n\neq0}$, because the roots of the numerator remain at the GR values ${r=\pm |n|}$, while the denominator is strictly positive. These roots correspond to the north and south poles, with $r$ playing the role of a latitudinal angular coordinate on the topological 2-sphere (${r=|n|\cos\theta}$). We can verify the expected regular behavior \eqref{eq:cond} at the poles ${r=r_{\pm}=\pm |n|}$, with the ${\phi}$-periodicity parameter set to 
\begin{equation}
    \nu=\frac{2|n|^3}{n^2+\ell^2}\;.
\end{equation}
A notable difference from the GR NHEK is in the Ricci scalar of the induced 2-dimensional metric at the horizon ${R_{\mathcal{H}}=-a''}$. In GR, it is positive at the equator, ${R_{\mathcal{H}}|_{r=0}=4/n^2>0}$, and negative at the poles, ${R_{\mathcal{H}}|_{r=\pm n}=-1/n^2<0}$. In this theory, however,
\begin{equation}
    R_{\mathcal{H}}|_{r=0}=\frac{4 \left(n^4-4 n^2 \ell ^2-8 \ell ^4\right)}{n^2\left(n^2+6 \ell ^2\right)^2}\;, \quad R_{\mathcal{H}}|_{r=\pm n}=-\frac{(n^2-\ell^2)(n^2+2\ell^2)}{n^6}\;,
\end{equation}
the horizon Ricci scalar can have different signs. In particular, $R_{\mathcal{H}}|_{r=0}$ becomes negative for ${n^2<2 \left(\sqrt{3}+1\right) \ell ^2}$, while $R_{\mathcal{H}}|_{r=\pm n}$ can be positive for ${n^2<\ell^2}$, see Fig.~\ref{fig:plot3plot4}. This signals a possible nontrivial distortion of the horizon for small~$|n|$.

\begin{figure}
    \centering
    \includegraphics[width=0.45\linewidth]{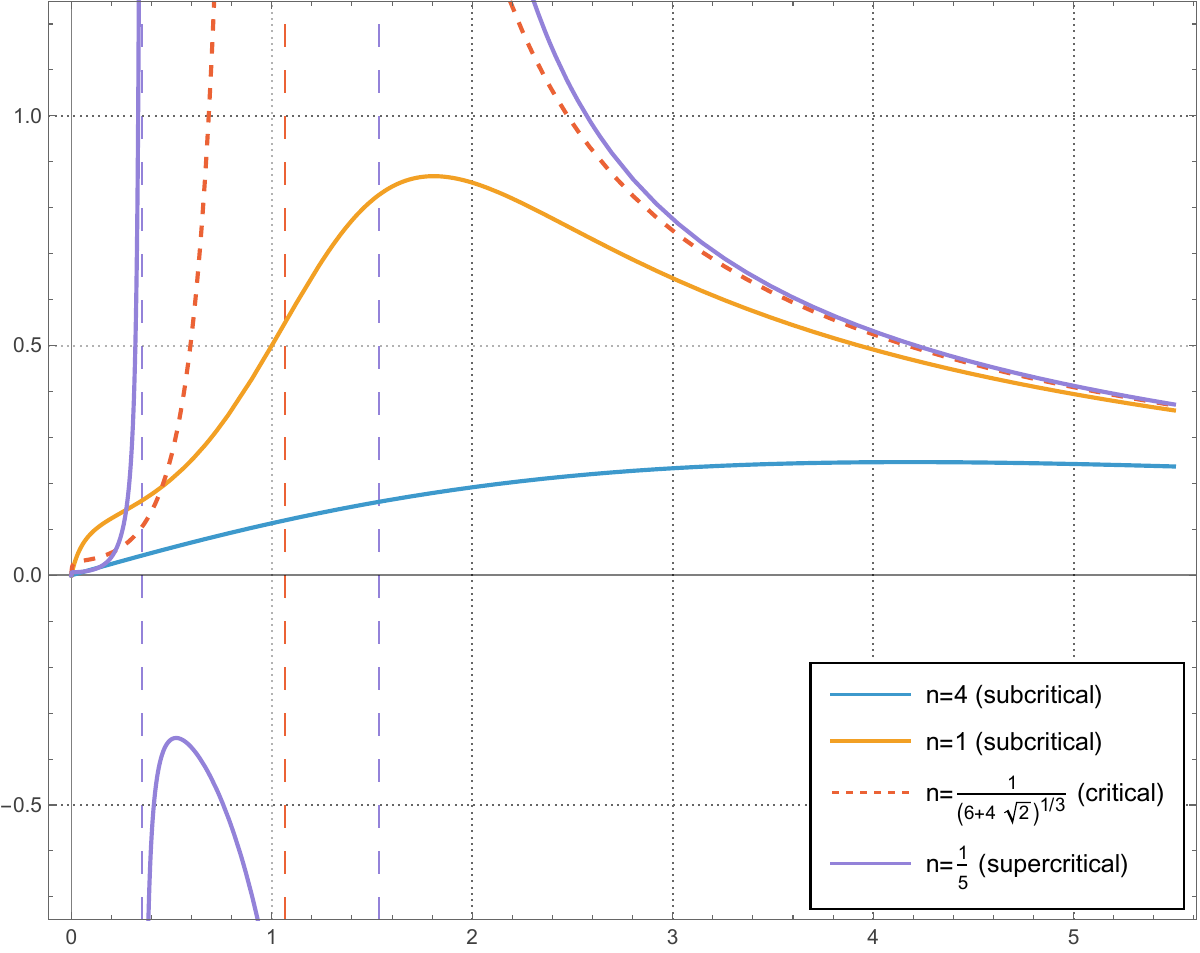}
    \caption{Swirling solution given by \eqref{eq:swirlsolsimpleth} for the first-derivative QTG-TNT \eqref{eq:simpletheory}. The plot is shown for ${\ell=1}$ and ${m=-1}$. The solution is regular for ${|n|^3>-m\ell^2/(6+4\sqrt{2})}$. Below this critical value it describes three distinct singular spacetimes (depending on the range of $r$).}
    \label{fig:plot5}
\end{figure}

The swirling-type solution corresponds to \eqref{eq:swirl} with \eqref{eq:gensol_simpletheory} where we set ${k=0}$. Focusing on non-compact topology of the ${\tau=x=0}$ surface, we require a root ${a(r_0)=0}$ and smooth positive ${a>0}$ at ${r>r_0}$ with positive derivative at the root. This corresponds to the condition on the p. Taub--NUT admitting a non-extreme horizon with stationary regular region reaching infinity. Setting ${\Lambda=0}$, for simplicity, we have
\begin{equation}\label{eq:swirlsolsimpleth}
    a=\frac{-2 m r \left(r^2+n^2\right)^2}{\left(r^2+n^2\right)^3+4\ell ^2 m r  \left(r^2-3 n^2\right)}\;,
\end{equation}
recovering the familiar swirling-universe solution in the GR limit ${\ell\to0}$. The graph is shown in Fig.~\ref{fig:plot5}. Observe that we need ${m<0}$ because $a$ must remain positive at ${r\to\infty}$. Also, the numerator has a root at ${r=0}$, which is our candidate symmetry axis. Recall that $r$ plays a role of a radial cylindrical coordinate now.  However, for sufficiently small $|n|$ the denominator acquires a root at some ${r>0}$, signaling a curvature singularity. This is clear as the denominator is negative at ${r=2|n|}$ for ${|n|<2 (-m)^{1/3} \ell ^{2/3}/5}$ while remaining positive for ${r\to\infty}$. On the other hand, the determinant is positive for all ${r>0}$ if ${|n|^3=-m\ell^2/(6+4\sqrt{2})}$.
With this in mind, we can check that the regular behavior \eqref{eq:cond2} at the symmetry axis ${r=r_0=0}$ is satisfied if the $q$-periodicity parameter is set to
\begin{equation}
    \nu=\frac{n^2}{-m}\;.
\end{equation}
Compared to the GR swirling, there is not a significant difference at large $r$,
\begin{equation}
    a=-\frac{2 m}{r}+\frac{2 m n^2}{r^3}+\frac{8 m^2 \ell ^2}{r^4}+O\left(\frac{1}{r^5}\right)\;,
\end{equation}
but a notable sub-leading term appears in the expansion at the symmetry axis ${r=0}$,
\begin{equation}
    a=-\frac{2 m r}{n^2}-\frac{24 m^2 \ell ^2}{n^6}r^2 +O\left(r^3\right)\;.
\end{equation}
Finally, recall the freedom ${t\to t/S}$, ${r\to Sr}$, ${\rho\to \rho/S}$, ${m\to S^3m}$, which allows us to set, e.g., ${m=-|n|/2}$, while maintaining ${m<0}$.

\subsection{Eguchi--Hanson --- \texorpdfstring{${\epsilon=0}$}{epsilon=0}}

Let us move on to the Euclidean solutions still within ${\mathcal{R}_4=0}$ but with ${b=-r^2/n^2}$ (i.e., ${\epsilon=0}$). The solution of \eqref{eq:feqalgeps0} for the simple theory \eqref{eq:simpletheory} then takes the form
\begin{equation}
    a = \frac{\left(r^2+n^2\right)  \left( k \left(r^2+n^2\right)^2- q-\frac{2}{3}\Lambda  \left(r^2+n^2\right)^3\right)\left(r^2+n^2-2 k \ell ^2\right)}{4 r^2 \left(\left(1+\frac{4}{3}\Lambda  \ell ^2\right) \left(r^2+n^2\right)^3-2\ell ^2 \left(k \left(r^2+n^2\right)^2-q\right)\right)}\;,
\end{equation}
or, equivalently, in the metric form \eqref{eq:EHmet} by ${j=1}$ and
\begin{equation}
    h=\frac{ \left( k y^4- q-\frac{2}{3}\Lambda  y^6\right)\left(y^2-2 k \ell ^2\right)}{4 \left(y^6 \left(1+\frac{4}{3}\Lambda  \ell ^2\right)-2 \ell ^2(k y^4- q )\right)}\;,
\end{equation}
which, in the limit ${\ell\to0}$, reproduce the GR solutions \eqref{eq:EH_GRab} and  \eqref{eq:EH_GRhj}, respectively. In a simple subcase, ${\Lambda=0}$ and ${k=1}$, we get
\begin{equation}
    h=\frac{\left(y^4-q\right) \left(y^2-2 \ell ^2\right)}{4 \left(y^6-2\ell ^2 (y^4- q )\right)}\;.
\end{equation}
The numerator has roots at the GR value ${y=q^{1/4}}$ and at the fixed position ${y=\sqrt{2}\ell}$. Assuming ${q>0}$ (e.g., to match GR asymptotics at ${y\to\infty}$), we denote ${y_0=\max(q^{1/4},\sqrt{2}\ell)}$. The denominator is an increasing function for ${y>{2 \ell}/{\sqrt{3}}}$, hence, certainly for ${y\geq y_0}$. Furthermore, it evaluates at ${y=y_0}$ to a positive value (either $q^{3/2}$ or $2q\ell^2$). As a consequence, the denominator must be strictly positive for ${y\geq y_0}$, resulting in regular curvature invariants. To match the expected regular behavior \eqref{eq:regularityEH}, we need to choose the $\psi$-periodicity parameter as
\begin{equation}
\nu=
\begin{cases} 
    \frac{2 \sqrt{q}}{\sqrt{q}-2 \ell^2}\;, &    q^{1/4}/\sqrt{2}>\ell\;,\\
    \frac{4 q}{4 \ell^4-q}\;, &   q^{1/4}/\sqrt{2}<\ell\;.
\end{cases}
\end{equation}

\subsection{Theories with non-SF solutions --- \texorpdfstring{${P=0}$}{P=0}}

Observe that $P$ is an analytic function in $a$ and $b$ which does not depend on their derivatives. This is analogous to what happens in s./h./p. symmetric Lovelock--Lanczos gravity.
The presence of branches of solutions satisfying ${P=0}$ can depend on the specific fine-tuning and constraints among the coupling constants $\alpha_p$ (and possibly $\Lambda$). Just like in Lovelock--Lanczos gravity, there are also degenerate cases of underdetermined solutions appearing for some critical values of the coupling constants. Although new solutions are often present at finite order $p$, they can be interpreted as truncation artifacts when ${P \neq 0}$ in the limit of an infinite tower of corrections. 

Consider first the theory given by \eqref{eq:simpletheory}. Since, $P$ is given by
\begin{equation}
    P=\frac{\left(r^2+n^2\right) \left(r^2+n^2-2 k \ell ^2\right)}{\left(2 \ell ^2  \left(r^2-3 n^2 b\right)a+\left(r^2+n^2\right) \left(r^2+n^2-2 k \ell ^2\right)\right)a }\;,
\end{equation}
which can never vanish, it implies that there are no other solutions than those satisfying ${\mathcal{R}_4=0}$ discussed above. In fact, this is a rather special case. If ${b= r^2/(3n^2)}$, then the equation ${P=0}$ becomes ${\mathcal{A}\left({2 k \ell ^2}/{r^2+n^2}\right)=-2}$ and there is no solution since we need ${\mathcal{A}(0)=0}$. Assuming ${b\neq r^2/(3n^2)}$, the equation ${P=0}$ in the general case can be recast to the form
\begin{equation}\label{eq:Pmod}
    -2k \ell ^2 \mathcal{A}'(x)+\left(r^2+n^2\right) \left(2+\mathcal{A}(x)+x \mathcal{A}'(x)\right)=0\;,
\end{equation}
where we eliminated $a(r)$ at the expense of ${x=x(r)}$\;. If ${x(r)=x_0}$ for some constant $x_0$, then for this equation to hold at some $r$, necessarily ${\mathcal{A}(x_0)=-2}$ and ${\mathcal{A}'(x_0)=0}$; which forces ${x_0\neq0}$. Rewriting ${x(r)=x_0}$ in terms $a$ and $b$ and imposing also the field equation \eqref{eq:varb}, we find that the solution exists only for ${\Lambda=k=0}$ and reads
\begin{equation}
    a=-\frac{x_0 \left(r^2+n^2\right)^2}{2 \ell ^2 \left(r^2-3 n^2 b\right)}\;,
\end{equation}
where $b$ is an arbitrary function, indicating that the solution is underdetermined. If $x$ is non-constant and ${k=0}$ then ${2+\mathcal{A}(x)+x \mathcal{A}'(x)=0}$ at some interval, but then again there is no theory satisfying ${\mathcal{A}(0)=0}$ condition. Assuming ${k\neq0}$ and ${2+\mathcal{A}(x)+x \mathcal{A}'(x)\neq0}$, we can further rewrite \eqref{eq:Pmod} as
\begin{equation}
    H(x(r))=\frac{r^2+n^2}{2k \ell ^2}\;,
\end{equation}
where we denoted
\begin{equation}
     H(x)=\frac{\mathcal{A}'(x)}{2+\mathcal{A}(x)+x \mathcal{A}'(x)}\;.
\end{equation}
If $H$ is invertible at a given $x$, we can write ${x(r)=H^{-1}\big((n^2+r^2)(2k \ell ^2)\big)}$. Clearly, the globally non-invertible case ${H(x)=H_0}$ for some constant $H_0$ is ${\mathcal{A}=2H_0x/(1-H_0x)}$. The constant ${H_0}$ can be set to ${+1}$ or ${-1}$ by rescaling of ${\ell}$, where the former is exactly the theory \eqref{eq:simpletheory}. Otherwise, excluding points of non-invertibility, the equation ${P=0}$ should admit a solution. Despite this fact, it can still happen that the other equation \eqref{eq:varb} is not satisfied; although we have not been able to solve all the cases, this seems to be the situation of the theories in Tab.~\ref{tab:regBH}. 
 
Let us also explore the finite order of $p$ starting with ${p=1}$,
\begin{equation}
    \mathcal{A}=x\;,
\end{equation}
where we already set ${\alpha_1=1}$ by rescaling of $\ell$ (without loss of generality for ${\alpha_1>0}$). Solving ${P = 0}$, we obtain the exact solution,
\begin{equation}
    a = \frac{\left(r^2+n^2\right) \left( r^2+n^2 + k \ell^2 \right)}{2 \ell^2 \left( r^2 - 3 n^2 b \right)}\;.
\end{equation}
It still contains $b$ which can be solved for from the remaining field equation that is also algebraic in $b$. If ${n\neq0}$, it yields a pair of solutions non-analytic in $n$,
\begin{equation}
    b = \frac{r^2}{3 n^2}  \, \frac{3 k^2 \ell^4 + 9 \left(1+\frac43 \ell^2 \Lambda \right)  \left(r^2+n^2\right)^2 \pm 2 \sqrt{3} \left(k \ell^2+ r^2+n^2\right) \sqrt{k^2 \ell^4 - 4 k \ell^2 \left(r^2+n^2\right)  + 6  \left(1+\frac43 \ell^2 \Lambda \right)  \left(r^2+n^2\right)^2 }}{k^2 \ell^4 - 10 k \ell^2 \left(r^2+n^2\right) +  \left(1+12 \ell^2 \Lambda \right) \left(r^2+n^2\right)^2}\;.
\end{equation}
Consider now the order ${p=2}$,
\begin{equation}
    \mathcal{A}=-x^2\;,
\end{equation}
where we set ${\alpha_2=-1}$. It admits the following underdetermined solutions for ${n=k=0}$ and ${\Lambda = \pm \sqrt{2/3}/\ell^2}$ given by 
\begin{equation}
    a = - \frac{\Lambda r^2 }{2} \;,
\end{equation}
and arbitrary $b$.

\section{Discussion}

In this work we have provided a complete classification of QTG-TNT theories. These are the 4-dimensional gravitational theories depending (analytically or non-analytically) only on the Riemann tensor which exhibit the integrability property for the full class of TNT metrics --- i.e., not only the geometries with symmetries of s./h./p. Schwarzschild but also s./h./p. Taub--NUT, their double-Wick rotations such as the B-metrics, NHEK, swirling, and the Eguchi--Hanson instanton. Using the principle of symmetric criticality and invariant representatives of the curvature, we have constructed the most general effective actions and identified all theories whose field equations reduce to TNT ansatz are zeroth-, first-, second-, or third-order (one less after the trivial integration of the left-over equation). We have shown that all theories with first and second order field equations necessarily require non-analytic dependence on the curvature, while polynomial representatives have been obtained up to quintic order in the Riemann tensor for third-order QTG-TNT.
Finally, within the unique theory admitting a first-order sector (i.e., algebraic field equations after one trivial integration --- the same integrability as in GR) we derived closed-form solutions for all TNT symmetries. These include regular static black holes, Taub--NUT, NHEK, swirling, and Eguchi-Hanson solutions.

In the process, we have identified a covariant algebraic definition of SF ansatz within TNT metrics, given by the condition ${\mathcal{R}_4=0}$, i.e., ${ \mathscr{I}_6 =  3 \mathscr{I}_{11}/\mathscr{I}_1}$, where $\mathscr{I}_k$ are the Zakhary--McIntosh invariants. This condition selects the Segre type of TF Ricci tensor as \{(1,1),(11)\} (i.e., the null alignment type D), while the Weyl tensor remains of Petrov type D. Naturally, the covariant properties of SF ansatz may help in identifying more general QTG theories, including those involving covariant derivatives of the Riemann tensor or matter fields.

Our results show that there is no 4-dimensional metric theory depending solely on the Riemann tensor which is QTG-TNT and admits a Wheeler-polynomial-like field equation in static s./h./p. symmetry. This can be understood from the fact that, although this is possible if the theory is assumed QTG only for the (static) s./h./p. symmetry (see for static \cite{Frolov:2024hhe} and \cite{Colleaux:2017ibe,Colleaux:2019ckh} for dynamical case), there is naturally less theories which are QTG for all TNT metrics.

\medskip

The exact solutions in all the first-order QTG-TNT theories have certain issues. For example, the static s. symmetric regular black holes become singular for low mass $m$. This is a consequence of the curious property that these black holes always feature (at least) two horizons, one of which being at ${r=2m}$, where $m$ is the ADM mass of general relativity, while the other is $m$-independent and fixed by the theory, i.e. $r\propto \ell$, where $\ell$ is the dimensionful length scale controlling the high-energy corrections to general relativity in the theory \eqref{eq:I1s4}. However, this could as well be seen as a concrete realisation of Bronstein's argument (see \cite{Rovelli:2014ssa} and references therein) in favor of a universal minimal length for black holes, the Planck length, in an otherwise diffeomorphism invariant theory. Furthermore, low mass black holes are expected to be highly quantum objects, owing to their large temperature, so that the semi-classical description of black holes as effective spacetimes should break down in this regime. In this sense, the low mass singularities are not strictly speaking an issue, especially because these black holes reach extremality (due to Hawking radiation) before reaching these masses, meaning ${m_{\text{sing.}} < m_{\text{extr.}} }$, so that the black holes (thermodynamically) stabilize as remnants and the region of the parameter space for which the metric is singular is excluded if the black hole forms with mass larger than $m_{\text{extr.}}$. Interestingly, the NHEK solutions remain always regular, although their horizons become severely deformed for small values of $|n|$. This again fits nicely with the interpretation that black holes with large temperature might not be possible to describe by effective classical geometries, while zero temperature objects, such as extremal black holes might. Regarding the Taub--NUT solutions, they are also singular for small NUT parameter $|n|$. Although these geometries are in general more difficult to interpret, considering the NUT as a magnetic mass (see \cite{DemianskiNewman1966,Dowker:1974znr,Lynden-Bell:1996dpw,Bicak:2000ea,Griffiths:2009dfa}) which decreases during the evaporation process\footnote{Notice that having a first law for Lorentzian Taub--NUT where the NUT can be independently varied (see \cite{Hennigar:2019ive}) can also be viewed as an argument in favor of NUT emission during evaporation. However, there is no known particle which could carry NUT charge, analogous to the hypothetic magnetic monopoles.} (see e.g. \cite{Chakraborty:2022ltc,Arevalo:2024kmo})
would again indicate that low NUT corresponds to highly quantum objects compared to large ones. 
 In order to clarify and possibly confirm these observations, a detailed study of the thermodynamical properties of our solutions should be carried out. However, notice that the previous interpretation does not seem to hold for our swirling solutions, which are also singular for small swirling parameter, while being horizonless, and so without temperature. This might be due to the fact that these solutions are not (locally) asymptotically flat.

Although QTG-TNT theories admitting regular black holes tend not to possess non-trivial solutions beyond the (covariant) SF ansatz, we identified others that do admit such solutions, although some remain underdetermined. Moreover, most topological QTG-TNT theories do not yield trivial field equations beyond SF ansatz. This therefore provides an interesting way to discriminate among these topological QTG-TNT models, depending on whether they admit exact solutions in this non-SF branch or not.

We restricted our search for exact solutions to the simplest first-derivative QTG-TNT theories. Naturally, it would be interesting to analyze the solutions in theories with second- and third-order field equations. Although more complicated than the first-order QTG-TNT field equations, they are still significantly simpler than the standard polynomial higher-order gravity \cite{Holdom:2002xy,Lu:2015cqa,Lu:2015psa,Hennigar:2016gkm,Bueno:2016lrh,Podolsky:2018pfe,Svarc:2018coe,Podolsky:2019gro,Pravda:2020zno,Daas:2023axu,Giacchini:2024exc,Giacchini:2025gzw}. This could enable one to understand the effect and possible interpretations of the additional integration constants appearing in these context in a much simpler setup.

\medskip

It would be natural to construct QTG-like theories also for other metric classes (more general symmetries, etc.), such as the stationary axisymmetric spacetimes. In this case, there might be some ambiguity regarding which integrability condition of GR one wishes to preserve. In QTG-TNT this is the SF subclass. However, for a more general ansatz in GR, many Lie point symmetries may exist. Even preserving some of them would be an achievement and could be sufficient to obtain QTG-like theories, for example, with exact rotating black holes. Furthermore, it is clear that the more integrability conditions are present in a theory, the more constrained it is. This could provide a way to construct non-analytic gravitational theories with certain uniqueness properties, unlike the large degeneracy in the choice of representatives existing in QTG (for TNT or even worse for dynamical s./h./p. symmetry). Regarding the construction of QTG-like theories for other ansatzes, one may be forced to abandon the reduction of the Lagrangian. This procedure is rigorously defined only for ansatzes given by most general symmetry-invariant metrics, and even then it may not reproduce all (or the correct) reduced field equations \cite{Fels:2001rv,Frausto:2024egp}. It remains unclear how one should proceed in constructing QTG-like theories under these circumstances.

One of the issues of theories that are non-analytic in curvature is that the maximally symmetric spacetimes (or any conformally flat metrics in our case) are not genuine solutions of the theory unless a specific (possibly dubious) limiting procedure is taken. This may pose a problem if one wishes to study the formation of regular black holes in these models, since modeling the interior with homogeneous and isotropic cosmologies would encounter precisely these difficulties. On the other hand, if the theories are considered as effective theories describing the strong-field regime, then the maximally-symmetric limit is not important as long as the models reduce to GR for vanishing couplings.

It is also natural to be concerned about the stability of maximally symmetric backgrounds. First, there are issues with the non-analyticity in the curvature, since the field equations cannot be evaluated on such backgrounds without taking appropriate limits. Second, even if the first-order perturbations coincide with those of GR,\footnote{For instance interesting 4D non-analytic representatives of the 2D Horndeski reduction of Lovelock gravity have been shown to have well-defined first order perturbations around maximally symmetric backgrounds in \cite{Bueno:2025zaj}, although one should restrict the set of admissible perturbations to, at least, non-conformally flat ones.}, this property is expected to be lost at the second order around these geometries, or already at the first order when expanding around more general backgrounds. This issue, however, is not specific to QTG-TNT theories, but likely generic to other QTG theories. 

We plan to come back to these issues in future work.

\begin{acknowledgments}
We thank Valeri Frolov, \'Angel Murcia, Robie Hennigar, David Kubiz\v{n}\'ak, and Breno L. Giacchini for stimulating and helpful discussions. A.C. and I.K. acknowledge financial support from the Primus grant PRIMUS/23/SCI/005 of Charles University and the Charles University Research Center grant UNCE24/SCI/016. T.M. is supported by the Czech Science Foundation (GA\v{C}R) grant No.~25-15544S.
\end{acknowledgments}

\appendix

\section{Curvature of TNT geometries --- spin coefficients, algebraic types, and invariants}\label{ap:CurvOfTNT}

We study the algebraic properties of the curvature tensor associated with the general metric \eqref{eq:ansatz} invariant under the symmetries of spherical, hyperbolic, and planar Taub--NUT and Schwarzschild geometries, labeled [4,3,1--6] in the Hicks notation, using the Newmann--Penrose formalism. This formalism is based on the introduction of a null tetrad that consists of two real null covectors ($\bs{l}$, $\bs{n}$) and a complex-conjugate pair of null covectors ($\bs{m}$, $\bar{\bs{m}}$). Here, we use the normalization convention $\bs{l} \cdot \bs{n}^\sharp = -1$, $\bs{m} \cdot \bar{\bs{m}}^\sharp = 1$, so that the metric can be expressed in the form  
\begin{equation}\label{eq:NPframemetric}
    \bs{g} = -\bs{l} \vee \bs{n} + \bs{m} \vee \bar{\bs{m}}\,.
\end{equation}
By comparing with equation $\eqref{eq:ansatz}$, we can identify a natural choice of null tetrad 
\begin{equation}\label{eq:NPtetrad}
\begin{aligned}
    \bs{l} &= \pm \sqrt{\frac{\pm ab}{2}} \left( \bs{\mathrm{d}}t + 2 n \bs{\omega}_k - \frac{d \pm \sqrt{d^2 + b}}{ab} \, \bs{\mathrm{d}}r \right)\; , \\
    \bs{n} &= \sqrt{\frac{\pm ab}{2}} \left( \bs{\mathrm{d}}t + 2 n \bs{\omega}_k - \frac{d \mp \sqrt{d^2 + b}}{ab} \, \bs{\mathrm{d}}r \right)\; , \\
    \bs{m} &= \sqrt{\frac{c}{2}} \left( \sqrt{\frac{1}{1 - k\rho^2}}\, \bs{\mathrm{d}}\rho + i \rho\, \bs{\mathrm{d}}\varphi \right)\; ,
\end{aligned}
\end{equation}
where the upper (lower) signs corresponds to the case where $a$ is positive (negative). Among the 12 complex spin coefficients in the NP formalism, only the following associated with the natural null tetrad are non-vanishing: 
\begin{equation}
\begin{aligned}
    \rho &= \pm \mu = \mp \frac{c'}{2 \, c} \sqrt{\frac{\pm ab}{2(d^2 + b)}} - i \frac{n}{c} \sqrt{\frac{\pm ab}{2}}\;, \quad
    \beta = - \alpha = \frac{1}{2\rho} \sqrt{\frac{1-k \rho^2}{2c}}\;, \\
    \epsilon &= \pm \gamma = \frac{(ab)'}{4} \frac{1}{\sqrt{\pm 2 ab (d^2 + b)}} - i \frac{n}{2c} \sqrt{\frac{\pm ab}{2}}\;.
\end{aligned}
\end{equation}
$\bs{l}^\sharp$, $\bs{n}^\sharp$ are tangent vectors to null geodesics ($\kappa = 0$, $\nu = 0$), but not affinely parametrized ($\epsilon + \bar\epsilon \neq 0$, $\gamma + \bar\gamma \neq 0$). $\sigma$ and $\lambda$ corresponds to shear of $\bs{l}^\sharp$ and $\bs{n}^\sharp$, respectively. $\rho$, $\mu$ encode the expansion and twist of $\bs{l}^\sharp$ and $\bs{n}^\sharp$, respectively. 

The Riemann curvature tensor $\bs{R}$ can be irreducibly decomposed into the Ricci scalar $R$, the TF Ricci tensor $\bs{S}$ and the completely TF Weyl tensor $\bs{C}$, as follows:
\begin{equation}\label{eq:irreddecomp}
    \bs{R} = \frac{R}{D(D-1)} \bs{g} \varowedge \bs{g} + \frac{2}{D-2} \bs{S} \varowedge \bs{g} + \bs{C}\;.
\end{equation}
Here $D$ denotes the spacetime dimension ($D=4$ in our case), and $\varowedge$ is the Kulkarni--Nomizu product of symmetric rank-2 tensors $\bs{s}$ and $\bs{t}$, defined by $(\bs{s} \varowedge \bs{t})_{abcd} = s_{a[c} t_{d]b} - s_{b[c} t_{d]a}$.
The five independent components of the curvature tensor $\bs{R}$ in spacetimes with hyperbolic, spherical, and planar Taub--NUT symmetries [4,3,\{2,4,5\}] are fully encoded in the Ricci scalar $R$, two real components of the TF Ricci tensor $\bs{S}$, namely $\Phi_{00} = \Phi_{22}$ and $\Phi_{11}$, and a single complex component $\Psi_2$ of the Weyl tensor $\bs{C}$. Thus, the trace-free Ricci tensor is of Segre type \{1,1(11)\} and of null alignment type G, while the Weyl tensor is of the Petrov type D. Specifically,
\begin{equation}\label{eq:SdecompTaubNUT}
    \bs{S} = 2\Phi_{00}\, \bs{n}^2 + 2 \Phi_{11}\,(\bs{l} \vee \bs{n} + \bs{m} \vee \bar{\bs{m}}) + 2 \Phi_{22}\, \bs{l}^2
\end{equation}
and
\begin{equation}\label{eq:CdecompTaubNUT}
    \bs{C} = \Psi_2 \left((\bs{n} \wedge \bs{l} - \bs{m} \wedge \bar{\bs{m}})^2 - (\bs{n} \wedge \bs{m}) \vee (\bs{l} \wedge \bar{\bs{m}}) \right) + \text{c.c.}\;,
\end{equation}
where the scalars are given by
\begin{equation}
\begin{aligned}
    R &= \frac{ab}{c^2} \left( \frac{(c')^2}{2\gamma} + 2n^2 \right) - \frac{2\gamma(ab)'' - \gamma'(ab)'}{2\gamma^2} - \frac{2ab'c'}{\gamma c} + b \frac{ \gamma'(ac') - 2\gamma (ac')'}{\gamma^2 c} + \frac{2k}{c}\;, \\
    \Phi_{00} &= \Phi_{22} = \pm \frac{ab}{c^2} \left( \frac{(\gamma cc')' - 3\gamma cc''}{8\gamma^2} + \frac{n^2}{2} \right)\;, \\
    \Phi_{11} &= \frac{ab}{c^2} \left( -\frac{(c')^2}{16\gamma} + \frac{3n^2}{4} \right) + \frac{2\gamma (ab)'' - \gamma'(ab)'}{16 \gamma^2} + \frac{k}{4c}\;, \\
    \Psi_2 &= \frac{ab}{c^2} \left( \frac{(c')^2}{12\gamma} - \frac{2n^2}{3} \right) + \frac{2\gamma(ab)'' - \gamma'(ab)'}{24\gamma^2} - \frac{ab'c'}{12\gamma c} + b \frac{ \gamma'(ac') - 2\gamma (ac')'}{24\gamma^2 c} - \frac{k}{6c} \mp i \frac{n}{2\sqrt\gamma} \left(\frac{ab}{c}\right)'\;,
\end{aligned}
\end{equation}
with ${\gamma = b + d^2}$.
In the limit $n = 0$, corresponding to hyperbolic, spherical, and planar Schwarzschild symmetries [4,3,\{1,3,6\}], the number of independent components reduces to four, since the Weyl scalar $\Psi_2$ becomes real.

The invariants \eqref{eq:5Invariants} can be expressed in terms of the five NP scalars (recalling that $\Phi_{00} = \Phi_{22}$) as
\begin{equation}\label{eq:R_AinNP}
    \begin{aligned}
    \mathcal{R}_0 &= \frac{R}{3} + 2 (\Psi_2 + \bar{\Psi}_2)\;, \\
    \mathcal{R}_1 &= 2i (\Psi_2 - \bar{\Psi}_2)\;, \\
    \mathcal{R}_2 &= \frac{R}{6} - 4 \Phi_{11} - 2 (\Psi_2 + \bar{\Psi}_2)\;, \\
    \mathcal{R}_3 &= \frac{R}{6} + 4 \Phi_{11} - 2 (\Psi_2 + \bar{\Psi}_2)\;, \\
    \mathcal{R}^2_4 &= 4 \Phi_{00}\Phi_{22} = 4 \Phi_{00}^2\;.    
    \end{aligned}
\end{equation}
Clearly, the Segre type of the trace-free Ricci tensor specializes to \{(1,1)(11)\} and its null alignment type to type D if and only if $\Phi_{00} = \Phi_{22} = 0$, i.e., $\mathcal{R}_4 = 0$. In the gauge \eqref{eq:gauge}, this is moreover equivalent to the single-function ansatz defined by $b=1$, see Section \ref{sc:QTGpropdef}.

The double Wick rotations studied in \cite{Colleaux:2025uiw} and summarized in Sec.~\ref{sc:wickrot} interchanges the time and spatial coordinates. In the natural null frame \eqref{eq:NPtetrad} of the original metric, the 1-form of the time coordinate appears only in the null pair $\bs{l}$, $\bs{n}$, while the 1-form of the spacial coordinate being Wick-rotated contributes only to the pair $\bs{m}$, $\bar{\bs{m}}$. This allows us to choose a natural null tetrad for the Wick-rotated metric, $\bs{l}'$, $\bs{n}'$, $\bs{m}'$, $\bar{\bs{m}}'$, which is related to the natural null tetrad of the original metric by the transformation $(\bs{l},\bs{n},\bs{m},\bar{\bs{m}}) \to i(\bs{m}',\bar{\bs{m}}',\bs{n}',\bs{l}')$. The relations between the NP scalars for the original and Wick-rotated metrics follow immediately from their definitions, and the non-vanishing scalars of the Wick-rotated metric are
\begin{equation}
\begin{aligned}
    &\pi'=-i\bar\rho\;, \quad \tau'=-i\bar\mu\;, \quad \epsilon'=-i\bar\alpha\;, \quad \gamma'=-i\bar\beta\;, \quad \alpha'=-i\bar\epsilon\;, \quad \beta'=-i\bar\gamma\;, \\
    &\Phi'_{02} = \Phi'_{20} = - \Phi_{00} = - \Phi_{22} \;, \quad \Phi_{11}' = -\Phi_{11}\;, \qquad \Psi'_2 = \bar\Psi_2\;.
\end{aligned}
\end{equation}
The Wick-rotated metrics belong to the Kundt class of spacetimes, since $\bs{l}'^\sharp$ and $\bs{n}'^\sharp$ are tangent to the respective congruences of non-expanding, non-shearing, and non-twisting null geodesics. The Weyl type is preserved under this double Wick rotation, whereas the null alignment type of the trace-free Ricci tensor is of special type D in general, in contract to the original metrics, where it is generically of type G and specializes to type D only in the single-function case.

\section{Proof of non-analyticity for QTG-TNT with second- and first-order field equations}\label{ap:ProofNAN}

In this section, we demonstrate that the Lagrangian of any theory that depends linearly on the component $\mathcal{R}_4$ of the Riemann tensor, when evaluated for the general symmetry-invariant metric \eqref{eq:ansatz} of TNT geometries, cannot be analytic (and therefore cannot be polynomial) in the Riemann tensor. To prove this, we proceed by contraposition and show that if the Lagrangian were analytic in the Riemann tensor, then it could not depend linearly on $\mathcal{R}_4$. In fact, it follows that $\mathcal{R}_4$ appears only with even powers in any Lagrangian that is analytic in the Riemann tensor and its dual for the metric \eqref{eq:ansatz}.

Let us start by considering a scalar curvature invariant $\mathcal{I}$ constructed as a contraction of $n_S$ trace-free Ricci tensors, $n_C$ Weyl tensors, and $n_{\tilde{C}}$ dual Weyl tensors:
\begin{equation}
    \mathcal{I} = [\bs{S}^{\otimes n_S} \otimes \bs{C}^{\otimes n_C} \otimes \bs{\tilde{C}}{}^{\otimes n_{\tilde{C}}}]_{a_1 \ldots a_{2 n_g}}[(\bs{g}^{-1} )^{\otimes n_g}]^{a_{\sigma(1)} \ldots a_{\sigma(2n_g)}} \;.
\end{equation}
Here the tensor product $\bs{S}^{\otimes n_S} \otimes \bs{C}^{\otimes n_C} \otimes \bs{\tilde{C}}{}^{\otimes n_{\tilde{C}}}$ is a covariant tensor of rank $2n_S + 4 n_C + 4 n_{\tilde{C}}$, and $n_g = n_S +2 n_C + 2 n_{\tilde{C}}$ inverse metrics are used to perform all contractions, with the index contraction pattern encoded by the permutation $\sigma$.
Such invariants arise in the irreducible decomposition of scalar curvature-invariant polynomials (or analytic expressions) in the Riemann tensor and its dual, as determined by \eqref{eq:irreddecomp}. As we show below in this section, the invariant $\mathcal{I}$ for metrics of Weyl type D with a trace-free Ricci tensors of the special algebraic form \eqref{eq:SdecompTaubNUT} (having only one component each of boost-weight $-2$, 0, and $+2$ in its tetrad decomposition), which applies to the general symmetry-invariant metric \eqref{eq:ansatz}, can be written in terms of the NP components of the curvature tensors as
\begin{equation}\label{eq:SCInvarDecomp}
    \mathcal{I} = \sum_{p=0}^{\lfloor n_S/2 \rfloor} \sum_{q=0}^{n_C + n_{\tilde{C}}} \hat{c}(p, q) (\Phi_{00} \Phi_{22})^{p} (\Phi_{11})^{n_S - 2p} (\Psi_2 + \bar\Psi_2)^q (i \bar\Psi_2 - i \Psi_2)^{n_C + n_{\tilde{C}} -q} \;,
\end{equation}
with $\hat{c}(p,q)$ being real constants. Notice that the boost-weight $-2$ and $+2$ components, $\Phi_{00}$ and $\Phi_{22}$, always occur in pairs in scalar invariants, as these are necessarily of boost weight 0. Using the inverted relations from \eqref{eq:R_AinNP}, which express $\mathcal{R}_i$ through the tetrad components, the invariant $\mathcal{I}$ takes the form
\begin{equation}\label{eq:IuR}
    \mathcal{I} = \sum_{p=0}^{\lfloor n_S/2 \rfloor} \sum_{q=0}^{n_C + n_{\tilde{C}}} \check{c}(p, q) \mathcal{R}_4^{2p} (\mathcal{R}_3 - \mathcal{R}_2)^{n_S - 2p} (\mathcal{R}_0 - \mathcal{R}_2 - \mathcal{R}_3)^q \mathcal{R}_1^{n_C + n_{\tilde{C}} - q} \;,
\end{equation}
where $\check{c}(p,q) = (-2)^{q - n_C - n_{\tilde{C}}} 6^{-q} 8^{p-n_S} \hat{c}(p, q)$. Therefore, $\mathcal{R}_4$ enters any invariant $\mathcal{I}$ only in even powers, i.e., as $\mathcal{R}_4^{2p}$, which completes the proof.

In what follows, we summarize the derivation of \eqref{eq:SCInvarDecomp}, providing the details underlying the result used above. This allows us to restrict attention to scalar curvature invariants built from contractions of
$\bs{S}$, $\bs{C}$ and $\bs{\tilde C}$, which, due to the decompositions \eqref{eq:SdecompTaubNUT}, \eqref{eq:CdecompTaubNUT}, are polynomial in the frame components:
\begin{equation}
    [\bs{S}^{\otimes n_S} \otimes \bs{C}^{\otimes n_C} \otimes \bs{\tilde{C}}{}^{\otimes n_{\tilde{C}}}]_{a_1 \ldots a_{2 n_g}}[(\bs{g}^{-1} )^{\otimes n_g}]^{a_{\sigma(1)} \ldots a_{\sigma(2n_g)}} = P_{\sigma}(\Phi_{00}, \Phi_{11}, \Phi_{22}, \Psi_2, \bar{\Psi}_2) \;.
\end{equation}
Recall that the number of contractions, i.e., the number of inverse metrics required when all curvature tensors are expressed with covariant indices, $n_g = n_S + 2 n_C + 2 n_{\tilde{C}}$, is given by the number of $\bs{S}$ and twice the number of Weyl tensors.
Various possible configurations of contractions are encoded by $\sigma \in \mathrm{S}_{2n_g}$, an arbitrary permutation of the total $2 n_g$ indices of the inverse metrics. For a given configuration $\sigma$, unless it vanishes due to the contraction of a traceless tensor or by symmetries (such as contracting the symmetric tensor $\bs{S}$ with an antisymmetric pair of indices of $\bs{C}$), the polynomial $P_\sigma$ takes the form
\begin{equation}
    P_\sigma(\Phi_{00}, \Phi_{11}, \Phi_{22}, \Psi_2, \bar{\Psi}_2) = \sum_{p_i, q_j} c(p_i,q_j) \Phi_{00}^{p_1} \Phi_{11}^{p_2} \Phi_{22}^{p_3} \Psi_2^{q_1} \bar\Psi_2^{q_2} \;,
\end{equation}
where the $c(p_i, q_j)$ are generally complex constants (although the individual terms may be complex, the polynomial $P_\sigma$ as a whole is real). Obviously, the sum of the powers of $\Phi_{00}$, $\Phi_{11}$ and $\Phi_{22}$ in each term of the polynomial equals the number of $\bs{S}$, $\sum p_i=n_S$, and the sum of the powers of $\Psi_2$ and $\bar\Psi_2$ in each term is given by the number of $\bs{C}$ and $\bs{\tilde C}$, $\sum q_i=n_C+n_{\tilde{C}}$.

The frame decomposition of the metric \eqref{eq:NPframemetric} is invariant with respect to the 6-parameter Lorentz group $\mathrm{SO}(1,3)$ of frame transformations, consisting of spatial rotations in the plane $\mathrm{span}\{\bs{m},\bar{\bs{m}}\}$, null rotations about a fixed $\bs{l}$ or $\bs{n}$, and boosts in the plane $\mathrm{span}\{\bs{l}, \bs{n}\}$:
\begin{equation}
    \bs{l} \longmapsto A \, \bs{l}\;, \quad \bs{n} \longmapsto A^{-1} \, \bs{n}\;, \quad A>0 \;.
\end{equation}
A scalar quantity $q$ is said to have boost weight $b$ if it transforms under boosts as
\begin{equation}
    q \longmapsto A^b q \;.
\end{equation}
Scalar invariants must have zero boost weight. In other words, all occurrences of $\bs{l}$ in such an expression must be contracted with an equal number of $\bs{n}$ (and similarly, $\bs{m}$ with $\bar{\bs{m}}$). Consequently, the frame components $\Phi_{00}$ and $\Phi_{22}$, with boost weights 2 and $-2$, respectively, can only appear in pairs, since all the remaining components $\Phi_{11}$, $\Psi_2$, and $\bar{\Psi}_2$ have boost weight zero and there is no other possibility to compensate for the imbalance.

Under the duality operation on 2-forms, the Weyl tensor splits into its self-dual and anti-self-dual parts. Notice that in the frame decomposition \eqref{eq:CdecompTaubNUT}, the component $\Psi_2$ appears in the self-dual part, while its complex conjugate $\bar\Psi_2$ appears in the anti-self-dual part. Each $\bs{C}$ contributes one $\Psi_2$ and one $\bar\Psi_2$ to the scalar invariant, and analogously, each dual $\tilde{\bs{C}}$ contributes $-i\Psi_2$ and $i\bar\Psi_2$. Hence,
\begin{equation}
    P_\sigma = \sum_{k=0}^{n_C + n_{\tilde{C}}} P'_{k}(\Phi_{00}\Phi_{22}, \Phi_{11}) \Psi_2^k \bar\Psi_2^{n_C + n_{\tilde{C}}-k} \, ,
\end{equation}
where $P'_k$ is a complex-valued polynomial that collects all signs and imaginary units. Since $P_\sigma$ is real-valued, the reality condition $\bar{P_\sigma} = P_\sigma$ implies that $P'_k = \bar{P}'_{n_C + n_{\tilde{C}}-k}$ must necessarily hold. Any real-valued polynomial $Q(z, \bar{z})$ of a complex variable $z$ can be expressed as a polynomial $\hat{Q}(z+\bar{z}, i(\bar{z} - z))$ in the real and imaginary parts of $z$, and we can thus conclude that 
\begin{equation}
\begin{aligned}
    P_\sigma &= P_\sigma(\Phi_{00}\Phi_{22}, \Phi_{11}, \Psi_2 + \bar\Psi_2, i \bar\Psi_2 - i \Psi_2) \\
    &= \sum_{p=0}^{\lfloor n_S/2 \rfloor} \sum_{q=0}^{n_C + n_{\tilde{C}}} \hat{c}(p, q) (\Phi_{00} \Phi_{22})^{p} (\Phi_{11})^{n_S - 2p} (\Psi_2 + \bar\Psi_2)^q (i \bar\Psi_2 - i \Psi_2)^{n_C + n_{\tilde{C}} -q} \;,
\end{aligned}
\end{equation}
with $\hat{c}(p,q)$ being real constants. This completes the derivation of \eqref{eq:SCInvarDecomp}.

In the remainder of this appendix, we present certain contraction patterns of Weyl tensors that ensure that the resulting invariants do not mix the self-dual and anti-self-dual parts. Since such invariants appear in QTG–TNT theories, these contraction patterns will be used in Appendix \ref{ap:covrep} to construct covariant representatives.
The Weyl tensor and its dual can be viewed as operators on the space of 2-forms, $\bs{C}: \Lambda^2 \to \Lambda^2$, mapping a 2-form $\bs{\omega}$ as $\omega_{ab} \mapsto C_{ab}{}^{cd} \omega_{cd}$. Under the duality operation on 2-forms, the space $\Lambda^2$ splits orthogonally as $\Lambda^2 = \Lambda_+^2 \oplus \Lambda_-^2$ into the direct sum of the spaces of self-dual and anti-self-dual 2-forms. The Weyl tensor preserves this decomposition by acting separately on each subspace,  i.e., it splits into its self-dual and anti-self-dual parts, $\bs{C} = \bs{C}^+ + \bs{C}^-$.
One can define projection operators onto the subspaces $\Lambda^2_+$ and $\Lambda^2_-$ as
\begin{equation}
    P^\pm_{ab}{}^{cd} = \frac12 \left( \delta^{cd}_{ab} \pm \frac{i}{2} \varepsilon_{ab}{}^{cd} \right)\;,
\end{equation}
where the generalized Kronecker delta $\delta_{ab}^{cd}$ acts as the identity operator on the space of 2-forms $\Lambda^2$. These projectors $\bs{P}^\pm$ are indeed idempotent, orthogonal and complete on $\Lambda^2$: 
\begin{equation}
    \bs{P}^\pm \circ \bs{P}^\pm = \bs{P}^\pm, \quad \bs{P}^\pm \circ \bs{P}^\mp = 0, \quad \bs{P}^+ + \bs{P}^- = \bs{1}_{\Lambda^2} \;.
\end{equation}
Note that the composition $\circ$ is understood as index contraction, i.e.\ $(\bs{P}^\pm \circ \bs{P}^\pm)_{ab}{}^{cd} = \bs{P}^\pm_{ab}{}^{ef} \bs{P}^\pm_{ef}{}^{cd}$.
It follows that the self-dual part of the Weyl tensor is given by $\bs{C}^+ = \bs{P}^+ \circ \bs{C}$ and contains only the Newman--Penrose components $\Psi_i$, while the anti-self-dual part $\bs{C}^- = \bs{P}^- \circ \bs{C}$ contains only their complex conjugates $\bar\Psi_i$. Analogously, for the dual tensor $\bs{\tilde C}^\pm = \bs{P}^\pm \circ \bs{\tilde C} = \mp i \bs{C}^\pm$, the components transform under duality as $\Psi_j \to -i \Psi_j$. Therefore, the dual Weyl tensor for the metric \eqref{eq:ansatz} decomposes as
\begin{equation}\label{eq:dualCdecompTaubNUT}
    \bs{\tilde{C}} = -i\Psi_2 \left((\bs{l} \wedge \bs{n} - \bs{m} \wedge \bar{\bs{m}})^2 - (\bs{l} \wedge \bs{m}) \vee (\bs{n} \wedge \bar{\bs{m}}) \right) + \text{c.c.}
\end{equation}

Now, employing the irreducible decomposition of the Weyl tensor and the fact that $\bs{P}^\pm$ act as identity operators on $\Lambda^2_\pm$, respectively, we can write 
\begin{equation}
    \bs{C} = \bs{C}^+ + \bs{C}^- = \bs{P}^+ \circ \bs{C}^+ \circ \bs{P}^+ + \bs{P}^- \circ \bs{C}^- \circ \bs{P}^- \;,
\end{equation}
Due to the properties of the projectors, when Weyl tensors are contracted over antisymmetric index pairs, their self-dual and anti-self-dual parts do not mix. In particular, consider the following composition involving Weyl and dual Weyl operators:
\begin{equation}\label{eq:Weylopcomposition}
    \underbrace{\bs{C} \circ \bs{C} \circ \cdots \circ \bs{C}}_\text{$k$ times} \circ \underbrace{\bs{\tilde C} \circ \bs{\tilde C} \circ \cdots \circ \bs{\tilde C}}_\text{$l$ times} = (-i)^l \, \underbrace{\bs{C}^+ \circ \bs{C}^+ \circ \cdots \circ \bs{C}^+}_\text{$k+l$ times} + \ i^l \, \underbrace{\bs{C}^- \circ \bs{C}^- \circ \cdots \circ \bs{C}^-}_\text{$k+l$ times}\;,
\end{equation}
where the left-hand side represents a general contraction pattern, since the operators $\bs{C}$ and $\bs{\tilde C}$ commute, $\bs{C} \circ \bs{\tilde C} = \bs{\tilde C} \circ \bs{C}$, as a direct consequence of the fact that the left and right duals of the Weyl tensor are equal, $\bs{\tilde C} = \star \bs{C} = \bs{C} \star$.
In the special case when the scalar invariant $P_\sigma$ factorizes into separate scalar terms and all Weyl tensors within such a factor are contracted over antisymmetric index pairs as described above, their contribution to this factor takes the form
\begin{equation}\label{eq:Weylcompsnotmixed}
    (-i)^l \left[\Psi_2^{k+l} + (-1)^l \bar\Psi_2^{k+l}\right]\;,
\end{equation}
with $k$ and $l$ denoting the numbers of Weyl tensors $\bs{C}$ and dual Weyl tensors $\bs{\tilde C}$, respectively, in this factor. 

Let us illustrate these general results with two explicit curvature invariants, each expressed first in terms of the real and imaginary parts of $\Psi_2$ (i.e.\ in the form of \eqref{eq:SCInvarDecomp}), and then expanded  as an explicit polynomial in frame components. An example of an invariant mixing the self-dual and anti-self-dual parts of the Weyl tensor, and consequently mixing $\Psi_2$ and $\bar\Psi_2$ in the frame-component polynomial, is
\begin{equation}
\begin{aligned}
    S^{b}{}_{f} S^{h}{}_{l} S_{ke} C^{a}{}_{h}{}^{kl} C_{ab}{}^{cd} C_{cd}{}^{ef} &= - 48\, \Phi_{00}\Phi_{11}\Phi_{22} \left(\Psi_2 + \bar\Psi_2\right)^3\left[\left(\Psi_2 + \bar\Psi_2\right)^2 - \left(i\Psi_2 - i \bar\Psi_2\right)^2\right] \\ 
    &= - 96 \,\Phi_{00}\Phi_{11}\Phi_{22} \left( \Psi_2^3 + \Psi_2^2 \bar\Psi_2 + \Psi_2 \bar\Psi_2^2 + \bar\Psi_2^3 \right)\;.
\end{aligned}
\end{equation}
Whereas an invariant that does not mix the self-dual and anti-self-dual parts is, for example (Weyl tensors contracted in accordance with \eqref{eq:Weylopcomposition}),
\begin{equation}
\begin{aligned}
    S^{ah} S_{hg} \tilde{C}_{ab}{}^{cd} C_{cd}{}^{ef} C_{ef}{}^{gb}
    &= 24 \left(2 \Phi_{11}^2 + \Phi_{00} \Phi_{22}\right) \left(i \Psi_2 - i \bar\Psi_2\right) \left[3\left(\Psi_2 + \bar\Psi_2\right)^2 - \left(i \Psi_2 - i \bar\Psi_2\right)^2 \right] \\
    &= 96 i \left(2 \Phi_{11}^2 + \Phi_{00} \Phi_{22}\right)\left(\Psi_2^3 - \bar\Psi_2^3\right)\;,
\end{aligned}
\end{equation}
where the frame-component polynomial is manifestly of the form \eqref{eq:Weylcompsnotmixed} with $k=2$ and $l=1$.

\section{Covariant polynomial representatives of third-order Class II QTG–TNT}\label{ap:covrep}

Let us now focus on finding covariant polynomial representatives of Class II QTG–TNT theories \eqref{eq:LagGQTG} with \eqref{CouplingFunctTNUT} at a fixed order $p$ in the curvature tensors, expressed in terms of the irreducible components $R$, $\bs{S}$, $\bs{C}$ and $\tilde{\bs{C}}$. In view of \eqref{eq:RicciScalarinRi} and \eqref{eq:IuR}, we first rewrite the reduced theories into a form more convenient for this task,
\begin{equation}\label{eq:RTirrcompform}
    \sum_{\alpha+\beta+\gamma+\delta=p} c(\alpha, \beta, \gamma, \delta) (2\mathcal{R}_0 + \mathcal{R}_2 + \mathcal{R}_3)^\alpha (\mathcal{R}_3 - \mathcal{R}_2)^\beta (\mathcal{R}_0 - \mathcal{R}_2 - \mathcal{R}_3)^\gamma \mathcal{R}_1^\delta \;,
\end{equation}
as this makes manifest that a given term in the reduced theory arises only from curvature invariants of order $\alpha$ in the Ricci scalar $R$, order $\beta$ in the trace-free Ricci tensor $\bs{S}$, and order $\gamma+\delta$ in the Weyl tensor $\bs{C}$ and its dual $\tilde{\bs{C}}$. It also allows the independent construction of covariant representatives for individual summands with fixed $\alpha$ and $\beta$ when $\gamma + \delta=0$, and for combinations of summands with fixed $\alpha$, $\beta$, $\gamma+\delta$ otherwise. A specific covariant representative is then chosen heuristically among the possible scalar invariants constructed from the corresponding numbers of $R$, $\bs{S}$, $\bs{C}$, and $\tilde{\bs{C}}$ with different index-contraction patterns, so that it reduces exactly to the desired form for the SF ansatz \eqref{eq:SFansatz}. Note that there are two sources of ambiguity in this procedure. On one hand, different index-contraction patterns do not necessarily lead to independent scalar curvature invariants, due to the index symmetries and tracelessness of the trace-free Ricci and Weyl tensors. On the other hand, independent invariants can still reduce, for the SF ansatz, to equivalent expressions up to an overall constant factor.

At order $p=2$, there are two QTG–TNT theories, one even-parity and one odd-parity,
\begin{equation}
    \mathcal{L}_{(2)}^{(II)} = \zeta_2 \mathcal{L}_{(2)}^\text{even} + \xi_2 \mathcal{L}_{(2)}^\text{odd} \;,
\end{equation}
where $\zeta_2$, $\xi_2$ are real coupling constants defined in \eqref{eq:couplingconsts} and
\begin{equation}
    \mathcal{L}_{(2)}^\text{even} = \frac12 \mathcal{R}_0^2 - \frac32 \mathcal{R}_1^2 + \mathcal{R}_2 \mathcal{R}_3\;,
    \qquad \mathcal{L}_{(2)}^\text{odd} = \mathcal{R}_1 (\mathcal{R}_0 -  \mathcal{R}_2 -  \mathcal{R}_3)\;.
\end{equation}
These theories can be cast into the form \eqref{eq:RTirrcompform} and then, using \eqref{eq:R_AinNP}, rewritten in terms of the NP components as
\begin{equation}
    \mathcal{L}_{(2)}^\text{even} = \frac{R^2}{12} - 16 \Phi_{11}{}^2 + 12 (\Psi_2{}^2 + \bar{\Psi}_2{}^2)\;,
    \qquad \mathcal{L}_{(2)}^\text{odd} = 12i (\Psi_2{}^2 -  \bar{\Psi}_2{}^2)\;.
\end{equation}
The scalars $R^2$, $S_{ab}S^{ab}$, $C_{abcd}C^{abcd}$, $\tilde{C}_{abcd}C^{abcd}$ form a complete basis for quadratic scalar curvature invariants, since any contraction of $\bs{C}\bs{C}$ or $\tilde{\bs{C}}\tilde{\bs{C}}$ can be expressed, up to an overall constant factor, as $C_{abcd}C^{abcd}$, while any contraction of $\tilde{\bs{C}}\bs{C}$ can similarly be expressed as $\tilde{C}_{abcd}C^{abcd}$. Therefore, for $p=2$, no ambiguity arises from different independent curvature invariants reducing to the same reduced theory, and the unique covariant representatives read 
\begin{equation}
\begin{aligned}
    \mathcal{L}_{(2)\text{rep}}^\text{even} &= \frac{R^2}{12} - S_{ab} S^{ab} + \frac12 C_{abcd} C^{abcd} = - \frac18 \delta_{abcd}^{efgh} R^{ab}{}_{ef} R^{cd}{}_{gh}\;, \\
    \mathcal{L}_{(2)\text{rep}}^\text{odd} &= -\frac12 \tilde{C}_{abcd} C^{abcd} = -\frac14 \varepsilon_{abef} R^{ef}{}_{cd} R^{abcd}\;,
\end{aligned}
\end{equation}
which correspond to the Gauss–Bonnet and Chern–Pontryagin topological invariants, respectively, as already mentioned in Section \ref{sc:GRandtopolinv}.
Note that while the odd-parity covariant representative $\mathcal{L}_{(2)\text{rep}}^\text{odd}$ reduces to the same reduced theory $\mathcal{L}_{(2)}^\text{odd}$ also for the general TNT metric \eqref{eq:ansatz}, the reduction of the even-parity covariant representative $\mathcal{L}_{(2)\text{rep}}^\text{even}$ on the general TNT metric differs from its reduction on the SF ansatz by the term quadratic in $\mathcal{R}_4$,
\begin{equation}
     \mathcal{L}_{(2)\text{rep}}^\text{even}\big|_\text{TNT} = \mathcal{L}_{(2)}^\text{even} - 2 \mathcal{R}_4^2\;, \qquad \mathcal{L}_{(2)\text{rep}}^\text{odd}\big|_\text{TNT} = \mathcal{L}_{(2)}^\text{odd}\;.
\end{equation}

Analogously, for $p=3$, one finds two theories
\begin{equation}
    \mathcal{L}_{(3)}^{(II)} = \xi_3 \mathcal{L}_{(3)}^\text{even} + \zeta_3 \mathcal{L}_{(3)}^\text{odd}\;,
\end{equation}
with
\begin{equation}
\begin{aligned}
    \mathcal{L}_{(3)}^\text{even} &= -\frac14 (\mathcal{R}_0^3 - 11 \mathcal{R}_0 \mathcal{R}_1^2) + \frac14 (\mathcal{R}_0^2 - 5 \mathcal{R}_1^2)(\mathcal{R}_2 + \mathcal{R}_3) - \mathcal{R}_0 \mathcal{R}_2 \mathcal{R}_3 \\
    &= -\frac{R^3}{108} + \frac{16R}{3} \Phi_{11}{}^2 + 32 \Phi_{11}{}^2 (\Psi_2{} + \bar{\Psi}_2{}) - 4 R (\Psi_2{}^2 + \bar{\Psi}_2{}^2) - 56 (\Psi_2{}^3 + \bar{\Psi}_2{}^3)\;, \\
    \mathcal{L}_{(3)}^\text{odd} &= \frac54 \mathcal{R}_0^2 \mathcal{R}_1 - \frac74 \mathcal{R}_1^3 - \frac32 \mathcal{R}_0 \mathcal{R}_1 (\mathcal{R}_2 + \mathcal{R}_3) + \mathcal{R}_1 \mathcal{R}_2 \mathcal{R}_3 \\
    &= 32i \Phi_{11}{}^2 (\bar{\Psi}_2 - \Psi_2) + 4i R (\Psi_2{}^2 - \bar{\Psi}_2{}^2) + 56i (\Psi_2{}^3 - \bar{\Psi}_2{}^3)\;.
    \end{aligned}
\end{equation}
Each term of the reduced theories arises uniquely as the reduction of an independent invariant, except for the terms cubic in the Weyl tensor. In this case, two independent parity-even invariants exist, $C_{ab}{}^{cd} C_{cd}{}^{ef} C_{ef}{}^{ab}$ and $C_a{}^c{}_b{}^d C_c{}^e{}_d{}^f C_e{}^a{}_f{}^b$, together with two independent parity-odd invariants, $\tilde{C}_{ab}{}^{cd} C_{cd}{}^{ef} C_{ef}{}^{ab}$ and $\tilde{C}_a{}^c{}_b{}^d C_c{}^e{}_d{}^f C_e{}^a{}_f{}^b$. However, in four dimensions, they are related by the dimension-dependent identities $C_{ab}{}^{cd} C_{cd}{}^{ef} C_{ef}{}^{ab} = 2 C_a{}^c{}_b{}^d C_c{}^e{}_d{}^f C_e{}^a{}_f{}^b$ and $\tilde{C}_{ab}{}^{cd} C_{cd}{}^{ef} C_{ef}{}^{ab} = 2 \tilde{C}_a{}^c{}_b{}^d C_c{}^e{}_d{}^f C_e{}^a{}_f{}^b$. 
We therefore choose the covariant representatives to take the form
\begin{equation}\label{eq:covrep3}
\begin{aligned}
    \mathcal{L}_{(3)\text{rep}}^\text{even} &= -\frac{R^3}{108} + 12 R S_{ab}S^{ab} - 36 S^{ac} S^{bd} C_{abcd} - 6 R C_{abcd} C^{abcd} + 42 C_{ab}{}^{cd} C_{cd}{}^{ef} C_{ef}{}^{ab}\;, \\
    \mathcal{L}_{(3)\text{rep}}^\text{odd} &= - \frac{R}3 \tilde{C}^{abcd} C_{abcd} - 2 S^{ac} S^{bd} \tilde{C}_{abcd} + \frac73 \tilde{C}_{ab}{}^{cd} C_{cd}{}^{ef} C_{ef}{}^{ab}\;.
\end{aligned}
\end{equation}
Einsteinian cubic gravity is recovered from the covariant representative $\mathcal{L}_{(3)\text{rep}}^\text{even}$ upon adding a term cubic in the TF Ricci tensor
\begin{equation}
    8 R_a{}^b R_b{}^c R_c{}^a - 12 R^{ac}R^{bd}R_{abcd} + R_{ab}{}^{cd} R_{cd}{}^{ef} R_{ef}{}^{ab} + 12 R_a{}^c{}_b{}^d R_c{}^e{}_d{}^f R_e{}^a{}_f{}^b = \frac16 \mathcal{L}_{(3)\text{rep}}^\text{even}\big|_{\alpha=\frac{3}{21}} + 8 S_a{}^b S_b{}^c S_c{}^a\;,  
\end{equation}
whose reduction
\begin{equation}
    S_a{}^b S_b{}^c S_c{}^a\big|_\text{TNT} = \frac32 (\mathcal{R}_2 - \mathcal{R}_3)\mathcal{R}_4^2
\end{equation}
yields a topological QTG-TNT theory.
Note that the TNT reductions of the covariant representatives again reproduce the reduced theories modulo terms quadratic in $\mathcal{R}_4$,
\begin{equation}
     \mathcal{L}_{(3)\text{rep}}^\text{even}\big|_\text{TNT} = \mathcal{L}_{(3)}^\text{even} + \frac94 (R - \mathcal{R}_0 + 2 (\mathcal{R}_2 - \mathcal{R}_3)) \mathcal{R}_4^2\;, \qquad \mathcal{L}_{(3)\text{rep}}^\text{odd}\big|_\text{TNT} = \mathcal{L}_{(3)}^\text{odd} - \frac18 \mathcal{R}_1 \mathcal{R}_4^2\;.
\end{equation}

The QTG-TNT theories at quartic order ($p=4$) take the form
\begin{equation}
    \mathcal{L}_{(4)}^{(II)} = \zeta_4 \mathcal{L}_{(4)}^\text{even} + \xi_4 \mathcal{L}_{(4)}^\text{odd}\;,
\end{equation}
where
\begin{equation}
\begin{aligned}
    \mathcal{L}_{(4)}^\text{even} &= - \frac{5}{24} \mathcal{R}_0^4 + \frac{17}4 \mathcal{R}_0^2 \mathcal{R}_1^2 - \frac{15}{8} \mathcal{R}_1^4 + \frac13 (\mathcal{R}_0^3 - 9 \mathcal{R}_0 \mathcal{R}_1^2) (\mathcal{R}_2 + \mathcal{R}_3)  - (\mathcal{R}_0^2 - \mathcal{R}_1^2) \mathcal{R}_2 \mathcal{R}_3
     \\    
     &= -\frac{R^4}{648} + \frac{4 R^2}{9} \bigl( 4 \Phi_{11}{}^2 - 3 (\Psi_2{}^2 +  \bar{\Psi}_2{}^2)\bigr)
     + \frac{16R}{3} \bigl(4 \Phi_{11}{}^2 (\Psi_2{} + \bar{\Psi}_2{}) - 7 (\Psi_2{}^3 + \bar{\Psi}_2{}^3)\bigr) \\
    &\quad 
     + 128 \Phi_{11}{}^2 (\Psi_2{}^2 + \bar{\Psi}_2{}^2) - 240 (\Psi_2{}^4 + \bar{\Psi}_2{}^4)\;, \\
    \mathcal{L}_{(4)}^\text{odd} &= - \frac32 \mathcal{R}_0^3 \mathcal{R}_1 + \frac{29}{6} \mathcal{R}_0 \mathcal{R}_1^3 + \frac{2}{3} (3 \mathcal{R}_0^2 \mathcal{R}_1 - 2 \mathcal{R}_1^3) (\mathcal{R}_2 + \mathcal{R}_3) - 2 \mathcal{R}_0 \mathcal{R}_1 \mathcal{R}_2 \mathcal{R}_3 \\
    &= - \frac{4R^2}{3} i (\Psi_2{}^2 - \bar{\Psi}_2{}^2)
    + \frac{16R}{3} i \bigl( 4 \Phi_{11}{}^2 (\Psi_2{} - \bar{\Psi}_2{}) - 7 (\Psi_2{}^3 - \bar{\Psi}_2{}^3) \bigr)
    + 128 i \Phi_{11}{}^2 (\Psi_2{}^2 -  \bar{\Psi}_2{}^2) \\
    &\quad - 240i (\Psi_2{}^4 - \bar{\Psi}_2{}^4)\;.
\end{aligned}    
\end{equation}
Unlike the previous cases, quartic and higher orders may exhibit another kind of ambiguity due to the growing number of independent invariants that reduce to equivalent expressions for the general TNT metric \eqref{eq:ansatz}, especially those involving Weyl tensors.
We therefore do not attempt to determine the most general covariant representatives in terms of a complete set of independent invariants. Instead, we explicitly construct a single representative reproducing each reduced theory.
Since the Weyl contributions entering the reduced theories take the form \eqref{eq:Weylcompsnotmixed}, the corresponding covariant representatives are sought directly in the form \eqref{eq:Weylopcomposition}. Explicitly, we find
\begin{equation}\label{eq:covrep4}
\begin{aligned}
    \mathcal{L}_{(4)\text{rep}}^\text{even} &= - \frac{R^4}{648}
    + \frac{R^2}{9} S_{a}{}^{b} S_{b}{}^{a}
    - \frac{R^2}{18} C_{abcd} C^{abcd}
    - \frac{2R}{3} S^{ac} S^{bd} C_{abcd}
    + \frac{7R}{9} C_{ab}{}^{cd} C_{cd}{}^{ef} C_{ef}{}^{ab} \\
    &\quad
    + 4 S^{ab} S^{cd} C_{acef} C_{bd}{}^{ef}
    - \frac{5}{6} C_{ab}{}^{cd} C_{cd}{}^{ef} C_{ef}{}^{gh} C_{gh}{}^{ab}\;, \\
    \mathcal{L}_{(4)\text{rep}}^\text{odd} &= \frac{R^2}{18} \tilde{C}^{abcd} C_{abcd}
    + \frac{2R}{3} S^{ac} S^{bd} \tilde{C}_{abcd}
    - \frac{7R}{9} \tilde{C}_{ab}{}^{cd} C_{cd}{}^{ef} C_{ef}{}^{ab}
    - 2 S^{ab} S_{cd} \tilde{C}^{cedf} C_{aebf} \\
    &\quad + \frac56 \tilde{C}_{ab}{}^{cd} C_{cd}{}^{ef} C_{ef}{}^{gh} C_{gh}{}^{ab}\;.
\end{aligned}
\end{equation}

Similarly, at order $p=5$, the even- and odd-parity QTG-TNT theories
\begin{equation}
    \mathcal{L}_{(5)}^{(II)} = \xi_5 \mathcal{L}_{(5)}^\text{even} + \zeta_5 \mathcal{L}_{(5)}^\text{odd}
\end{equation}
are given by
\begin{equation}
\begin{aligned}
    \mathcal{L}_{(5)}^\text{even}
    &= \frac15 \mathcal{R}_0^5 - 6 \mathcal{R}_0^3 \mathcal{R}_1^2 + 7 \mathcal{R}_0 \mathcal{R}_1^4
    - \frac18 (3 \mathcal{R}_0^4 - 42 \mathcal{R}_0^2 \mathcal{R}_1^2 + 11 \mathcal{R}_1^4) (\mathcal{R}_2 + \mathcal{R}_3)
    + (\mathcal{R}_0^3 - 3 \mathcal{R}_0 \mathcal{R}_1^2) \mathcal{R}_2 \mathcal{R}_3 \\
    &= \frac{R^5}{3240} - \frac{4}{27} R^3 \bigl(4 \Phi_{11}{}^2 - 3 (\Psi_2{}^2 + \bar{\Psi}_2{}^2)\bigr) - \frac83 R^2 \bigl(4 \Phi_{11}{}^2 (\Psi_2{} + \bar{\Psi}_2{}) - 7 (\Psi_2{}^3 + \bar{\Psi}_2{}^3)\bigr) \\
    &\quad - 16 R \bigl(8 \Phi_{11}{}^2 (\Psi_2{}^2 + \bar{\Psi}_2{}^2) - 15 (\Psi_2{}^4 + \bar{\Psi}_2{}^4)\bigr) - \frac{128}{5} \bigl(20 \Phi_{11}{}^2 (\Psi_2{}^3 + \bar{\Psi}_2{}^3) - 39 (\Psi_2{}^5 + \bar{\Psi}_2{}^5)\bigr)\;, \\
    \mathcal{L}_{(5)}^\text{odd} &= -\frac74 \mathcal{R}_0^4 \mathcal{R}_1 + \frac{19}{2} \mathcal{R}_0^2 \mathcal{R}_1^3 - \frac{39}{20} \mathcal{R}_1^5
     + \frac12 (5 \mathcal{R}_0^3 \mathcal{R}_1 - 9 \mathcal{R}_0 \mathcal{R}_1^3) (\mathcal{R}_2 + \mathcal{R}_3)
     - (3 \mathcal{R}_0^2 \mathcal{R}_1 - \mathcal{R}_1^3) \mathcal{R}_2 \mathcal{R}_3 \\
    &= - \frac49 i R^3 (\Psi_2{}^2 -  \bar{\Psi}_2{}^2)
     + \frac83 R^2 \bigl(4i \Phi_{11}{}^2 (\Psi_2{} -  \bar{\Psi}_2{}) - 7i (\Psi_2{}^3 -  \bar{\Psi}_2{}^3)\bigr) \\
    &\quad + 16 R \bigl(8i \Phi_{11}{}^2 (\Psi_2{}^2 -  \bar{\Psi}_2{}^2) - 15i (\Psi_2{}^4 -  \bar{\Psi}_2{}^4)\bigr)
     + 512i \Phi_{11}{}^2 (\Psi_2{}^3 -  \bar{\Psi}_2{}^3) - \frac{4992}{5} i (\Psi_2{}^5 -  \bar{\Psi}_2{}^5)\;.
\end{aligned}
\end{equation}
We again construct only a single covariant representative for each theory, using curvature invariants where the Weyl tensors are contracted over antisymmetric pairs of indices only, as motivated by \eqref{eq:Weylcompsnotmixed}. This yields
\begin{equation}\label{eq:covrep5}
\begin{aligned}
    \mathcal{L}_{(5)\text{rep}}^\text{even} &= \frac{R^5}{3240}
    - \frac{R^3 }{27} S_{ab} S^{ab}
    + \frac{R^3 }{54} C_{ab}{}^{cd} C_{cd}{}^{ab}
    + \frac{R^2}{3} S^{ac} S^{bd} C_{abcd}
    - \frac{7R^2}{18} C_{ab}{}^{cd} C_{cd}{}^{ef} C_{ef}{}^{ab} \\
    &\quad
    - \frac{R}{3} S_{gh} S^{gh} C_{ab}{}^{cd} C_{cd}{}^{ab}
    + \frac{5R}{6} C_{ab}{}^{cd} C_{cd}{}^{ef} C_{ef}{}^{gh} C_{gh}{}^{ab}
    + \frac23 S_{gh} S^{gh} C_{ab}{}^{cd} C_{cd}{}^{ef} C_{ef}{}^{ab} \\
    &\quad -  \frac{26}{25} C_{ab}{}^{cd} C_{cd}{}^{ef} C_{ef}{}^{gh} C_{gh}{}^{ij} C_{ij}{}^{ab}\;, \\
    \mathcal{L}_{(5)\text{rep}}^\text{odd} &= \frac{R^3 }{54} \tilde{C}^{abcd} C_{abcd}
    + \frac{R^2}{3} S^{ac} S^{bd} \tilde{C}_{abcd}
    - \frac{7 R^2}{18} \tilde{C}_{ab}{}^{cd} C_{cd}{}^{ef} C_{ef}{}^{ab}
    - \frac{R}{3} S_{gh} S^{gh} \tilde{C}_{ab}{}^{cd} C_{cd}{}^{ab} \\
    &\quad
    + \frac{5R}{6} \tilde{C}_{ab}{}^{cd} C_{cd}{}^{ef} C_{ef}{}^{gh} C_{gh}{}^{ab}
    + \frac23 S_{gh} S^{gh} \tilde{C}_{ab}{}^{cd} C_{cd}{}^{ef} C_{ef}{}^{ab}
    - \frac{26}{25} \tilde{C}_{ab}{}^{cd} C_{cd}{}^{ef} C_{ef}{}^{gh} C_{gh}{}^{ij} C_{ij}{}^{ab}\;.
\end{aligned}
\end{equation}


%

\end{document}